\documentclass{iopjournal}
\usepackage[utf8]{inputenc}

\usepackage{graphicx}
\usepackage{amsmath}
\usepackage[version=4]{mhchem}
\usepackage{siunitx}
\usepackage{longtable,tabularx}
\usepackage{subcaption}
\usepackage{tikz}
\usepackage{comment}
\usepackage{makecell}
\usepackage{ragged2e}
\usepackage{soul}
\usepackage{dirtytalk}
\usepackage[dvipsnames]{xcolor}
\usetikzlibrary{calc} 

\usepackage[english]{babel}
\begin{document}
\justifying
\articletype{Paper} 

\title{Flow behind the Imperial Front Wing: comparison of results from volumetric PTV experiment and Nektar++ simulations}
\footnote{Preprint submitted to Measurement Science and Technology; currently under review.}
\author{Isabella Fumarola$^{1}$\orcid{0000-0002-5503-0229}, Alexandra I. Liosi$^1$\orcid{0009-0006-0296-1712}, Parv Khurana$^{1,}$\orcid{0000-0002-7857-814X}, Adam Meziane$^1$, Isaac Balbolia$^{1}$, Sherwin J. Spencer$^{1}$\orcid{0000-0001-7681-2820}, Jonathan F. Morrison$^{1}$\orcid{0000-0003-3688-9860} }

\affil{$^1$Department of Aeronautics, Imperial College London, London, SW7 2BX, United Kingdom}


\email{isabella.fumarola12@imperial.ac.uk}


\begin{abstract}
\justifying
High-fidelity simulations are increasingly adopted, due to advances in computational power and methods such as Direct Numerical Simulation (DNS) and hybrid Large-Eddy Simulation (LES). These approaches are particularly valuable for unsteady flows around complex geometries at high Reynolds numbers; however they still require careful experimental validation. Planar and stereoscopic Particle Image Velocimetry (PIV) are widely used for measurements, yet they are limited by measurement-plane selection and their ability to capture "vortices" shapes and trajectories. This motivates the growing interest in volumetric techniques, historically difficult to implement in industrial settings. Recent advances in Particle Tracking Velocimetry (PTV) for measuring flows over large volumes make this approach suitable for validating numerical simulations of complex flows. This study compares volumetric PTV measurements against high-fidelity LES to assess the capabilities and limitations for industrial flows. The aim is to establish a benchmark PTV dataset for motorsport aerodynamics using the Shake-The-Box algorithm. The experiment was carried out in the 10x5 wind tunnel at Imperial College London equipped with a rolling road system for ground effect simulation and capable of testing up to 50\% scale Formula 1 models. Volumetric PTV measurements were performed downstream of the open-source Imperial Front Wing (IFW) at a Reynolds number $Re_c$ = 74896. Results are compared with planar PIV studies and with implicit LES simulation using spectral h/p elements in Nektar++. This work addresses open questions in the literature concerning the wake behind the IFW. Good quantitative agreement is observed in the mean wake topology, including the rolling-road boundary layer and secondary vorticity generated by the vortices in ground effect. A previously unreported vortex is identified, influencing the downstream evolution of the primary vortices by preventing the merging of dominant structures. These results demonstrate the suitability of PTV and STB for industrial applications while providing a benchmark dataset for the IFW. 
\end{abstract}
\section*{Nomenclature}
{\renewcommand\arraystretch{1.0}
\noindent\begin{longtable*}{@{}l @{\quad=\quad} l@{}}
 $X,Y,Z$ & streamwise, spanwise and wall-normal coordinates in  full scale in [mm].\\  
$x,y,z$ & streamwise, spanwise and wall-normal coordinates in the half scale in [mm].\\
$x^+,y^+,z^+$ & non dimensional coordinates with respect to $\nu$ and $u_\tau$.\\
$u,v,w$ & "instantaneous" velocity components, non dimensional with respect to $U_\infty$.\\  
$U,V,W$ & average velocity components, non dimensional with respect to $U_\infty$.\\  
$\Omega$ & non-dimensional streamwise vorticity component $\Omega=(\partial W/\partial y-\partial V/ \partial z) \cdot U_\infty/c$.\\
$X^*$ & non-dimensional absolute distance from the reference point ($X^*=(X+300)/c$).\\
$c$ & chord length of the mainplane, $c=0.25$ m full scale.\\ 
$C_p$ & pressure coefficient.\\
$C_L$ & lift coefficient.\\
$C_D$ & drag coefficient.\\
$\Gamma$ & non dimensional circulation.\\
$Re_c$ & Reynolds number based on the chord.\\
$\nu$ & Kinematic viscosity of the fluid [$m^2$/s].\\
$\tau_w$ & wall shear-stress.\\
$u_{\tau_w}$ & friction velocity.\\
\end{longtable*}}
\setcounter{table}{0}
%
\section{Introduction}
The motorsport and automotive industries express a growing interest in high-fidelity representations of high Reynolds number ($Re$) flows. This interest is driven by the need to understand transient, non-linear dynamics through the use of advanced experimental and numerical models \cite{Katz2021}. The current state-of-the-art optical experimental methods in motorsport is two dimensional (2D) Particle Image Velocimetry (PIV). However, this technique is limited by the need to choose the experimental plane, which poses challenges when attempting to capture the evolution of streamwise vortical structures over long distances. Although there have been successful attempts to automate 2D PIV systems, using motorised laser and cameras to capture multiple planes sequentially, these setups are often complex and require a long time to configure. Recent developments of Particle Tracking Velocimetry (PTV) for measuring the flow within large volumes, make this technique particularly suitable for assessing the flow evolution in space. Furthermore, it can be utilised for validating numerical models of complex flows. Finally, the measured volume field can be employed to calibrate the free coefficients of turbulence models specifically designed for this purpose, such as GEKO \cite{Klavaris2024, Menter2025}.

The primary limitation of tomographic PIV lies in the dimension of the achievable measurement volume. In fact, using a typical high-power Nd/YAG laser for PIV (approximately 30 W), the maximum measurable linear dimension is typically of the order of a few centimetres \cite{grille2024scalability}. This restriction arises from the need for intense illumination to scatter light from commonly used small seeding particles (diameter $\approx 1 \mu m$), such as Poly Ethylene Glycol (PEG), Di-Ethyl-Hexyl-Sebacic-Acid-Ester (DEHS), or olive oil. A significant breakthrough in achieving volumetric measurements over large flow domains has been the introduction of pulsed high-power LEDs in combination with Helium-Filled Soap Bubbles (HFSB) \cite{grille2024scalability}. LEDs are relatively inexpensive and easily scalable since they can be stacked to expand the illuminated volume and increase light intensity. Advancements in LED technology now allow high-frequency pulsing (order of kHz) to deliver sufficient energy within the illumination period. However, unlike lasers, LEDs emit diffused, incoherent light. Consequently, the use of LEDs for particle imaging in air has only become feasible by moving from micron-sized droplets to larger, more reflective tracers like HFSB. These bubbles, with an average diameter of approximately 0.3 mm, are made neutrally buoyant by filling them with helium. Their larger size makes them significantly brighter than traditional seeding particles, enabling flow measurements over volumes several orders of magnitude larger than those used in conventional PIV.
In parallel to the hardware advancement, the enabler for the success of volumetric PTV has been the advancement in novel image processing techniques. In particular, the Shake-the-Box (STB) algorithm \cite{schanz2014shake} enables precise tracking of individual particles from tomographic imaging in space at a reduced computational cost. As a result, volumetric PTV has largely superseded tomographic PIV. In some setups, permanent installations of LED arrays and camera systems allow flexible adjustment of the measurement volume, enabling researchers to shift the area of interest between experiments without the need for complex realignment procedures. This flexibility greatly enhances efficiency in wind tunnel testing, making the technique particularly attractive for applications in race-car aerodynamics, in particular for Formula 1 (F1) where wind tunnel running time is limited.

\begin{figure}[!ht]
    \centering
\begin{tikzpicture}
    \node[inner sep=0pt] (img) at (0,0)
{\includegraphics[width=0.6\linewidth]{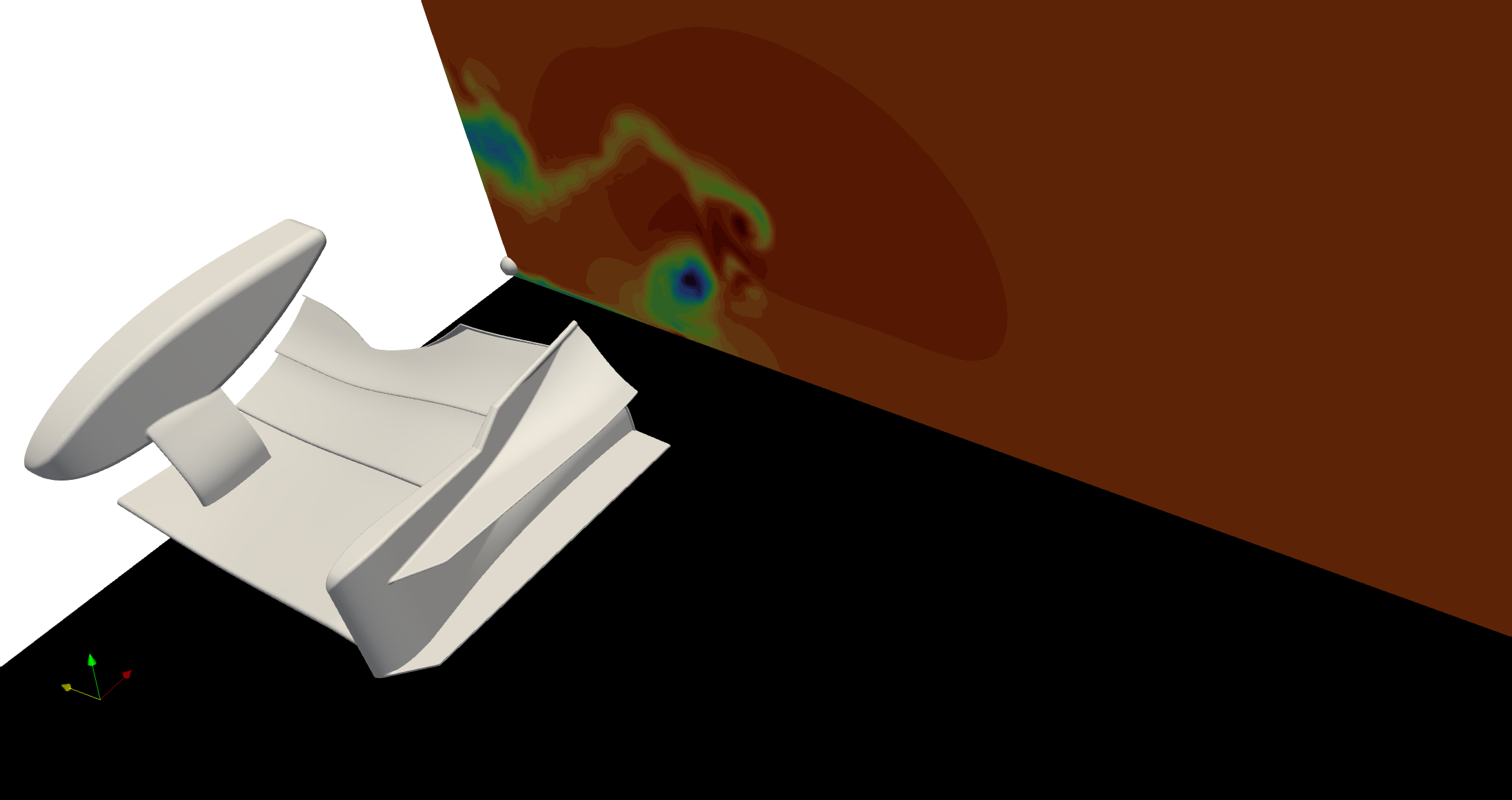}};
    \draw[->, thick, red] (-1.5,0.75) -- (-1,1.2) node[pos=1, below] {\small X};
    \draw[->, thick, red] (-1.5,0.75) -- (-1.7,1.5) node[pos=1, left] {\small Z};
    \draw[->, thick, red] (-1.5,0.75) -- (-2,1) node[pos=1, below] {\small Y};
    \draw [<->, thick, red] (0,0.19) -- (-0.5,-0.3) node[midway, sloped, below] {\small 300 mm};
    \node at (-3.5,0.25) {\small 1};
    \node at (-3.3,-0.35) {\small 2};
    \node at (-2.75,-0.75) {\small 3};
    \node at (-2.25,-0.25) {\small 4};
    \node at (-1.75,0.25) {\small 5};
    \node at (-1.72,-1.2) {\small 6};
    \node at (-2.25,-1.2) {\small 7};
    \node at (-1.2,0) {\small 8};
\end{tikzpicture}
    \caption{Representation of half-body of the Imperial Front Wing and the coordinate system as defined by Buscariolo et al. \cite{Buscariolo2022} (flow left to right). The elements of the body are: 1) Nose Cone, 2) Hanger, 3) Mainplane, 4) First Flap, 5) Second Flap, 6) Footplate, 7) Endplate, 8) Canard, Pink: Rolling road, Blue: Symmetry Plane. The contour plane shows the pressure at the spanwise/wall-normal plane from the LES results.}
    \label{fig:experiments_datum}
\end{figure}
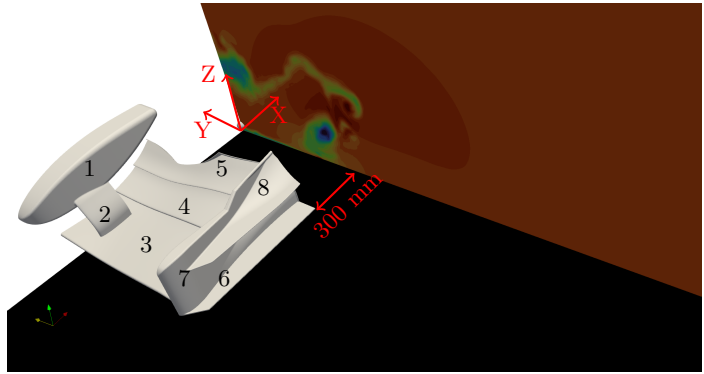

To understand and explore best practices for leveraging the STB method for industrial geometries, we used the Imperial Front Wing (IFW) as a representative benchmark for flow problems in race-car aerodynamics. The IFW is an open-source model of an F1 front wing based on the McLaren 17D race car front end. The detailed CAD is available at \cite{IFWSource}. The geometry consists of a three-element wing with an endplate attached to a simplified nose cone operating in ground effect, as shown in Figure \ref{fig:experiments_datum}. The focus of this work is on on high $Re$ incompressible flows within the context of race-car aerodynamics, mainly because it uses a F1 open source geometry, but the methodology can be extended to any vortex dominated flows with applications to different sectors such as aviation, urban flows or turbomachinery.

Pegrum \cite{Pegrum2006} introduced the IFW test-case and conducted an experimental survey for the vortical system evolution emerging from the IFW wake, an isolated wheel, and their combination using hot-wire anemometry, 2D PIV and laser-smoke visualisation. He described how the complex flow developing from the IFW generates a series of interacting vortical structures in a time-averaged representation. The IFW became open-source in 2018, Buscariolo et al. \cite{Buscariolo2022} repeated an experimental survey only on the IFW using 2D PIV with increased acquisition frequency of 250 Hz. They also simulated the flow around the wing using the high-order Spectral/hp element method and validated their simulation against the time-averaged experimental data. Good qualitative agreement was observed with the largest differences showing between the wing and the rolling road, but information on the transient behaviour of the flow was not available. O'Sullivan et al. \cite{Sullivan2023} investigated different turbulence modelling approaches from low to high fidelity and reported their comparison to the experiments from Buscariolo et al. \cite{Buscariolo2022}.
Follow-up work on the simulations include \cite{Khurana2025, Lombard2017, OSullivan2025}. Liosi et al. \cite{Liosi2024, Liosi2026} repeated the numerical configuration from \cite{Buscariolo2022} and performed a grid-dependency study to establish the requirements for efficiently performing implicit Large Eddy Simulations (iLES) around complex geometries. Good quantitative agreement with the experimental results of \cite{Buscariolo2022} was reported in terms of spanwise and wall-normal velocity components. However, a boundary layer on the rolling road was observed that was absent in \cite{Buscariolo2022}; this remains an open question in the literature. The authors also noted that increasing the sampling frequency of both the simulation field variables and the experimental measurements would be beneficial for analysing the transient behaviour of the wing.

Practitioners aim to simulate the flow around more complex geometries and capture the resulting flow phenomena, such as laminar to turbulent flow transition or separation, vortex shedding, jets and more. Turbulence models are often calibrated for a subset of these flow mechanisms yet struggle to accurately predict the entire range of turbulent behaviour. This limitation has motivated the increasing adoption of Large Eddy Simulation (LES) and hybrid RANS-LES approaches, which resolve the energy-containing turbulent scales directly and are less dependent on empirical calibration than conventional RANS closures. Among LES strategies, implicit LES (iLES) resolves the larger turbulent structures while relying on the numerical scheme's inherent dissipation to model the smallest, unresolved scales rather than resolving them directly, as in DNS \cite{Sagaut2009}.
 The growing use of LES and hybrid LES methods has, in turn, increased the demand for high quality, spatially resolved experimental datasets suitable for validating these approaches.

In this work, we apply for the first time STB to an experimental survey on the wake of the IFW, a representative benchmark of industrial external aerodynamics applications. The experimental results are compared to an equivalent iLES computational model. The front wing configuration is identical to that of Buscariolo et al. \cite{Buscariolo2022} at reduced Re, and the numerical model is validated against the experimental results through a systematic analysis of the time-averaged flow development over an extended streamwise domain. Relative to the configuration of Buscariolo et al. \cite{Buscariolo2022}, the interrogation domain has been expanded to encompass the endplate vortex system and the near-wall region above the rolling road. In addition, the results identify a new vortical structure and describe its role in the flow development.

The paper is organised with the following sections: methodology, results, discussion and conclusion. The methodology section \ref{sec:methods} covers the details of both the simulation and the experiment with information about the flow conditions for the tests. The results section \ref{sec:results} is structured in two parts: the first gives a detailed description on the formation and development of the vortical system that populates the wake of the IFW; the second a quantitative comparison between the simulation and the experimental results. The results demonstrate a good agreement between experiment and simulation throughout the domain. The largest discrepancy is found in the region near the ground, which would require further sensitivity study. In particular, a new vortical structure has been identified in both experiment and simulation results. Its formation and role is briefly described in the result section, and further analysed in the discussion section. The good agreement in terms of mean flow quantities and vortices downstream development between the results offers a benchmark for future study and demonstrates how the two techniques can be applied as complementary methods for several industrial aerodynamic studies.

\section{Methodology}
\label{sec:methods}

\begin{figure}[h]
\centering
\begin{subfigure}[t]{0.44\textwidth}
\includegraphics[width=\textwidth]{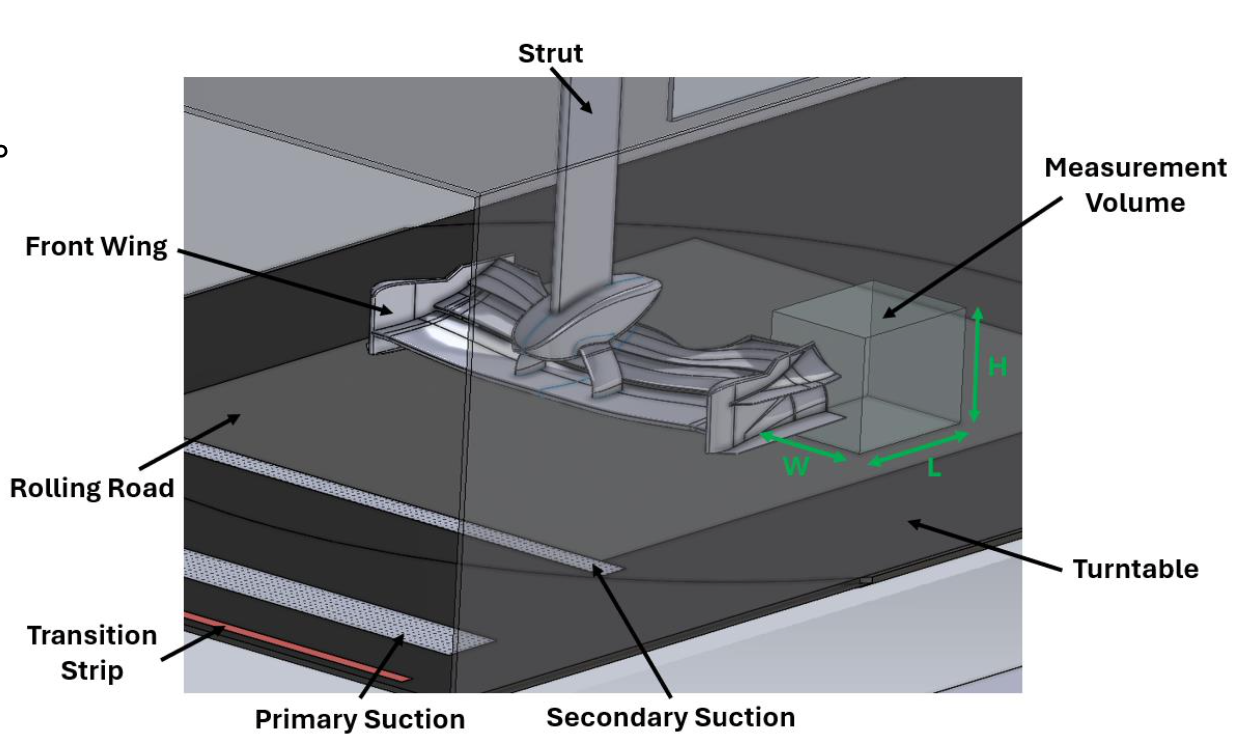}
\caption{}
\end{subfigure}
\begin{subfigure}[t]{0.33\textwidth}
\includegraphics[width=\textwidth]{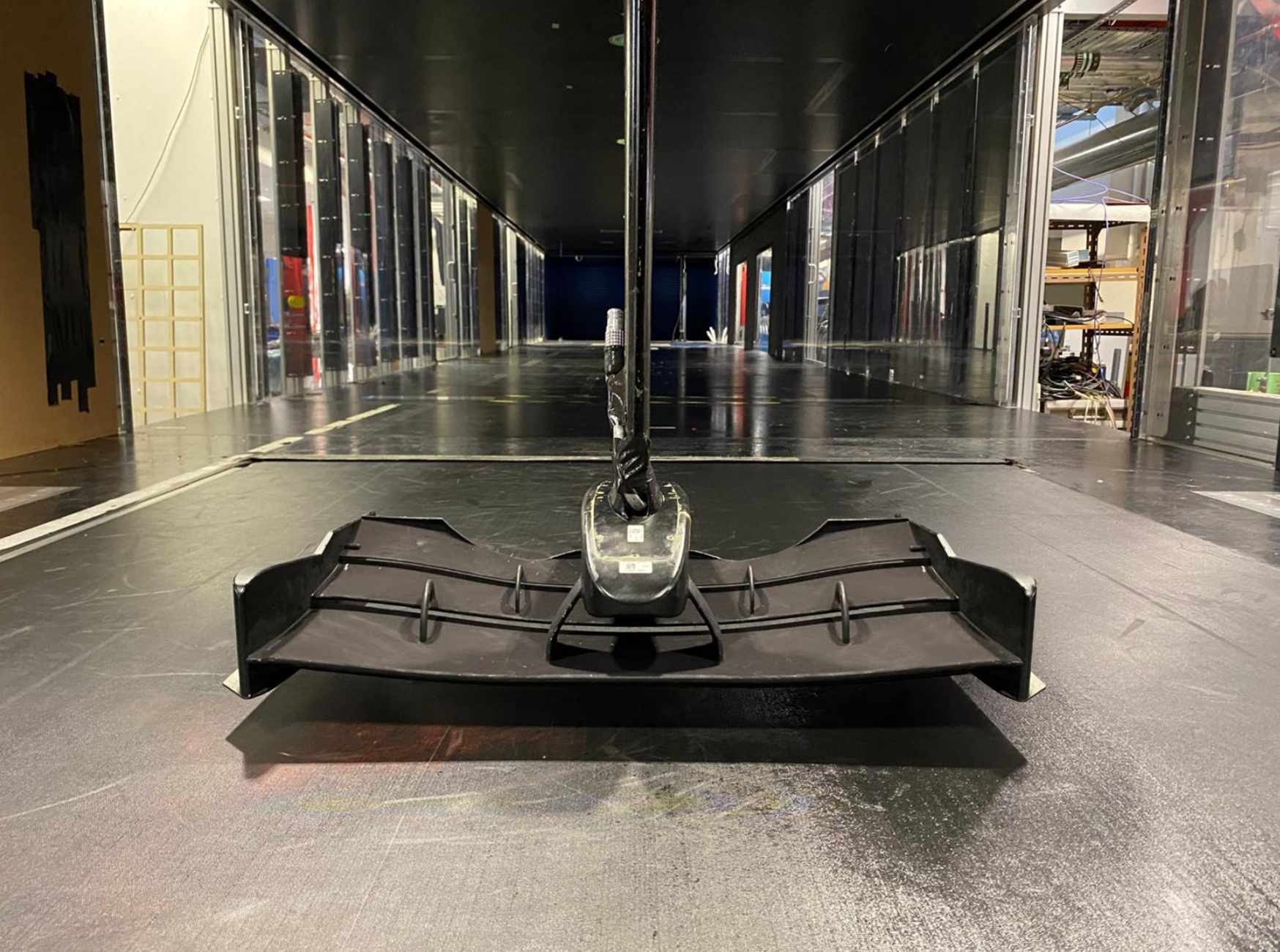}
\caption{}
\end{subfigure}
\begin{subfigure}[t]{0.2\textwidth}
\includegraphics[width=\textwidth]{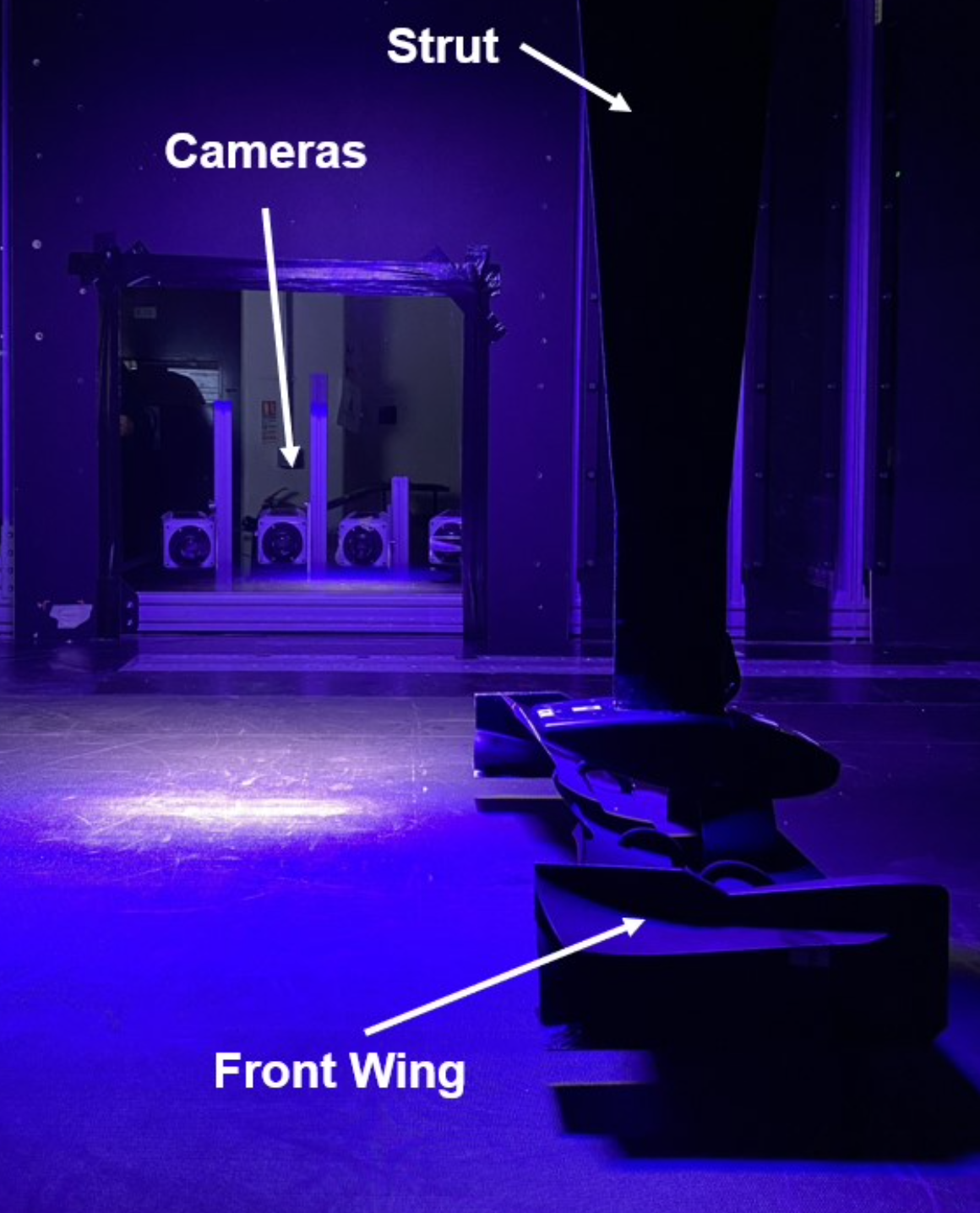}
\caption{}
\end{subfigure}
\caption{Experimental setup of front wing: a) schematic of the setup and volume location (volume not in scale), b) Imperial Front Wing in the wind tunnel, c) cameras and illumination of the measurement volume.}
\label{fig:Exp_setup}
\end{figure}

The flow around the IFW is examined at a $Re_c = 74896$. In the simulation, this corresponds to a freestream velocity $U_\infty = 4.5$ m/s and a characteristic length equal to the chord of the mainplane of $c = 0.25$ m at full scale. In the experiment the wing is a 50\% scale model and therefore to match the Reynolds number the freestream velocity is $U_\infty = 9$ m/s. In both cases, the model of the wing includes a simplified nose. The strut that holds the wing in the experimental setup, as shown in figure \ref{fig:Exp_setup} is not present in the simulation, while the moving wall, used to simulate the ground effect, is defined in both experiment and simulation as discussed later. 
\graphicspath{{./Figures/}}

\subsection{Numerical Configuration}
The flow is governed by the unsteady Navier-Stokes equations for incompressible fluids. The computational model for performing implicit LES was developed using the high-order Spectral/hp element method via \emph{Nektar++} \cite{MoxeyEtAl2020}. \emph{Nektar++} is a framework for solving unsteady, partial differential equations using the high-order Spectral/hp element method. 
High-Order FE methods have proven their ability to accurately simulate complex industrial geometries in previous works \cite{Beck2014, Goc2021,  Khurana2025, Mengaldo2021, Merzari2020, Min2022, Slaughter2022}. They offer superior geometrical flexibility as they combine linear cell refinement, similar to their low-order counterparts, with high-order curvature refinement around the geometry \cite{Kirilov2026}. They can include turbulence modelling or solve directly the conservation laws of unsteady fluid motion to resolve different turbulence scales. The high-order approximation for the field variables, such as pressure and velocity, contributes to reduced numerical diffusion and dispersion errors \cite{Karniadakis2005}. Finally, the inherent nature of the formulation permits the development of efficient implementations for numerical differentiation, interpolation and other operators inside the code, which could potentially enhance its computational performance and parallelism \cite{Vos2010}.

The high-order Spectral/hp element method is employed to discretize this system of equations in space following the continuous Galerkin projection \cite{Karniadakis2005}. The pressure and the velocity fields are approximated by high-order polynomials. The polynomial order of the velocity is four and for pressure is three, ensuring improved stabilization properties for the simulation \cite{Boffi2013}. The high-order velocity correction scheme \cite{Karniadakis1991} is utilized to decouple the velocity from the pressure due to its easier implementation and reduced order of the splitting error. In particular, the \emph{sub-stepping} scheme \cite{Maday1990, Sherwin2003, Wustenberg2026}, or a semi-Lagrangian discretisation was adopted for the time-stepping to enable larger timesteps for the simulation. The pressure field at the next timestep $p^{n+1}$ is found by

    \begin{equation}
        \nabla^2{p^{n+1}} = \frac{1}{\Delta t} \nabla \cdot u^{*}
    \end{equation}
    
\noindent where $u^*$ is an auxiliary, divergence-free field. Similarly, the velocity at the next timestep $ u^{n+1}$ is evaluated by

    \begin{equation}
        \nabla^2 u^{n+1} - \frac{1}{\nu \Delta t} u^{n+1} = \frac{1}{\nu}\nabla{p^{n+1}} - \frac{1}{\nu \Delta t} u^{*}
    \end{equation}
    
\graphicspath{{./Figures/}}
\begin{figure}[h]
    \centering
    \begin{subfigure}[t]{0.44\textwidth}
        \includegraphics[width=\textwidth]{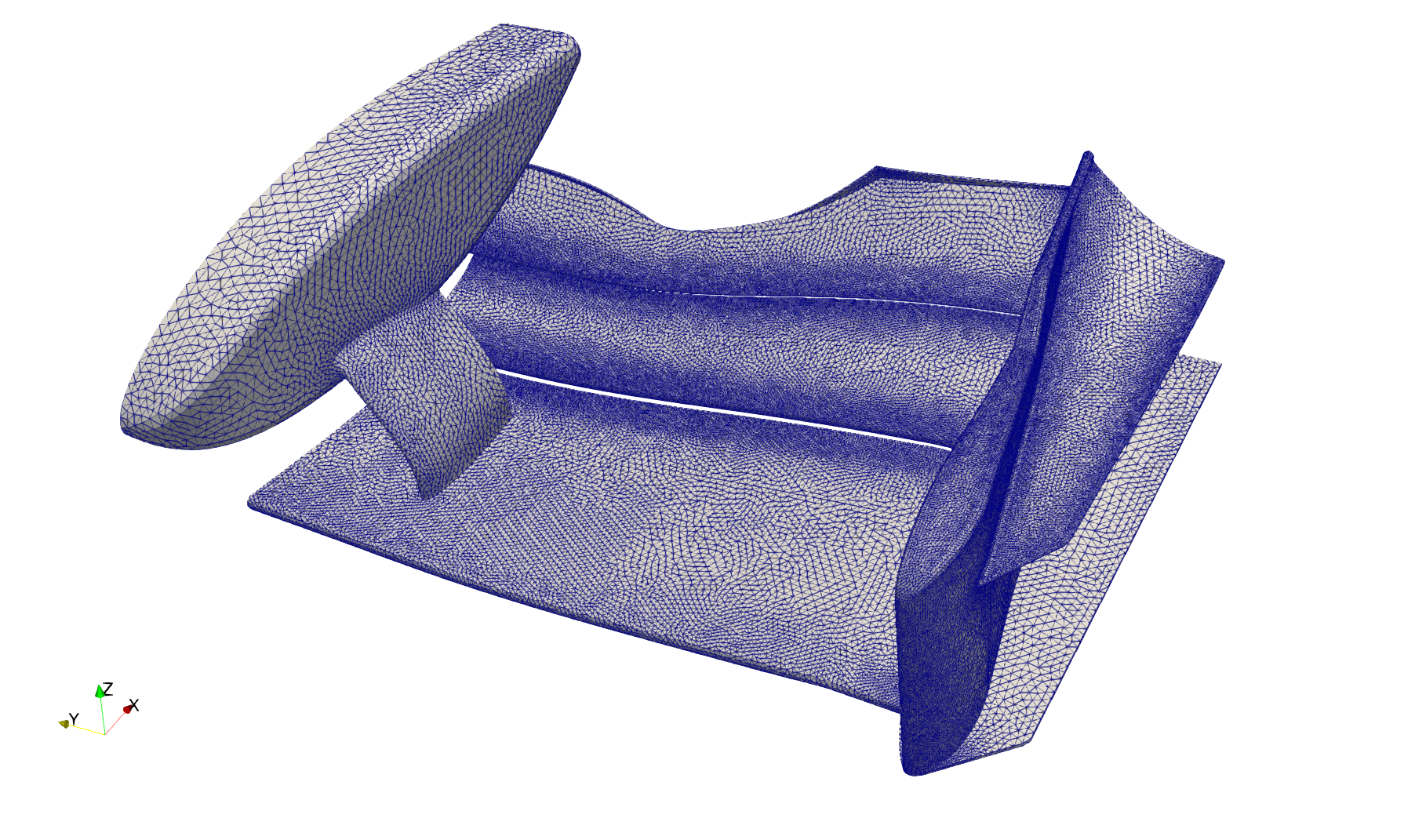}
        \caption{}
    \end{subfigure}
    \begin{subfigure}[t]{0.44\textwidth}
        \includegraphics[width=\textwidth]{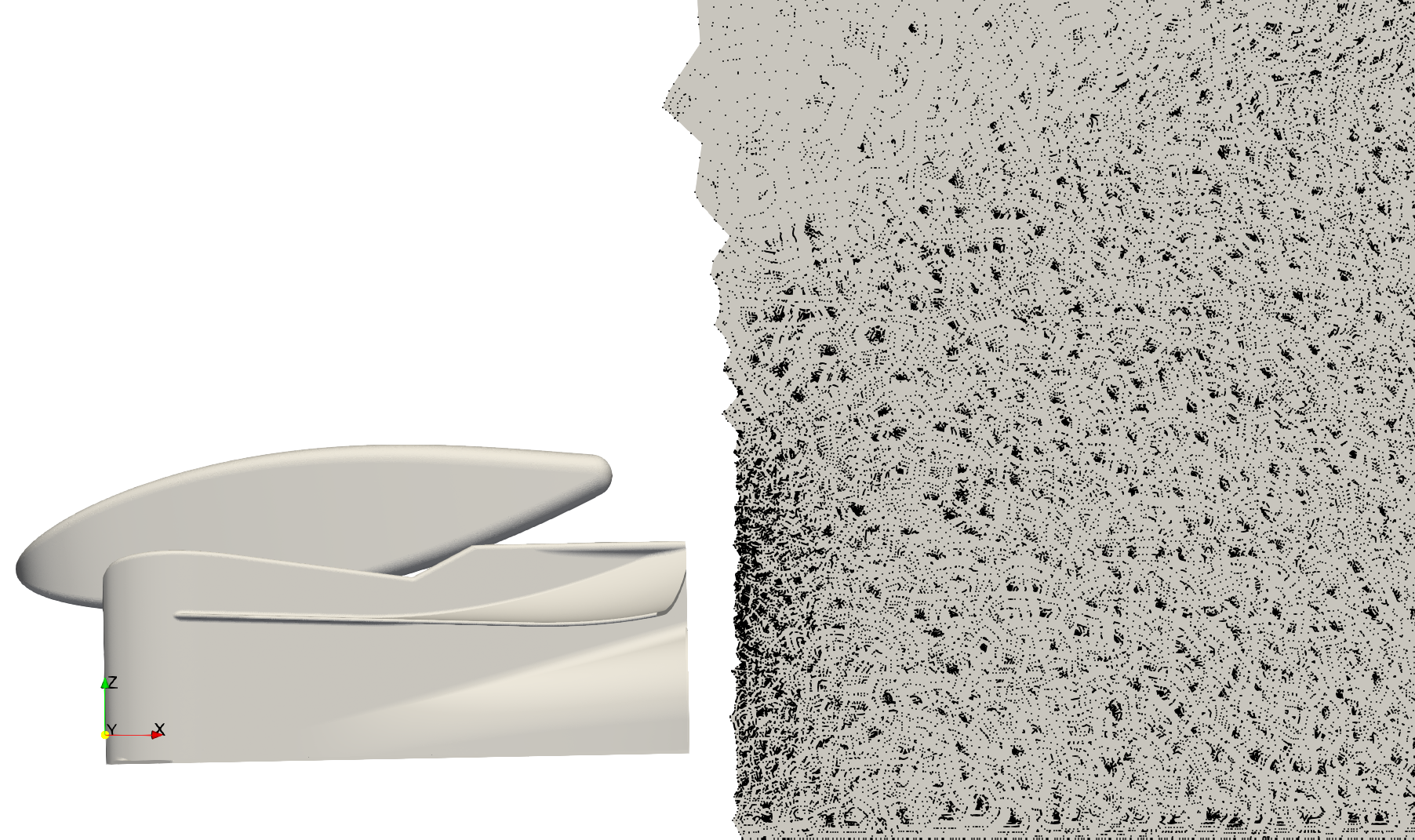}
        \caption{}
    \end{subfigure}
    \caption{Computational mesh of IFW: a) Surface resolution of the linear mesh, b) Volume refinement in the wake of the wing (black points mark the degrees of freedom).}
    \label{fig:mesh}
\end{figure}

\begin{figure}[h]
\centering
\includegraphics[width=\textwidth]{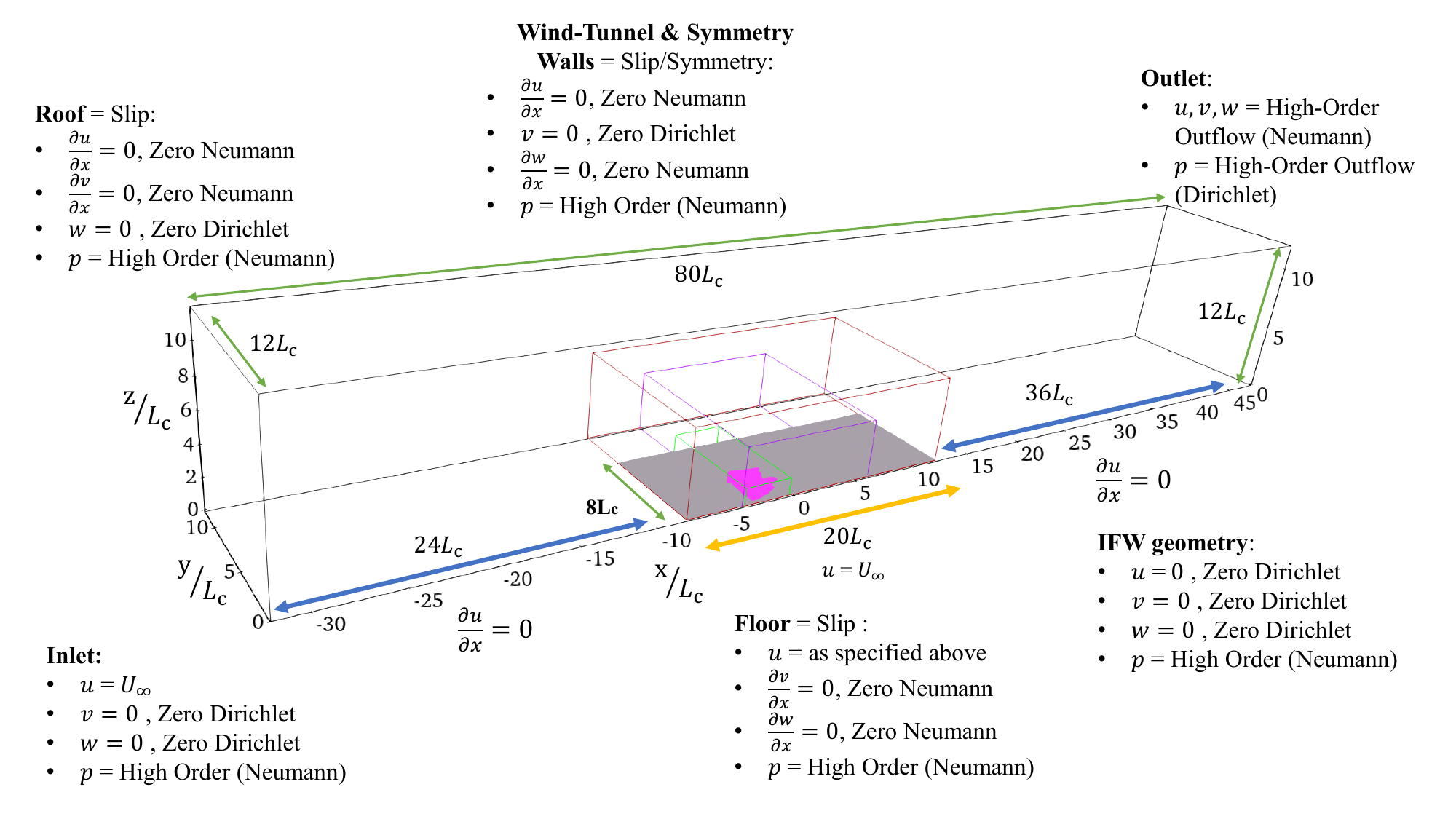}
\caption{Schematic of the computational domain with the boundary conditions}
\label{fig:computational_domain}
\end{figure}

The computational grid is created in two stages, see figure \ref{fig:mesh}. First, the linear mesh is generated by importing the geometry file in .STEP format \cite{IFWSource} to a traditional Finite Volume mesh generator, that produces an unstructured, straight-sided and conformal mesh. The linear mesh includes only prismatic and tetrahedral elements, where a single \say{macro prism-layer} of 2.5 mm total thickness is surrounding the geometry. To ensure better volume mesh quality and smoother transition between different refinement levels, the expansion ratio of the volume mesh was 1.1 to 1.3. The total number of linear mesh elements is $5.3 \cdot 10^{6}$. Second, the linear mesh is converted to a compatible format for Nektar++, using \emph{NekMesh} \cite{Green2023}, the mesh generation utility of Nektar++. During the second stage, the \emph{Industrial Pipeline} is utilized to obtain the high-order mesh \cite{Kirilov2026}. 

In the \emph{Industrial Pipeline}, the first step is to deform the boundaries of the linear mesh using high-order polynomials to match the original geometry expressed in .STEP format. The boundaries of the curved mesh are expressed using fourth order polynomials, which is a good compromise between geometrical accuracy \cite{Kirilov2026} and numerical stability, as the polynomial order of the geometry needs to be equal, or lower than the polynomial order of the solution to avoid geometrical aliasing errors. The quality of the deformed mesh is assessed based on the scaled Jacobian of the mapping between the ideal and curved elements.

The computational domain, illustrated in figure \ref{fig:computational_domain}, is constructed to emulate the IFW configuration within the wind tunnel. Its dimensions are adopted directly from the wind-tunnel test section and are described in Section \ref{experiment}. The freestream air enters the domain and is applied as a uniform Dirichlet boundary condition at the inlet. The ceiling and the side walls are treated as slip walls where the gradient of the planar velocity components is zero, and the normal velocity component is set to zero to prevent cross-flow development. A high-order stable form of zero Neumann boundary condition at the outlet is used for velocity, and the pressure is fixed to the ambient pressure \cite{Dong2014}. The floor is split into three parts. The first and the third parts are modelled as slip walls to prevent the boundary layer build-up from the inlet. The second part corresponds to the rolling road below the geometry and is a moving wall with the same velocity as the freestream.

The simulation is initialised from an existing velocity field obtained via the two-equation k-$\epsilon$ RANS model, and the pressure is set to zero.  The discretized system of equations for velocity and pressure is solved numerically using an iterative method, the Conjugate Gradient (CG) algorithm. It involves applying static condensation to reduce the number of algebraic degrees of freedom \cite{ELWilson1974}. An absolute tolerance of $1 \cdot 10^{-4}$ is set for all the flow variables to ensure algebraic convergence. A second-order accurate time-integration scheme is utilized with a constant time-step of $1 \cdot 10^{-4}$ s. The total degrees of freedom for the streamwise velocity component is $198.5 \cdot 10^{6}$.

The simulations were performed on ARCHER2 \cite{ARCHER2}, the UK National Supercomputing Service. ARCHER2 is an HPE Cray EX system comprising 5860 compute nodes, each equipped with dual 64-core AMD EPYC 7742 processors and either 256 GB or 512 GB of DDR4 RAM. The compute nodes are interconnected via an HPE Cray Slingshot interconnect arranged in a dragonfly topology. The simulations were executed on 48 fully populated compute nodes, employing 128 tasks per node. The average computational cost per timestep was $83.68$ s, corresponding to an average Courant–Friedrichs–Lewy (CFL) number of 13.

The flow of air around the IFW was simulated for a total duration of 3 s, or 12 \emph{Convective Time Unit} (CTU). The characteristic time, or CTU, is defined as the time required for the flow to travel one characteristic length with the freestream velocity. The temporal convergence of the flow is assessed by monitoring the evolution of the integral values over time, as illustrated in the appendix (see figure \ref{fig:forcesTraces}). To determine the starting time of this converged state (end of initial transients) an additional criterion is applied, known as the Mean Squared Error Rule (MSER) \cite{Bergmann2022}. The time-averaging and data-sampling of the simulation begins at T = 8.5 CTUs, making the duration of the data-acquisition is 3.5 CTUs, or 0.875 s.

\subsection{Experiment}
\label{experiment}
The experiment was carried out in the lower test section of the 10x5 wind tunnel at Imperial College London \cite{Imperial10x5}. The facility has a lower test section of 3 m x 1.5 m, 20 m in length, equipped with a rolling road, which is essential to simulate the effect of the ground proximity. Two suction systems ahead of the rolling road, shown in figure \ref{fig:Exp_setup}, are used to remove the boundary layer developing from the contraction. The wind tunnel and rolling road are able to operate up to a maximum speed of 40 m/s.

The model of the IFW in 1:2 scale was mounted on top of the rolling road at the desired height of $h/c = 0.36$ from the ground. That is the ride height ($h$) measured from the ground to the footplate trailing edge and non dimensionalised based on the main element chord ($c$) as specified in previous studies \cite{Pegrum2006}. The wing is further characterized by a pitch angle of $1.094^\circ$ to reproduce the configuration used in previous experiments \cite{Buscariolo2022}. The configuration is matched by the simulation.

To capture the flow over a large volume behind the IFW, a combination of LED lights and Helium-Filled Soap Bubbles (HFSB) were used to carry out volumetric PTV. The setup is shown in figure \ref{fig:Exp_setup}. Two LaVision LED-Flashlight 300 were mounted side-by-side on the roof, each of them consisting of 72 high-power LEDs with a small divergence angle of $±5^\circ$. Helium-Filled Soap Bubbles (HFSB) were inserted upstream of the model using seven Linear Nozzle Arrays (LNA), each of them producing about 280,000 bubbles/second. The particles size was between 9 and 13 $\mu$m with a relaxation time $\tau_p\approx 10 \mu$m \cite{scarano2015use}. Considering the time scaled of the flow defined as $\tau_f=c/2U_{\infty,exp}$, where $c/2$ the chord of the half scale model and $U_{\infty,exp}$ the freestream speed of the experiment, the Stokes number of the particle results $St=\tau_p/\tau_f\approx7.9 \times 10^{-4}$ which confirms excellent tracer fidelity for the flow \cite{scarano2015use}. Due to the large wind tunnel contraction and the presence of the suction system used to remove the boundary layer generated on the floor of the wind tunnel, the LNA could not be mounted in the contraction of the wind tunnel, but they were mounted immediately downstream of the primary suction system. Measurements of the freestream flow at the model location with the LNA in the empty test section are available. To account for errors due to the LNA blockage the effective freestream speed at the measurements location has been used to non-dimensionalise the following data.

Four high-speed Phantom v641 cameras were mounted in line outside of the wind tunnel test section, at a working distance of approximately 1 m. Each camera was equipped with Nikon 50 mm f/1.4D lens. The system provides a measurement volume of 510 mm ($x$ streamwise) $\times$ 585 mm ($y$ vertical) $\times$ 304 mm ($z$, spanwise). The experiment was run at 9 m/s in single-frame mode using a sampling frequency of $f_s=1454.5 Hz$ for a total sampling time of 2 s, see section \ref{sec:convergence} for a typical convergence plot. A Pitot tube was mounted in front of the model in the freestream to monitor the wind tunnel speed, which was matched by the rolling road. The suction systems were kept on for the whole duration of the experiment to guarantee the absence of a boundary layer on the ground.

The geometrical configuration between this experiment and the previous from Buscariolo et al.\cite{Buscariolo2022} is identical. The latter used stereo PIV in five discrete streamwise locations, but published only spanwise and wall-normal velocity components. In the present work, not only three-velocity components are presented in the full volume, but also the field of view has been extended in the wall-normal direction and the acquisition frequency is $\times$5.8 higher with respect to Buscariolo et al. \cite{Buscariolo2022} to resolve, in time, a larger portion of the turbulence spectrum. The duration of the data-acquisition was short due to limitations in the available hardware memory. The $Re_c$ is also lower compared to Buscariolo et al., but it was matched by the new simulation.\\
\subsection{Data processing}

The calibration, acquisition and data processing were carried out using the LaVision software DaVis 10.
The calibration used the live calibration option with a two-dimensional target moved around the field of view. Care was taken to capture enough images to cover the entire measurement volume. A 3rd order polynomial fit model was applied, which resulted in a fit error on the order of 0.02 px.
Each image was pre-processed with a sequence of filters to improve the particle detectability and suppress the background reflections. In particular, the background was effectively reduced by subtracting the minimum over time, considering a window of 27 images, and subtracting a sliding minimum of 5 pixels per image. A normalisation with local average helped equalising the particle intensity, although the illumination was generally quite uniform with the exception of the region next to the rolling road. Finally, a combination of Gaussian smoothing and sharpening filter improved the particle appearance and each pixel was multiplied by a factor of 10 to improve the dynamic range.

\begin{figure}[!ht]
    \centering
\begin{subfigure}[t]{0.45\textwidth}
    \includegraphics[height=0.8\textwidth]{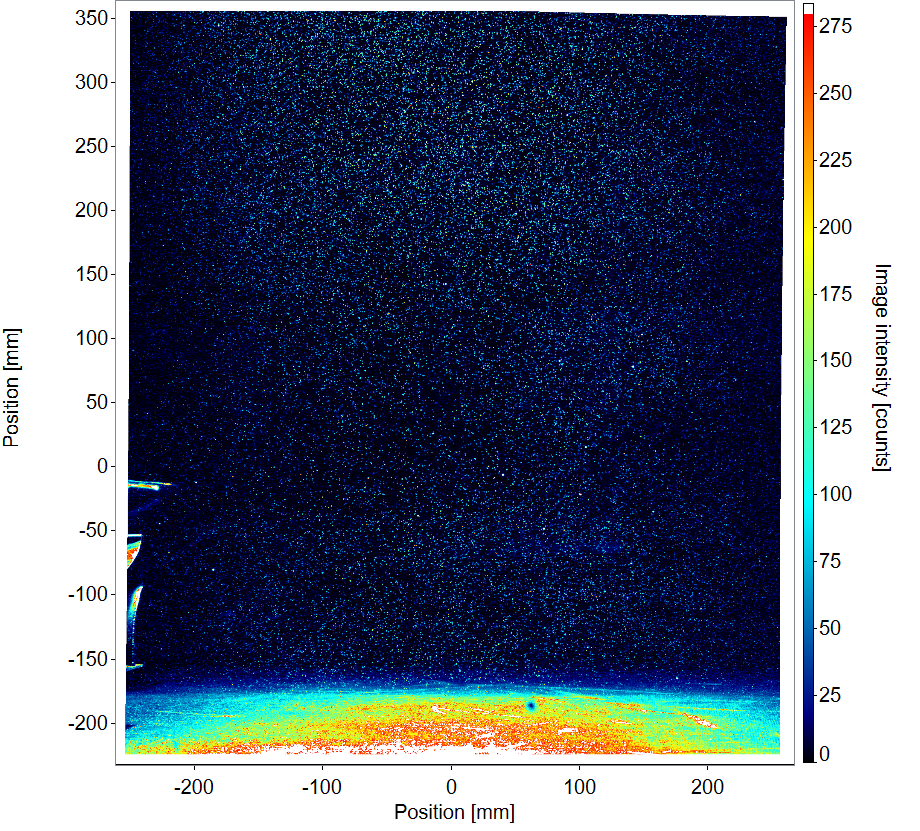}
\caption{ }
\label{fig:FW60001.bmp}
\end{subfigure}
\begin{subfigure}[t]{0.45\textwidth}    
    \includegraphics[height=0.8\textwidth]{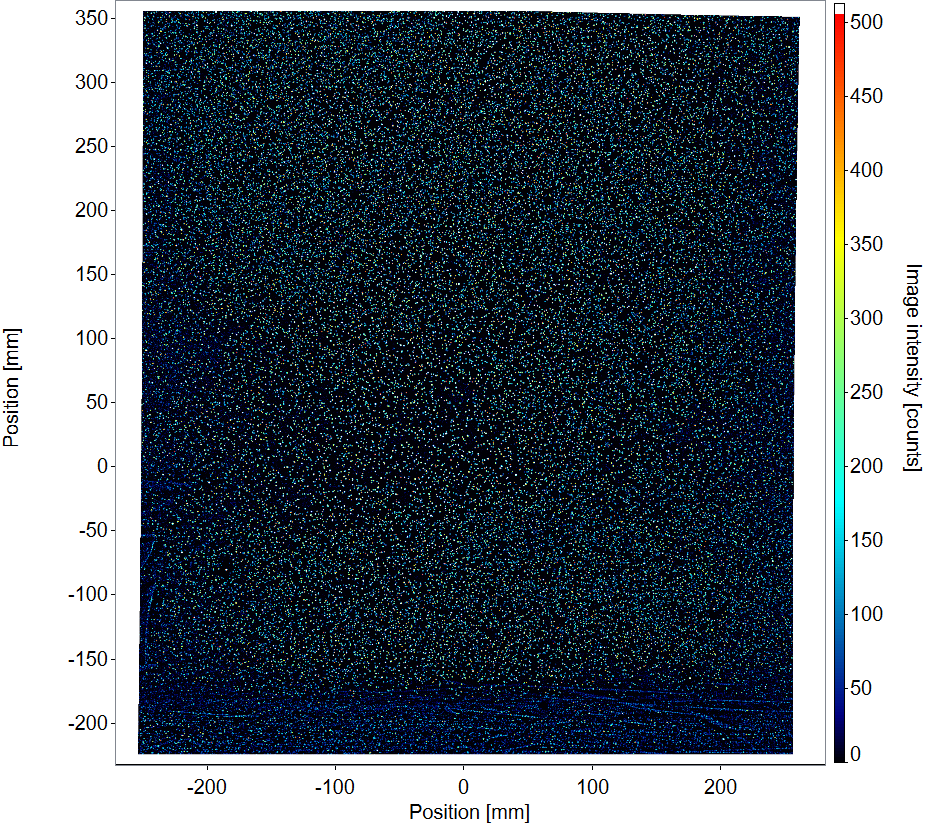}
    \caption{ }
    \label{fig:FW6_process0001.bmp}
\end{subfigure}
    \caption{Example of an image a) before and b) after the pre-processing to enhance the particles appearance. The images were corrected perspective and distortion errors. The position on the coordinates is for reference only. It corresponds to the first location of the calibration target and it has not been transformed to the full scale coordinate system used in the paper.}
        \label{fig:FW6_img_exp}
\end{figure}

Figure \ref{fig:FW6_img_exp} shows an example of a raw and a pre-processed image corrected for perspective and distortion errors in a calibration coordinate system. The approach was clearly effective in creating a homogeneous picture with sharp particles and removing the reflections from the rolling road and the trailing edge of the IFW, which were visible in the raw image.
To further improve the calibration and eliminate the remaining inaccuracies in the particle reconstruction a volume self-calibration was carried out. For this step, the same pre-processing procedure described above was applied to about 200 sample images and disparity vectors were iteratively calculated until the disparity error was less than 0.1 voxel. Details about the method can be found in \cite{wieneke2018improvements} and information regarding the algorithm in DaVis 10 can be found in \cite{wieneke2008volume}. The Optical Transfer Function (OTF) was calculated following the volume self-calibration. The OTF enables characterisation of the intensity distribution and effective size of particle images as a function of position and viewing angle for each camera, hence improving accuracy of particle detection.
Finally, the Shake-the-Box (STB) algorithm was used to reconstruct the particle tracks from the pre-processed images.

An important step was to crop the images in the vertical direction to exclude the floor of the tunnel before calculating the tracks. This step was key to prevent the data, and in particular the standard deviation, to be contaminated by spurious vectors caused by the reflections of the moving ground.
A standard STB procedure with only 1 pass, with three sub-passes, and 1.0 voxel allowed triangulation error was conducted. These settings result in $\sim 3.9 \times 10^6$ tracks in the entire volume, with the majority, around $ \sim 2.4 \times 10^6$, in the first sub-pass. 
Finally, to convert the sparse tracks into a regular grid an averaged binning method was applied. The FoV was divided into subvolumes of 32 voxels per side with an 87.5\% overlap, which led to a grid resolution of 1.17 mm. Since the focus of this paper is on the average quantities, the binning was applied considering all tracks in the specific subvolume across the different time snapshots. A polynomial of the 2nd order with a filter length of 5 time steps was applied.

\subsection{Coordinate system}
\label{sec:coordinate system}

All the measurements were converted to the full-scale coordinate system in accordance with the numerical model to ensure a genuine comparison. The full-scale coordinate system follows the convention set by the Automotive SAE, where $X$ represents the longitudinal direction with positive values from the origin towards the outlet of the domain, the spanwise direction is $Y$, which is positive from the geometry toward the symmetry plane (left-hand side of the full-model), and the vertical direction is $Z$, with positive values from the origin towards the roof. The origin of the coordinate system is located within the wing's wake to follow previous conventions \cite{Buscariolo2022}.The reference point on the wing is the rearmost, lowest and inboard corner of the footplate. 
Their streamwise distance is 300 mm and their spanwise distance is such that the final coordinate system is centred around the middle of the wing body.

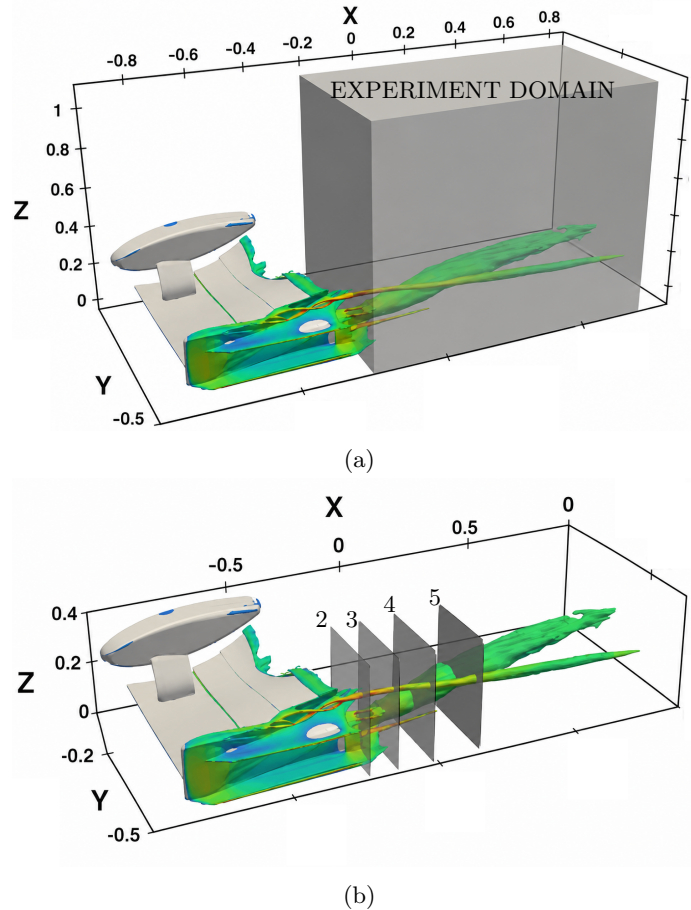
\begin{figure}[!ht]
    \centering
\begin{subfigure}[t]{0.6\textwidth}
    \begin{tikzpicture}
    \node[inner sep=0pt] (img) at (0,0)
    {\includegraphics[trim=5.5cm 1cm 5.5cm 0, clip, width=\textwidth]{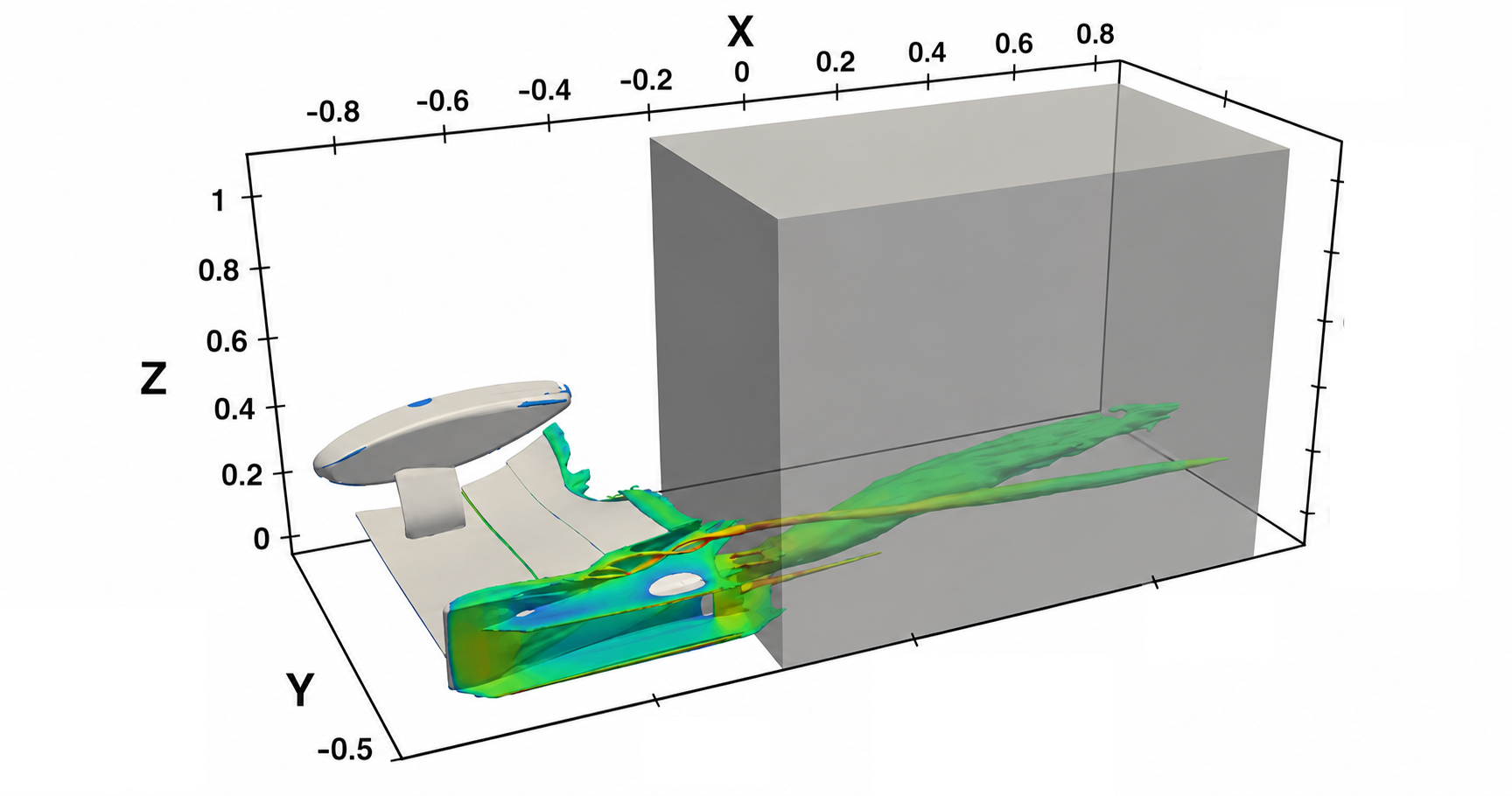}};
    \node at (1.5,1.65) {\small EXPERIMENT DOMAIN};
    \end{tikzpicture}
\caption{ }
\label{fig:Exp_volume}
\end{subfigure}
\begin{subfigure}[t]{0.6\textwidth}    
    \begin{tikzpicture}
    \node[inner sep=0pt] (img) at (0,0)
    {\includegraphics[trim=4cm 1.5cm 3cm .5cm, clip, width=\textwidth]{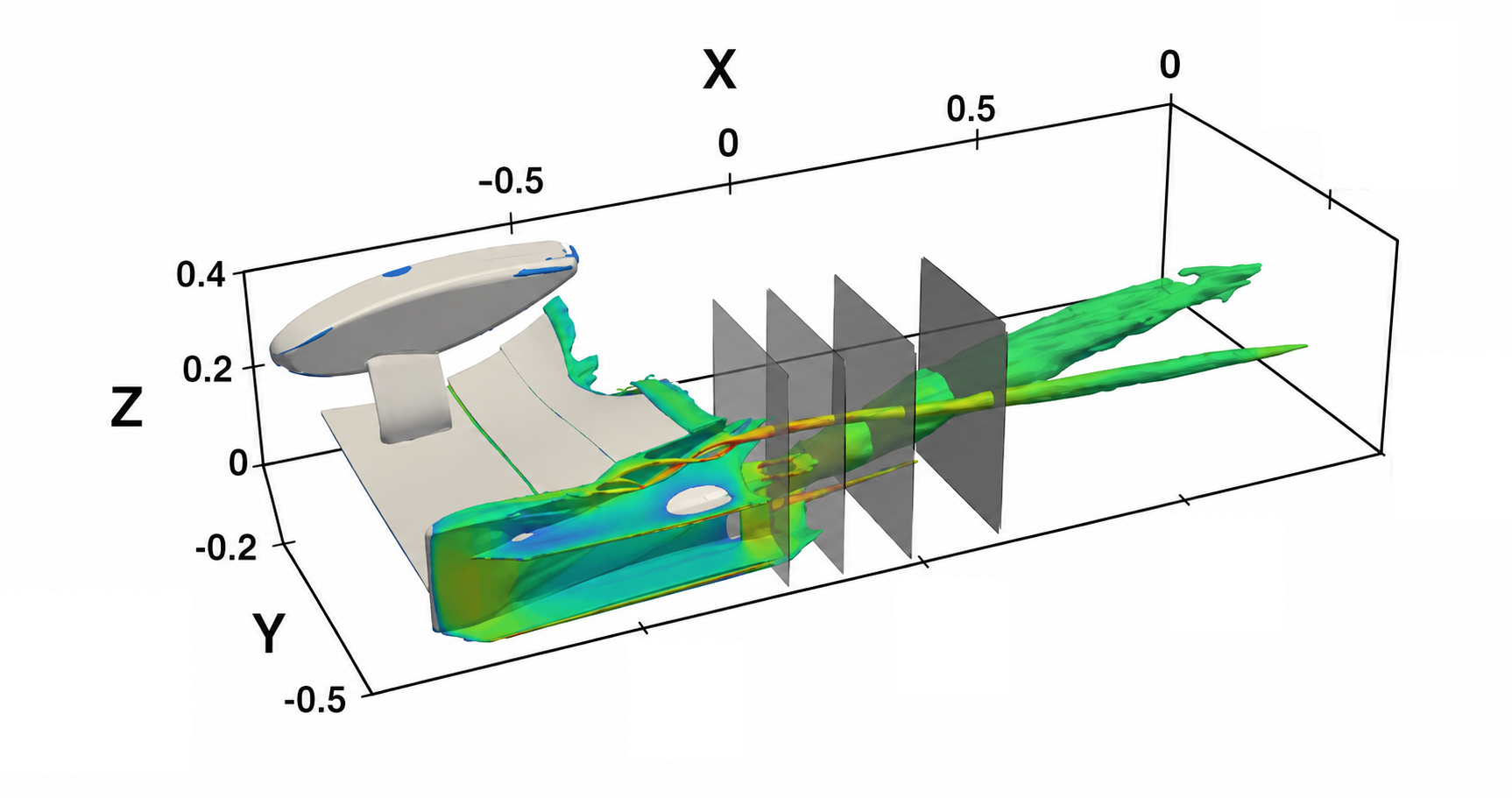}};
    \node at (-0.5, 0.7) {\small 2};
    \node at (-0.1, 0.7) {\small 3};
    \node at (0.4, 0.9) {\small 4};
    \node at (1, 1) {\small 5};
    \end{tikzpicture}
    \caption{ }
    \label{fig:planes_Buscariolo}
\end{subfigure}
    \caption{a) Relative position of the data-acquisition for the experiments with respect to the wing and b) the selected PIV Planes from \cite{Buscariolo2022}. The contours are the isosurfaces using the $\lambda_2$ criterion coloured by the streamwise velocity from the simulation results. The coordinates are in meter.}
    \label{fig:rel_placement}
\end{figure}

To enable point-by-point comparison between the numerical and the experimental data are expressed on the same structured grid. For this reason, both planar datasets are interpolated onto a structured grid with cell size of 2 mm. Careful selection of the cell size is essential to reduce the interpolation error and extreme smoothing of the measurements. Following a sensitivity study on the cell size, it was observed that the maximum cell size preventing these issues is 2 mm. It is possible to use a smaller cell size, provided that the researcher is not limited by the available memory of their computer. 
All the plots will be shown in the full scale coordinate system using the same coordinate system defined by \cite{Buscariolo2022} to enable comparison to previous studies.

Figure \ref{fig:Exp_volume} shows the location of the experiment volume relative the wing body overlie to the results from the simulation. The results are well-resolved from a distance (in full-scale coordinates) of 70 mm from the reference point on the footplate of the wing.
The comparison between the experimental and the simulation data is evaluated examining the same streamwise locations of the PIV planes collected by Buscariolo et al. \cite{Buscariolo2022}, as summarised in table \ref{tab:piv_location}. Unfortunately, in the present experiment the data at the edges of the measurement volume are not well resolved; hence comparison is made only for planes 2 to 5 keeping the nomenclature from Buscariolo et al. \cite{Buscariolo2022}. 

The location and size of each plane relative to wing is shown in figure \ref{fig:planes_Buscariolo}. In this work the planes examined in section \ref{sec:main_flow} have larger area when compared to \cite{Buscariolo2022} and the base dimensions for each plane are $0.3$ m $\times 0.35$ m. In addition, the non-dimensional coordinate $X^*= (X+300)/c$ is used to quantify the absolute distance in the wake from the footplate reference point. In the following all quantities are in non dimensional units unless specified. Generally, the dimensional coordinates are retained to enable easy comparison to previous studies.

\begin{table}[htb!]
 \centering
 \begin{tabular}{cccc}
 \hline
Plane Number & $X$ (mm) & $X/c$ & $X^*$ \\\hline
2            & -250  & -1.000 & 0.2 \\
3            & -150  & -0.600  & 0.6 \\
4            & -24   & -0.096 & 1.1 \\
5            & 150   &  0.600 & 1.8 \\
 \hline
\end{tabular}
\caption{Location of selected planes following Buscariolo et al. \cite{Buscariolo2022}. $X$ is the streamwise coordinate as defined in figure \ref{fig:experiments_datum}, while $X^*$ is the absolute distance from the reference point on the footplate non-dimensional by the chord of the wing.}
\label{tab:piv_location} 
\end{table}

\section{Results}
\label{sec:results}

The flow around the IFW is examined in two main areas. First, the evolution of the vortical system in space is described observing the merging process and the role of individual structures. Second, the numerical and experimental results are quantitatively compared using primary flow variables such as the measured velocity components and vorticity. 

This work has three additional objectives: to describe the shape and strength of the footplate vortex and its interaction with the main vortex; to compare and discuss whether the secondary structure peeling off from the floor is captured by the new experimental survey and by the simulation; to enhance the current understanding of the vortical system behind the IFW. 

To confirm that the near-wall mesh resolution was sufficient to resolve the boundary layer within the LES, the dimensionless wall distance $z^{+}$ was computed on the wing surface. The wall-normal units are defined as:

\begin{equation}
   \Delta z^{+} = \Delta z \frac{u_{\tau}}{\nu}
\end{equation}

with $ u_{\tau} = \sqrt{\frac{\tau_w}{\rho}} $ the friction velocity, $ \tau_w $ is the wall shear-stress and $z$ is the wall-normal direction of the flow. Building upon the seminal work of Piomelli et al. \cite{Piomelli2002}, Georgiadis et al. \cite{Georgiadis2010} proposed specific ranges for the spatial resolution of the computational grid in the wall-normal and wall-tangential directions to ensure adequate LES or DNS resolution. These guidelines have been widely adopted in the literature \cite{Beck2014, OSullivan2024, Vivarelli2025} for external aerodynamic flows.

\begin{equation}
 50 \leq \Delta x^{+} \leq 150 \hspace{20 mm}  \Delta z^+ \leq 1 \hspace{20 mm} 15 \leq \Delta y^{+} \leq 40
\end{equation}

The maximum value of $z^{+}$ is above one only at the stagnation point of each element, followed by a rapid reduction downstream, safely below the limit \cite{Georgiadis2010}. For brevity, a figure shows $z^+$ on the wing surface is reported in the appendix, figure \ref{fig:yplus}. To evaluate $z^{+}$, it was assumed that the quadrature points are equispaced, when in reality they follow a distribution by a Lagrange polynomial and the distance of the first quadrature point from the wing boundary is lower than the equivalent distance from the equispaced quadrature points. For this reason, it is safe to infer that the mesh resolution is well within the acceptable limits for implicit LES \cite{Piomelli2002}.

\subsection{Main structures identification}
\begin{figure}[!ht]
    \centering
    \includegraphics[width=0.49\linewidth]{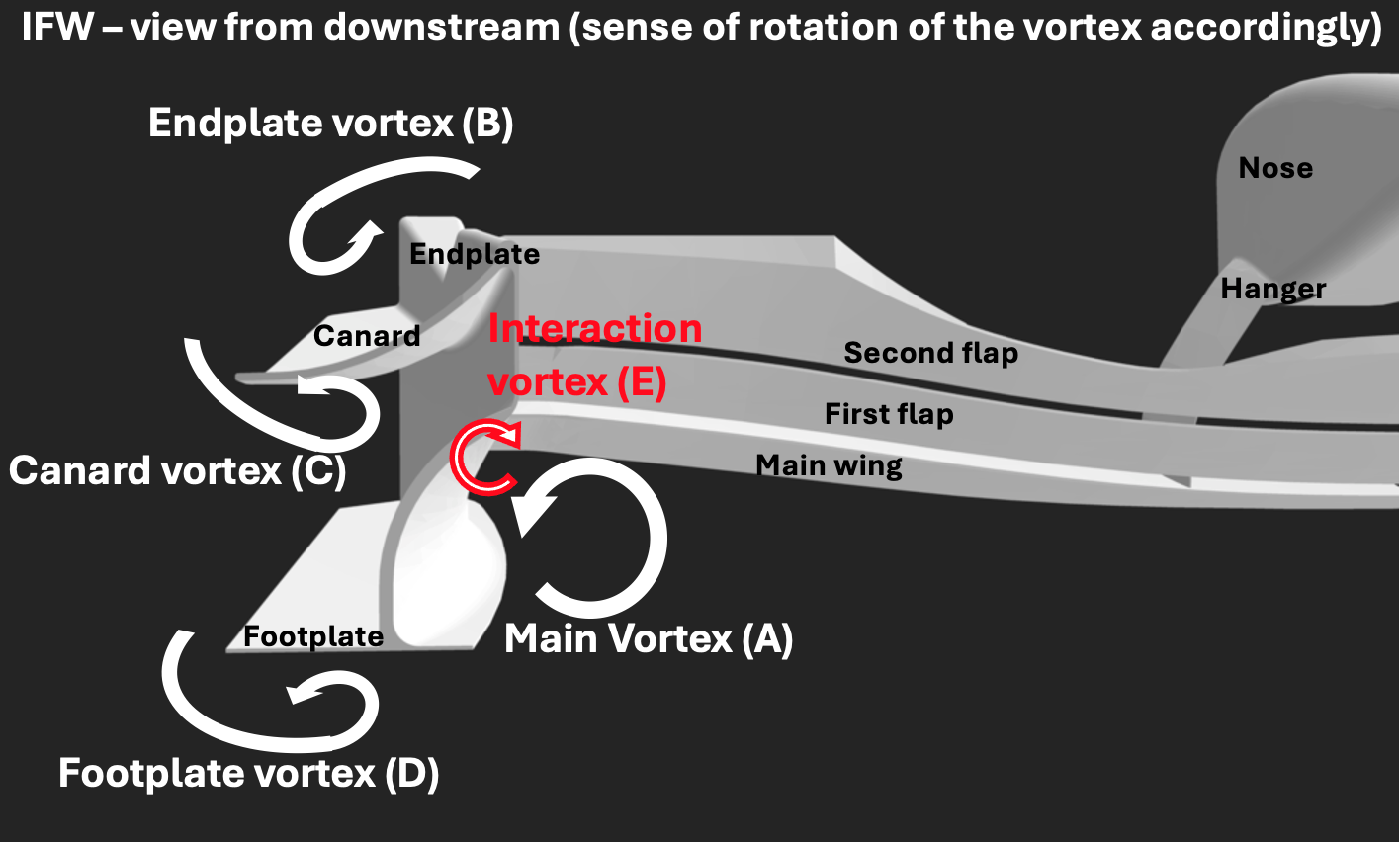}
    \caption{System of vortices observed in the IFW as developing from the elements of the IFW body.}
    \label{fig:ifw_vort}
\end{figure}

In previous works \cite{Buscariolo2022, Pegrum2006} four vortices were observed to propagate from the different elements of the imperial front wing. Following the schematic in figure \ref{fig:ifw_vort}, they were named according to the geometrical device that produces them: (A) the \textit{main vortex}, (B) the \textit{endplate vortex}, (C) the \textit{canard vortex}, (D) the \textit{footplate vortex}. In the current study a fifth vortex has been identified, situated between the main and canard vortices originating adiajent to the inboard wall of the endplate: we call this the \textit{interaction vortex} (E). It will be shown that this vortex is the only structure rotating in the opposite direction and plays a key role in preventing merging between the canard and the main vortices.

\begin{figure}[!ht]
   \centering
      \begin{subfigure}{0.49\textwidth}
      \includegraphics[width=\linewidth]{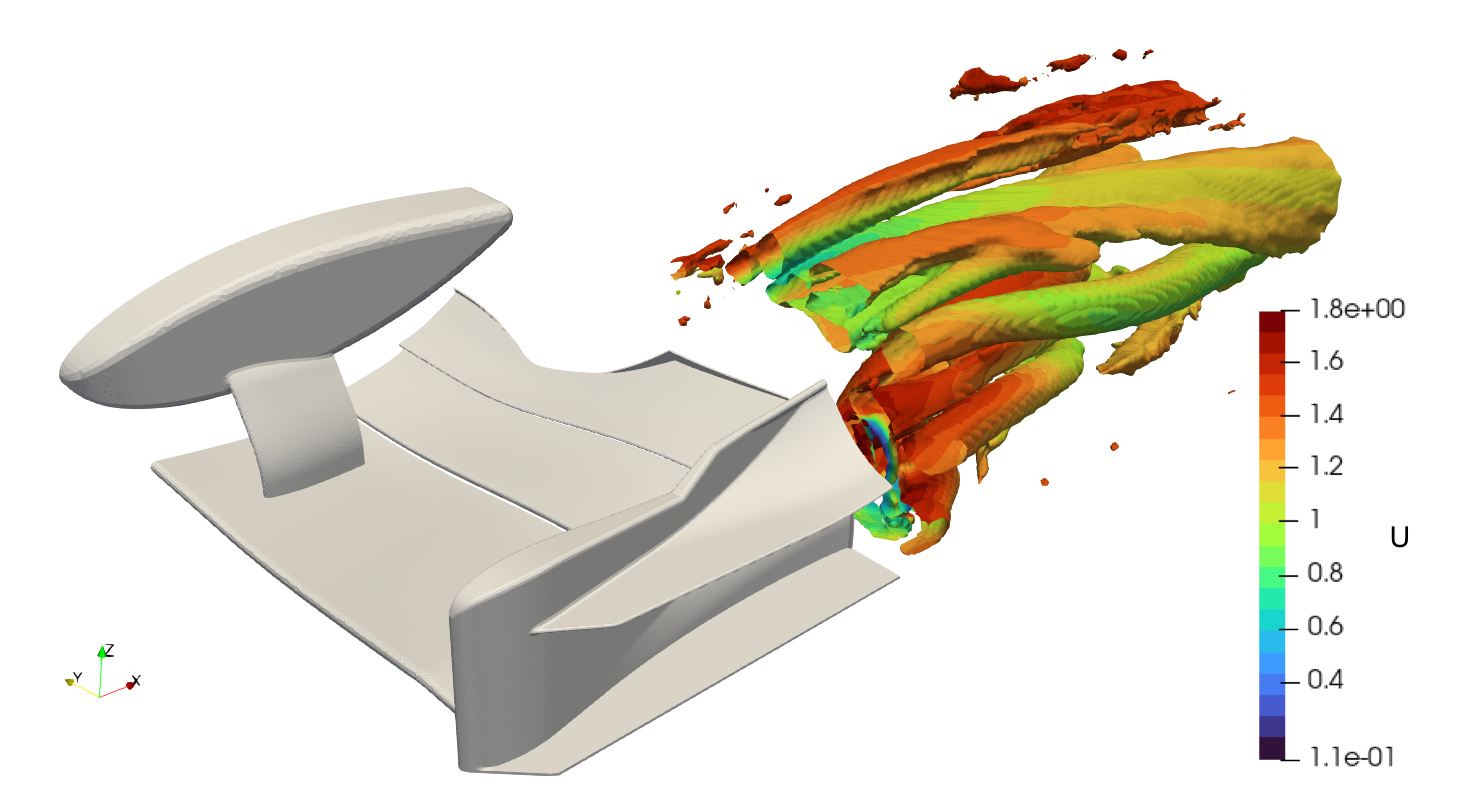}
      \caption{ PTV }
      \end{subfigure}
      \begin{subfigure}{0.49\textwidth}
      \includegraphics[width=\linewidth]{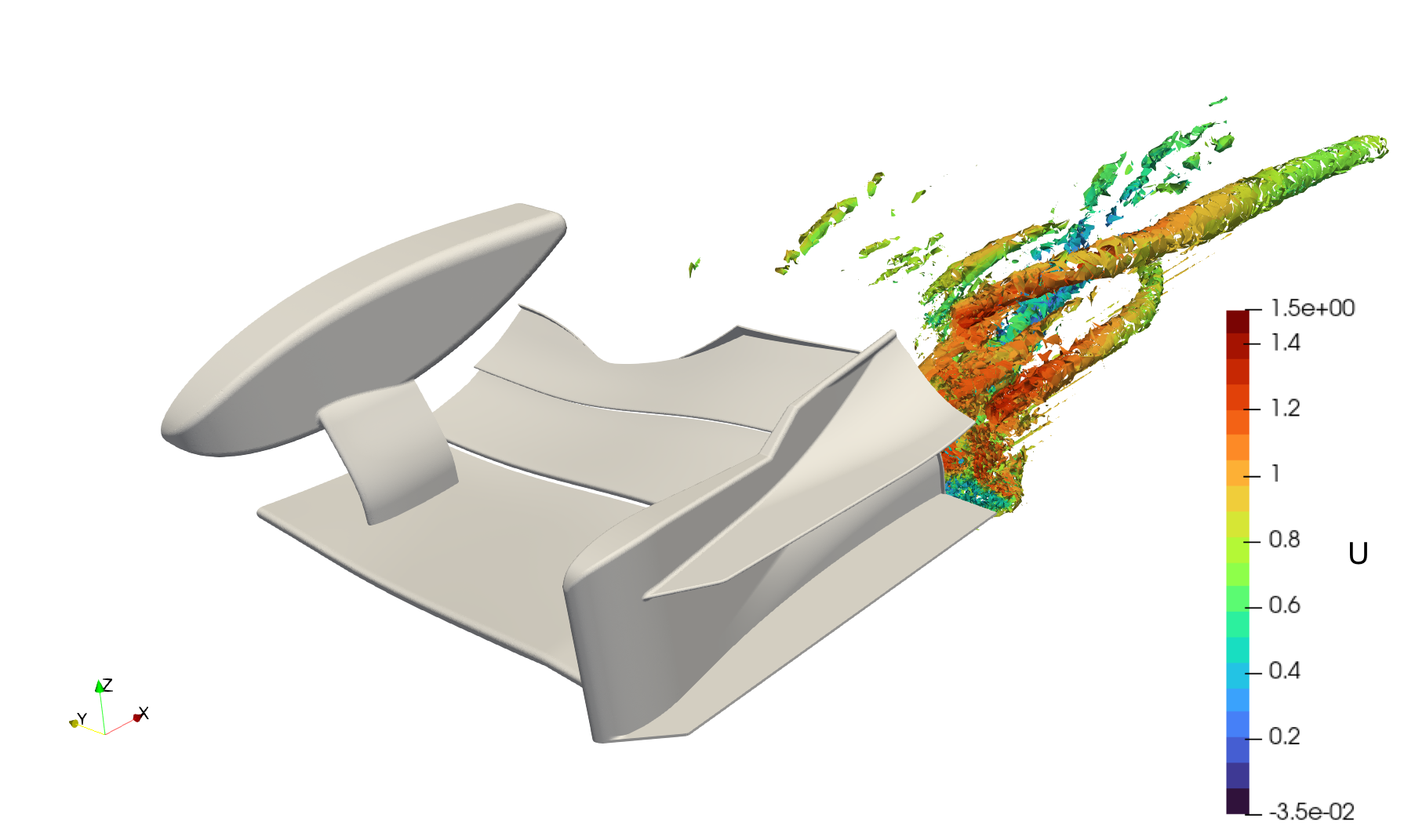}
      \caption{ LES}
      \end{subfigure}
      \caption{Comparison of $\lambda_2=-100$ contours between (a) experiment and (b) simulation at $Re= 74896$, coloured by the non-dimensional streamwise mean velocity $U$.}
   \label{fig:exp_vs_num_IFW_lambda2_contours}
\end{figure}

For both the experimental and the numerical results, the vortical structures downstream of the wing are identified using the $\lambda_2$ criterion \cite{Jeong1995} and the iso-contours with $\lambda_2 = -100$ ,as shown in figure \ref{fig:exp_vs_num_IFW_lambda2_contours}. The images demonstrate that both methods are able to capture the structure well in their downstream evolution. The contour from the simulation appears less defined and thinner given the $\lambda_2$ threshold chosen. The wake of the second flap seems to be more persistent in the experiment and to survive within the entire domain. The contour are coloured by the streamwise velocity which appears higher for the experiment, as discussed in section \ref{sec:results}. The results from the experiment and the simulation are aligned in the whole volume except for a small spanwise offset which will be quantified in the next section; this is probably due to small errors in the alignment of the experimental model.

\subsection{Vortex identification and development}
In the following, the downstream evolution of the vortices from their formation is described using the non-dimensional streamwise vorticity.
The different stages of evolution of the vortices are reported in figure  \ref{fig:Vort_evolution_num} and \ref{fig:Vort_evolution}. Since the results from the experiment are well resolved only from $X/c$=0.3 downstream of the reference point on the wing (figure \ref{fig:Vort_evolution}), the discussion of the initial stages of the vortex development is accompanied by results from the simulation at $X/c$=0.02 and 0.2 respectively, corresponding plane 1 and plane 2 (figure  \ref{fig:Vort_evolution_num}).

\begin{figure}[ht!]
    \centering
    \begin{subfigure}[b]{0.45\textwidth}
        \centering
    \begin{tikzpicture}
    \node[inner sep=0pt] (img) at (0,0)
    {\includegraphics[width=\linewidth]{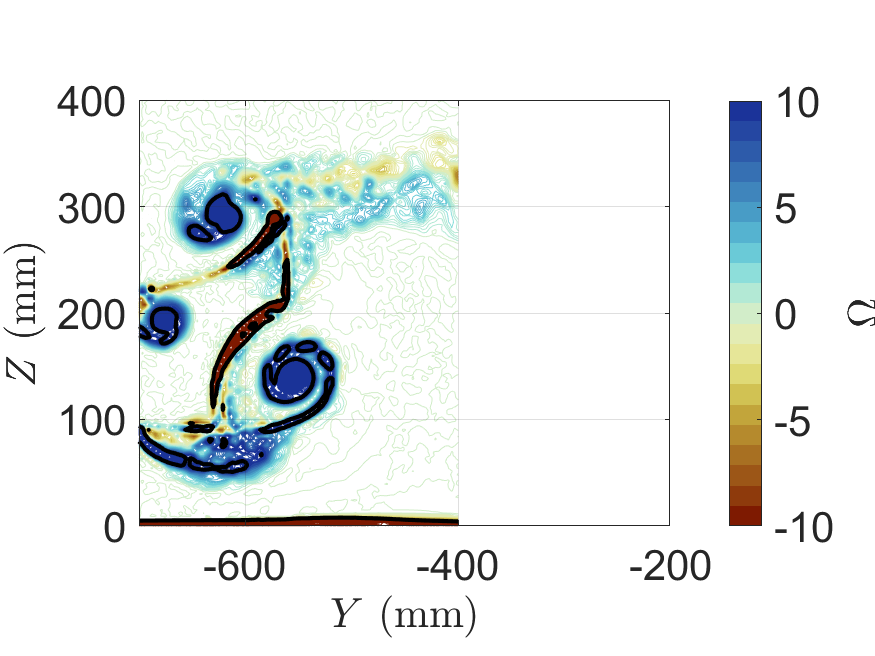}};
    \node at (-2,1.3) {\tiny\textbf{B}};
    \node at (-2,0.2) {\tiny\textbf{C}};
    \node at (-1,0.1) {\tiny\textbf{A}};
    \node at (-2,-1.4) {\tiny \textbf{$\textrm{D}$}};
\end{tikzpicture}
        \caption{LES $X$=-294 mm, $X^*=0.02$}
        \label{fig:Vort1sim}
    \end{subfigure}
    \begin{subfigure}[b]{0.45\textwidth}
        \centering
    \begin{tikzpicture}
    \node[inner sep=0pt] (img) at (0,0)
    {\includegraphics[width=\linewidth]{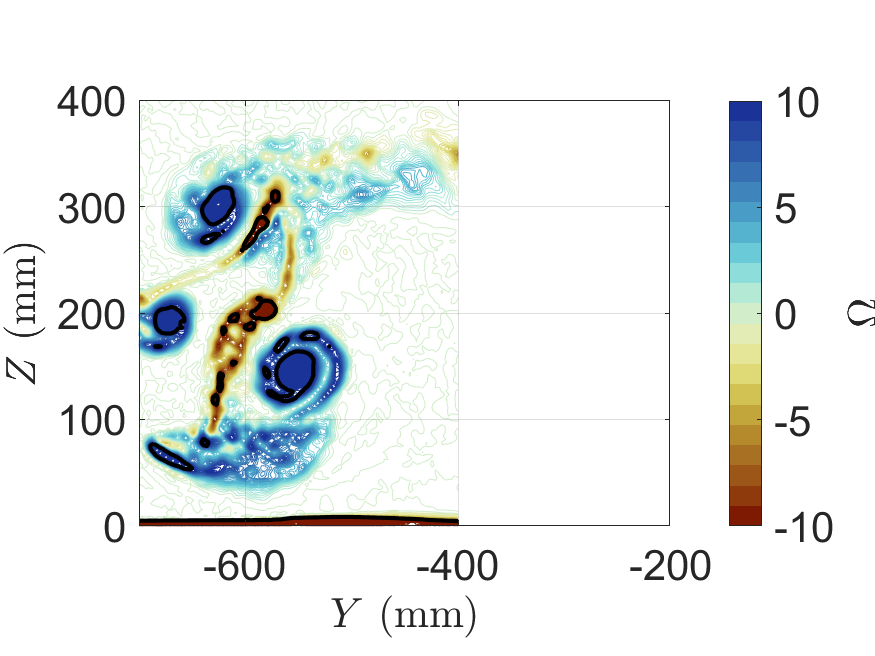}};    
    \node at (-2,1.3) {\tiny\textbf{B}};
    \node at (-2,0.2) {\tiny\textbf{C}};
    \node at (-1.7,0.3) {\tiny\textbf{E}};
    \node at (-1,0.1) {\tiny\textbf{A}};
    \node at (-2,-1.4) {\tiny \textbf{$\textrm{D}$}};
\end{tikzpicture}
        \caption{LES X=-250 mm, $X^*=0.20$}
        \label{fig:Vort2num}
    \end{subfigure}   
    \caption{Downstream evolution of the vortical structures: isocontourns of non-dimensional vorticity from the LES results at a) $X^*=0.02$ and $X^*=0.20$.} 
    \label{fig:Vort_evolution_num}
\end{figure}
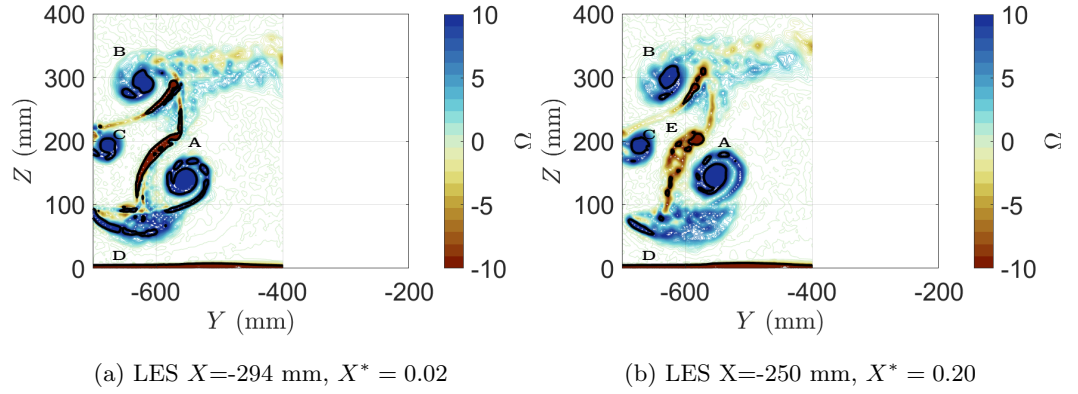
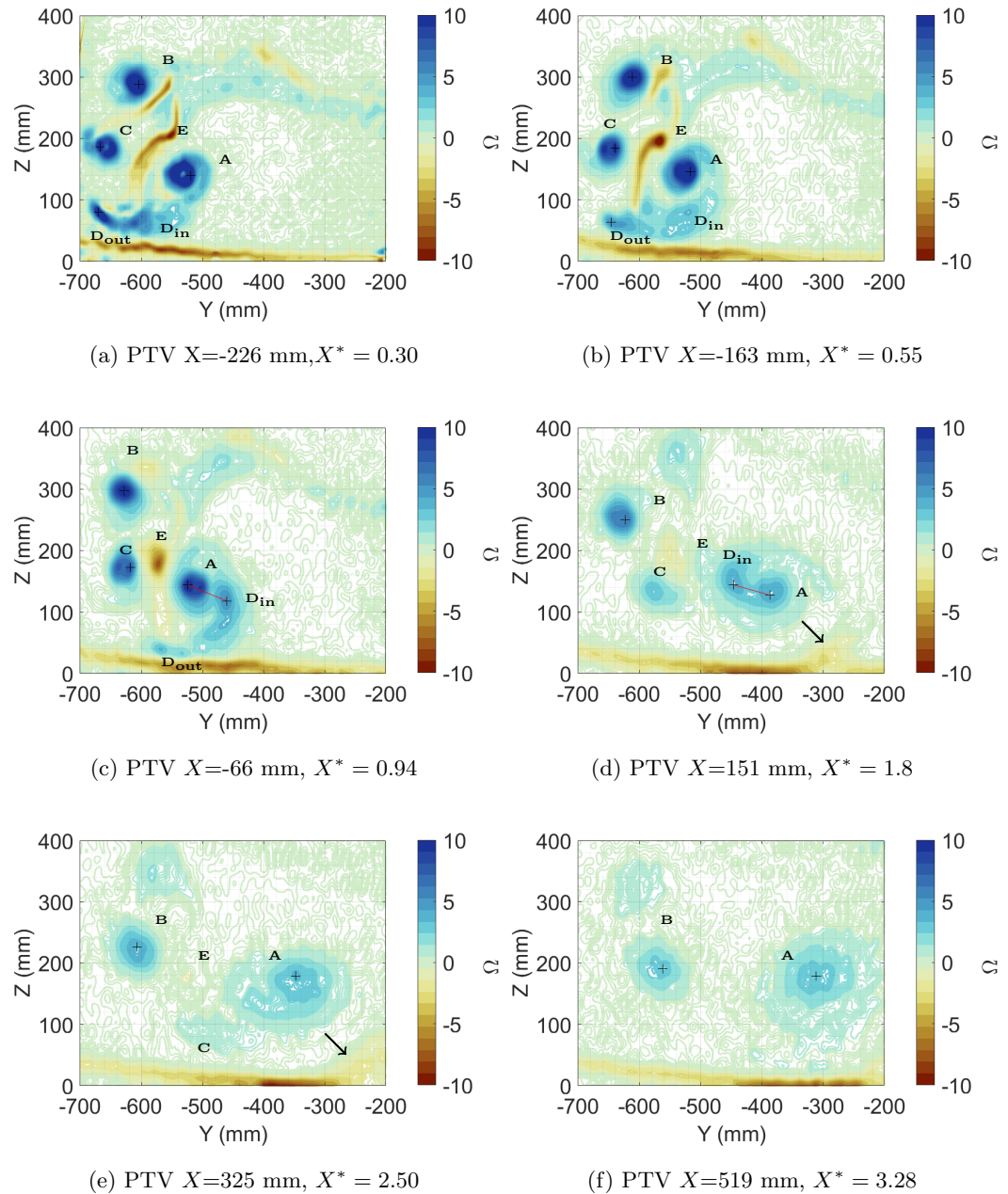
\begin{figure}[ht!]
    \centering
    \begin{subfigure}[b]{0.45\textwidth}
        \centering
    \begin{tikzpicture}
    \node[inner sep=0pt] (img) at (0,0)
    {\includegraphics[width=\linewidth]{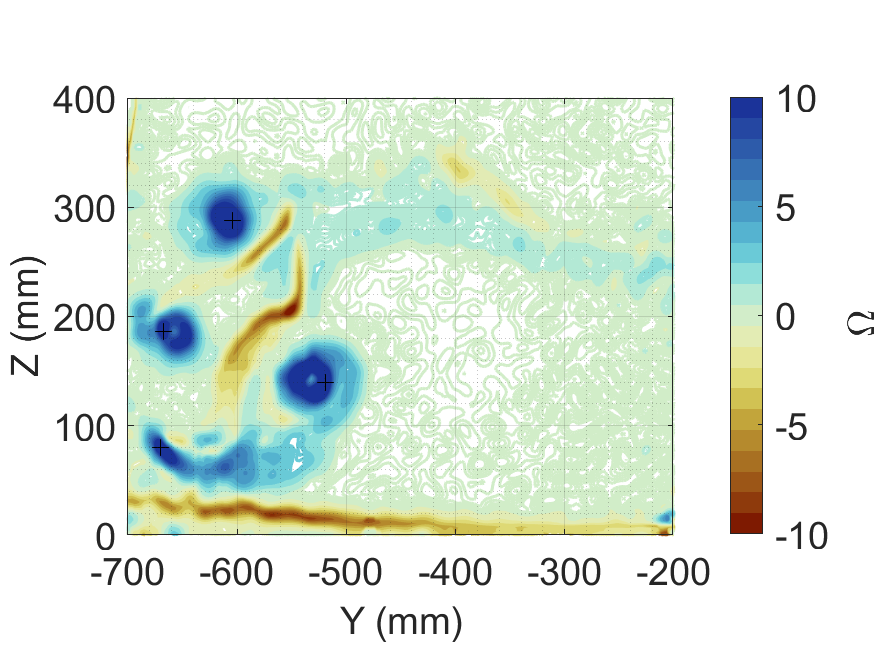}};
    \node at (-1.2,1.2) {\tiny\textbf{B}};
    \node at (-1.8,0.2) {\tiny\textbf{C}};
    \node at (-1,0.2) {\tiny\textbf{E}};
    \node at (-0.4,-0.2) {\tiny\textbf{A}};
    \node at (-2,-1.3) {\tiny \textbf{$\textrm{D}_\textrm{out}$}};
    \node at (-1.1,-1.2) {\tiny \textbf{$\textrm{D}_\textrm{in}$}};
\end{tikzpicture}
        \caption{PTV X=-226 mm,$X^*=0.30$}
        \label{fig:Vort1}
    \end{subfigure}
    \begin{subfigure}[b]{0.45\textwidth}
        \centering
    \begin{tikzpicture}
    \node[inner sep=0pt] (img) at (0,0)
{\includegraphics[width=\linewidth]{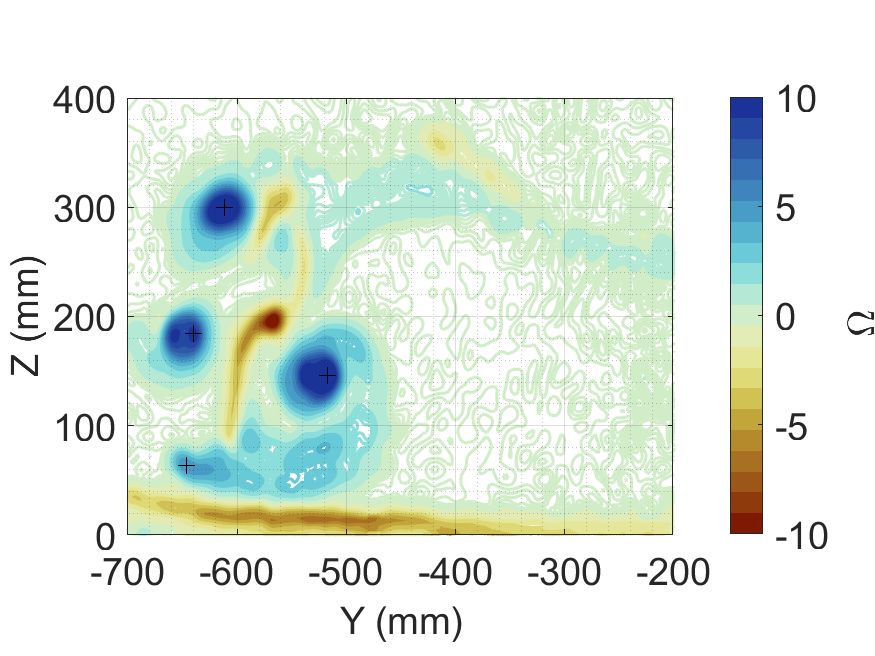}};   
    \node at (-1.2,1.2) {\tiny\textbf{B}};
    \node at (-2,0.3) {\tiny\textbf{C}};
    \node at (-1,0.2) {\tiny\textbf{E}};
    \node at (-0.5,-0.2) {\tiny\textbf{A}};
    \node at (-1.7,-1.3) {\tiny \textbf{$\textrm{D}_\textrm{out}$}};
    \node at (-0.6,-1.1) {\tiny \textbf{$\textrm{D}_\textrm{in}$}};
\end{tikzpicture}
        \caption{PTV $X$=-163 mm, $X^*=0.55$}
        \label{fig:Vort2}
    \end{subfigure} 
 \begin{subfigure}[b]{0.45\textwidth}
        \centering
    \begin{tikzpicture}
    \node[inner sep=0pt] (img) at (0,0)
    {\includegraphics[width=\linewidth]{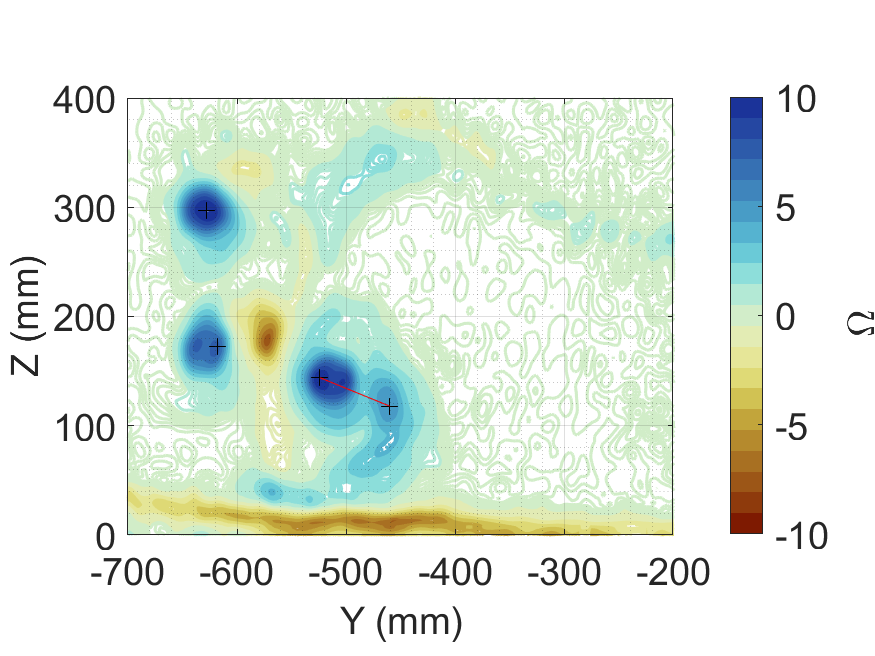}};
    \node at (-1.7,1.5) {\tiny\textbf{B}};
    \node at (-1.8,0.1) {\tiny\textbf{C}};
    \node at (-1.3,0.3) {\tiny\textbf{E}};
    \node at (-0.6,-0.1) {\tiny\textbf{A}};
    \node at (-1,-1.5) {\tiny \textbf{$\textrm{D}_\textrm{out}$}};
    \node at (0.1,-0.6) {\tiny \textbf{$\textrm{D}_\textrm{in}$}};
\end{tikzpicture}
        \caption{PTV $X$=-66 mm, $X^*=0.94$}
        \label{fig:Vort3}
    \end{subfigure}
    \begin{subfigure}[b]{0.45\textwidth}
        \centering
    \begin{tikzpicture}
    \node[inner sep=0pt] (img) at (0,0)
   {\includegraphics[width=\linewidth]{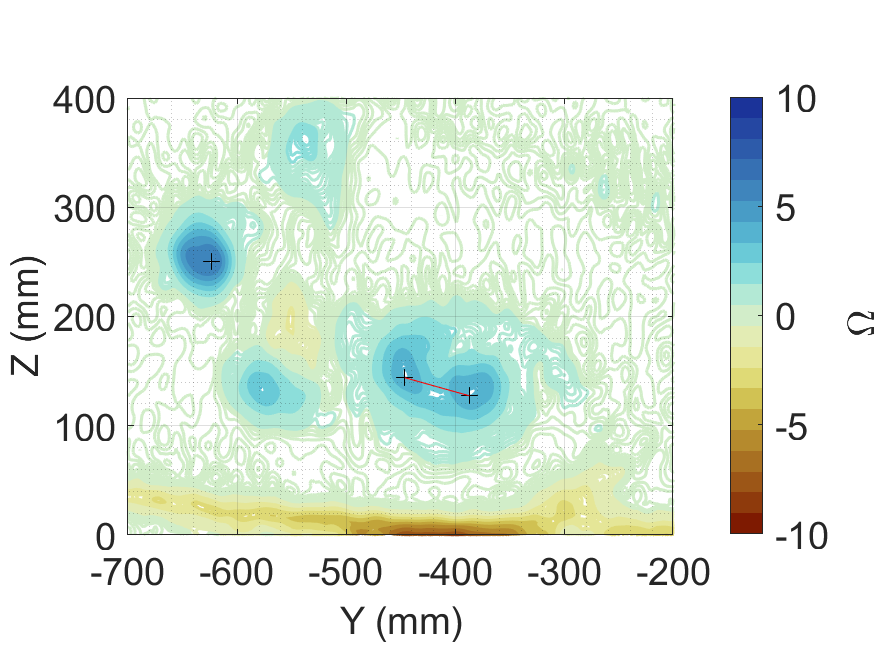}};  
    \node at (-1.3,0.8) {\tiny\textbf{B}};
    \node at (-1.3,-0.2) {\tiny\textbf{C}};
    \node at (-0.7,0.2) {\tiny\textbf{E}};
    \node at (-0.2,0) {\tiny\textbf{$\textrm{D}_\textrm{in}$}};
    \node at (0.7,-0.5) {\tiny \textbf{A}};
\draw[->, thick] (0.7,-0.9) -- (1,-1.2);
\end{tikzpicture}
        \caption{PTV $X$=151 mm, $X^*=1.8$}
        \label{fig:Vort4}
    \end{subfigure} 
    
    \begin{subfigure}[b]{0.45\textwidth}
        \centering
    \begin{tikzpicture}
    \node[inner sep=0pt] (img) at (0,0)
{\includegraphics[width=\linewidth]{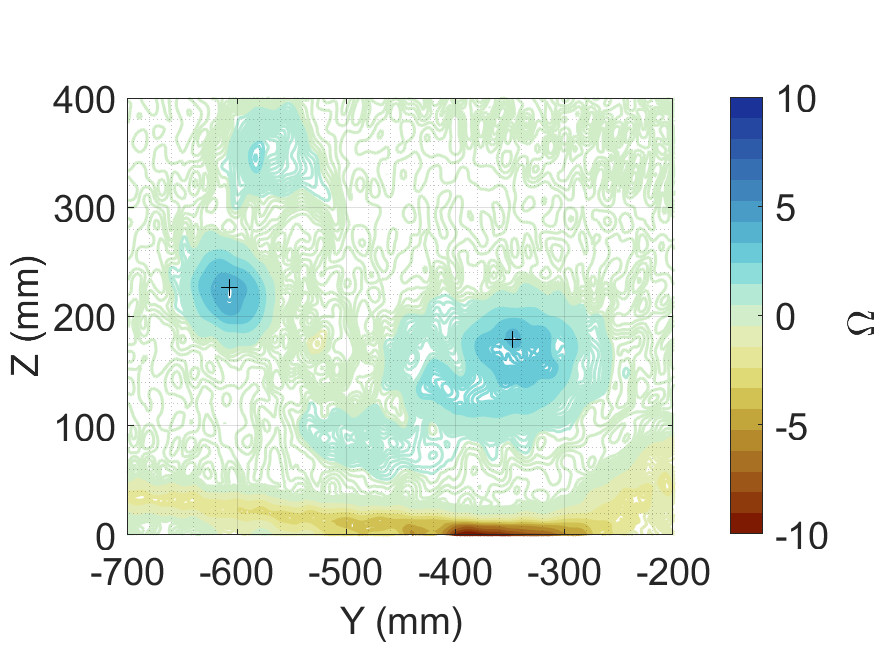}};
    \node at (-1.3,0.7) {\tiny\textbf{B}};
    \node at (-0.7,-1.1) {\tiny\textbf{C}};
    \node at (-0.7,0.2) {\tiny\textbf{E}};
    \node at (0.3,0.2) {\tiny\textbf{A}};
\draw[->, thick] (1,-0.9) -- (1.3,-1.2);
\end{tikzpicture}
        \caption{PTV $X$=325 mm, $X^*=2.50$}
        \label{fig:Vort5}
    \end{subfigure}
    \begin{subfigure}[b]{0.45\textwidth}
        \centering
    \begin{tikzpicture}
    \node[inner sep=0pt] (img) at (0,0)
{\includegraphics[width=\linewidth]{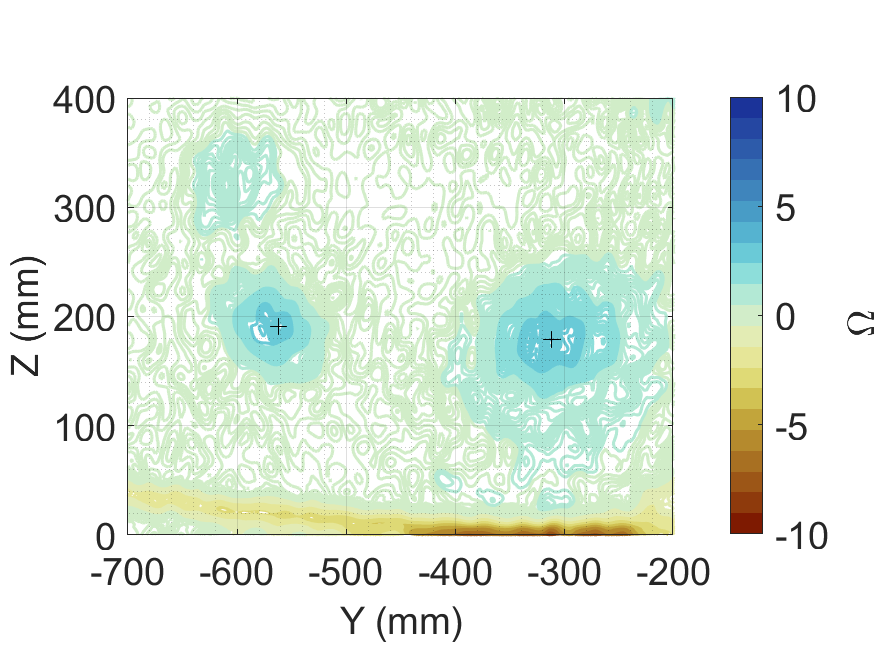}};
    \node at (-1.2,0.7) {\tiny\textbf{B}};
    \node at (0.5,0.2) {\tiny\textbf{A}};
\end{tikzpicture}
        \caption{PTV $X$=519 mm, $X^*=3.28$}
        \label{fig:Vort6}
    \end{subfigure}    
    \caption{Downstream evolution of the vortical structures: isocontours of non-dimensional vorticity field from the PTV from $X^*=0.3$ to $X^*=3.28$.} 
    \label{fig:Vort_evolution}
\end{figure}

\subsubsection*{Vortex formation}
\hfill

Although already unravelled by previous works \cite{Buscariolo2022, Pegrum2006}, the formation of the vortices A to D is here reviewed and amplified with a discussion on the origin of the newly discovered interaction vortex (E). The reader can refer to figure \ref{fig:ifw_vort} for the nomenclature of the different parts of the wing. The mechanism of origin of the vortex will be discussed in section \ref{sec:discussion}. 

\noindent The main vortex (A), is produced by the shear-layer that is formed at the leading edge of the endplate. This shear-layer rolls-up following an anti-clockwise trajectory based on the inboard shape of the endplate. The pressure difference between the pressure and the suction side of the mainplane element and the flaps is important in determining the strength of the main vortex as it impacts the flow acceleration at the shortest area below the mainplane, hence affecting the strength of the shear-layer that rotates around the inboard side of the endplate.

\noindent The endplate vortex (B), on its turn, is also produced at the upper tip of the endplate, where two helical structures are created due to the aerofoil-like profile of the endplate's vertical area. These helical structures, visible in figure \ref{fig:rel_placement}, rotate around each other and merge into a single vortex downstream.

\noindent On the canard, the pressure difference between its upper and lower surface creates flow circulation around its edges, which generates the canard vortex (C). The curved profile of the canard's geometry enhances this pressure difference, which contributes to powering up the vortex. 

\noindent Closer to the ground, on the footplate, two shear-layers are formed from its triangular leading edge. Each shear-layer rolls-up towards the inboard and the outboard side of the footplate respectively. They interact, pulling each other together until they merge into a single structure at the trailing edge of the footplate. The footplate vortex (D) which is naturally pushed spanwise, towards the main vortex, by the strong inwash present at that location (see section \ref{sec:main_flow} for details). It will be shown that although the footplate vortex appears as a unique structure at the trailing of the footplate, the separation in an outbound and an inbound structure actually persists downstream.

\noindent These four coherent vortical structures are visible from within 6 mm ($X/c$=0.024) of the trailing edge; while there is the fifth structure that emerges only downstream (figure \ref{fig:Vort1sim}). This last vortex has been named the interaction vortex (E). Its formation is due to the shear layer leaving the trailing edge of the endplate. A region of distributed negative vorticity is observed at the surface of the endplate probably due to the main vortex being very close to a stationary wall. This distributed vorticity region eventually rolls up on itself forming the clear rounded coherent structure observed from about 50 mm downstream (figure \ref{fig:Vort2num}). Its origin and the physical mechanism is further discuss in section \ref{sec:discussion}.

 \subsubsection*{The footplate vortex: separation, orbiting and merging}
 \hfill \\
\noindent Figure \ref{fig:Vort1} shows the first location of fully-resolved experimental data, only 24 mm downstream of figure \ref{fig:Vort2num}. The main difference is that the footplate vortex breaks into two distinct structures, one more outboard ($\textrm{D}_\textrm{out}$) which was already observable in figure \ref{fig:Vort2num} and one inboard ($\textrm{D}_\textrm{in}$). The black crosses in the figures represent the loci of the local maximum of positive vorticity and red line shows the distance between the main vortex and the inboard footplate vortex.\\
From this location onwards, the inboard vortex ($\textrm{D}_\textrm{in}$) completely detaches from the outboard one ($\textrm{D}_\textrm{out}$) and starts orbiting around the main vortex (figures \ref{fig:Vort2} to \ref{fig:Vort4}), which is clearly a much stronger vortex. In this process the vortex $\textrm{D}_\textrm{in}$ is lifted away from the ground and increases in intensity, in opposition to the main vortex that is reducing its strength. At $X=151$ mm ($X/c=1.8$), figure \ref{fig:Vort4}, the two structures are almost of the same size and intensity with comparable circulation. The orbiting process terminates around a relative angle of 330 degrees in agreement with literature on vortex merging process \cite{CERRETELLI_WILLIAMSON_2003, Leweke2016DynamicsAI} when $\textrm{D}_\textrm{in}$ is fully dissipated into the main vortex. The main vortex slowly return to a symmetric shape and continues travelling downstream. In figure \ref{fig:Vort4} and \ref{fig:Vort5} a lift-off of secondary vorticity due to the interaction between vortex A and $\textrm{D}_\textrm{in}$ can be observed as pointed by the black arrow.\\
Meanwhile, the outboard footplate vortex ($\textrm{D}_\textrm{out}$) is also swept towards the centre of the wing due to the strong inwash, but it is too far from the main vortex to enter in the merging process and it appears to dissipates quickly while approaching the ground, figures \ref{fig:Vort1} to \ref{fig:Vort3}, as expected if no other vortical interactions are in place \cite{CERRETELLI_WILLIAMSON_2003,Leweke2016DynamicsAI}.

 \subsubsection*{The role of the interaction vortex}
 \hfill \\
In this merging process the interaction vortex plays a key role. In fact, born in-between the main vortex and the canard vortex and rotating in opposite direction, it ensures that those two vortices are kept apart as shown in figure \ref{fig:Vort3}. Furthermore, it probably plays also a role further downstream (figure \ref{fig:Vort4}) in maintaining the endplate vortex B far enough from the carnard vortex C, letting the latter naturally decay as it approaches the wall (figure \ref{fig:Vort4}).

 \subsubsection*{Far wake}
\hfill \\
In the furthest downstream location capture by the experiment, $X=519$ mm, which corresponds to a distance $X/c=3.27$ from the wing, the vortical wake is still not fully dissipated. Figure \ref{fig:Vort6} show two persistent, although widely diffused, vortices: the endplate vortex B and the main vortex A. In addition, the secondary vorticity at the ground is still relatively strong and keeps travelling in-wash. Finally, another vortical structure above the vortex B, which appeared in figure \ref{fig:Vort4} is slowly dissipating.

\begin{figure}[ht!]
    \centering
    \begin{subfigure}[b]{0.29\textwidth}
\centering
    \begin{tikzpicture}
    \node[inner sep=0pt] (img) at (0,0)
    {\includegraphics[trim=0cm 0cm 3.8cm 0, clip, width=\textwidth]{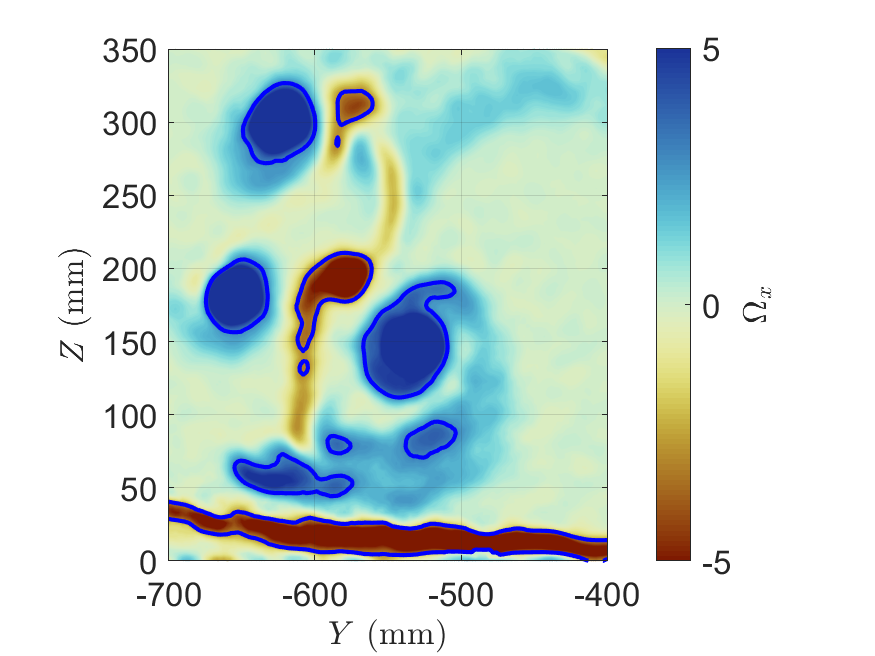}};
    \node at (-0.8,1.3) {\footnotesize\textbf{B}};
    \node at (-0.8,0.6) {\footnotesize \textbf{C}};
    \node at (-0.1,0.6) {\footnotesize \textbf{E}};
    \node at (0.1,-0.1) {\footnotesize \textbf{A}};
    \node at (-0.8,-1) {\footnotesize \textbf{D\textsubscript{out}}};
    \node at (1.2,-0.8) {\footnotesize \textbf{D\textsubscript{in}}};
\end{tikzpicture}
\caption{PTV - Plane 3}
    \label{fig:Omexp_Plane3}
    \end{subfigure} 
    \begin{subfigure}[b]{0.29\textwidth}
        \centering
    \begin{tikzpicture}
    \node[inner sep=0pt] (img) at (0,0)
    {\includegraphics[trim=0cm 0cm 3.8cm 0, clip, width=\linewidth]{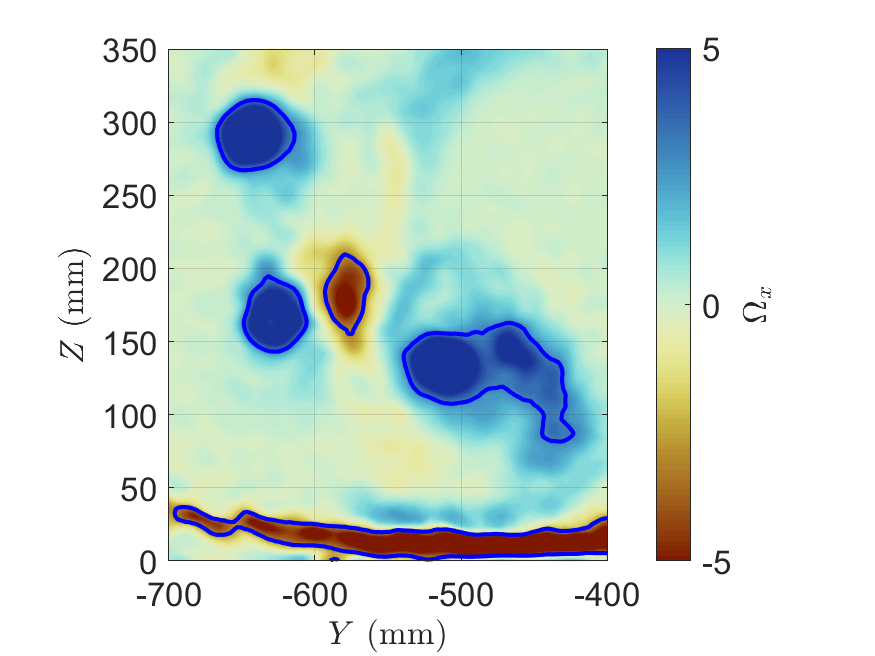}};
    \node at (-0.9,1.4) {\footnotesize \textbf{B}};
    \node at (-0.8,0.2) {\footnotesize \textbf{C}};
    \node at (-0.1,0.6) {\footnotesize \textbf{E}};
    \node at (0.8,0.2) {\footnotesize \textbf{A/\textbf{D\textsubscript{in}}}};
\end{tikzpicture}
        \caption{PTV - Plane 4}
        \label{fig:Omexp_Plane4}
    \end{subfigure}  
    \begin{subfigure}[b]{0.29\textwidth}
        \centering
    \begin{tikzpicture}
    \node[inner sep=0pt] (img) at (0,0)
    {\includegraphics[trim=0cm 0cm 3.8cm 0, clip, width=\linewidth]{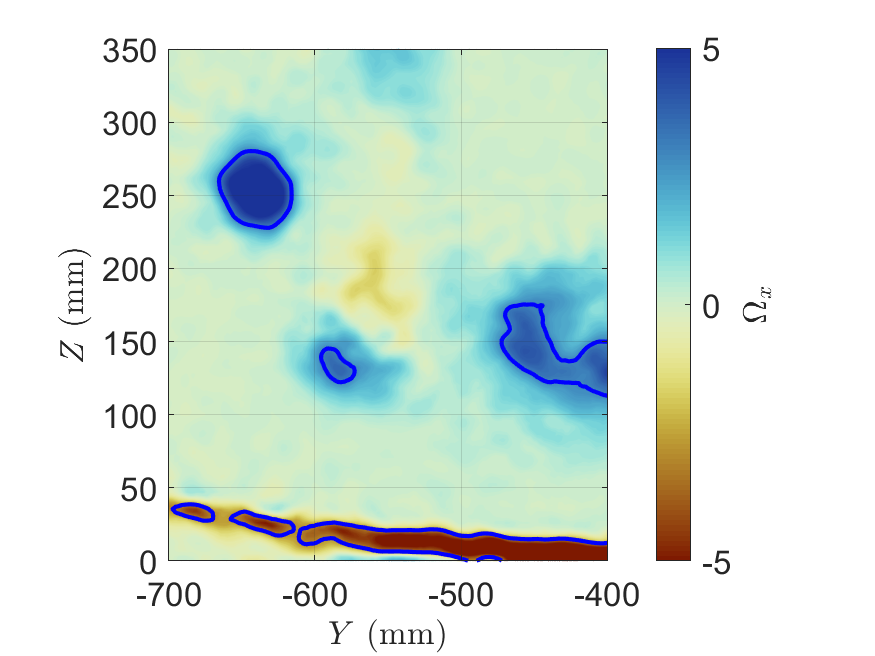}};
    \node at (-0.9,1.1) {\footnotesize \textbf{B}};
    \node at (-0.3,-0.3) {\footnotesize \textbf{C}};
    \node at (0,0.3) {\footnotesize \textbf{E}};
    \node at (1.2,0.3) {\footnotesize \textbf{A/\textbf{D\textsubscript{in}}}};
\end{tikzpicture}
        \caption{PTV - Plane 5}
        \label{fig:Omexp_Plane5}
    \end{subfigure}
     \begin{subfigure}[b]{0.08\textwidth}
        \centering
\raisebox{0.25\height}{\includegraphics[trim=11cm 0cm 0cm 0, clip, width=\linewidth]{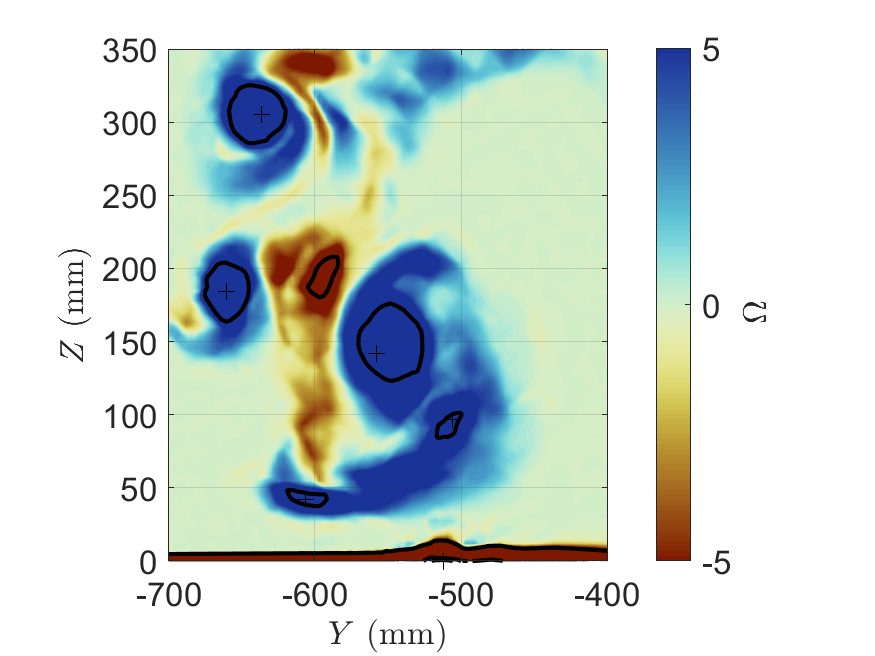}}
        \end{subfigure}
    \begin{subfigure}[b]{0.29\textwidth}
        \centering
    \begin{tikzpicture}
    \node[inner sep=0pt] (img) at (0,0)
    {\includegraphics[trim=0cm 0cm 3.8cm 0, clip, width=\linewidth]{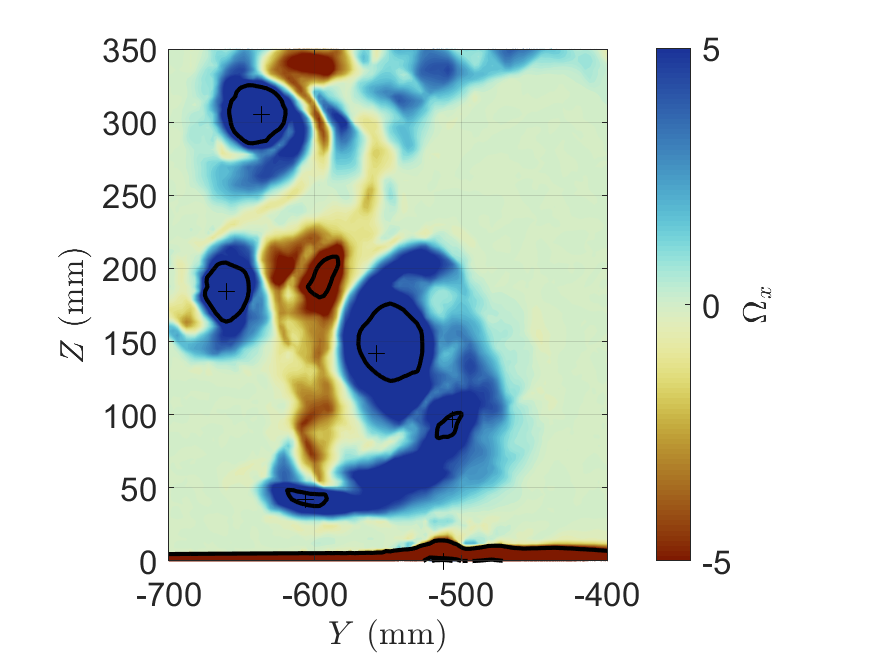}};
    \node at (-0.9,1.3) {\footnotesize \textbf{B}};
    \node at (-1,0.5) {\footnotesize \textbf{C}};
    \node at (-0.3,0.6) {\footnotesize \textbf{E}};
    \node at (0,-0.1) {\footnotesize \textbf{A}};
    \node at (-0.5,-1.3) {\footnotesize \textbf{D\textsubscript{out}}};
    \node at (1.4,-0.6) {\footnotesize \textbf{D\textsubscript{in}}};
\end{tikzpicture}
        \caption{LES - Plane 3}
        \label{fig:Omnum_Plane3}
    \end{subfigure}
    \begin{subfigure}[b]{0.29\textwidth}
        \centering
    \begin{tikzpicture}
    \node[inner sep=0pt] (img) at (0,0)
    {\includegraphics[trim=0cm 0cm 3.8cm 0, clip, width=\linewidth]{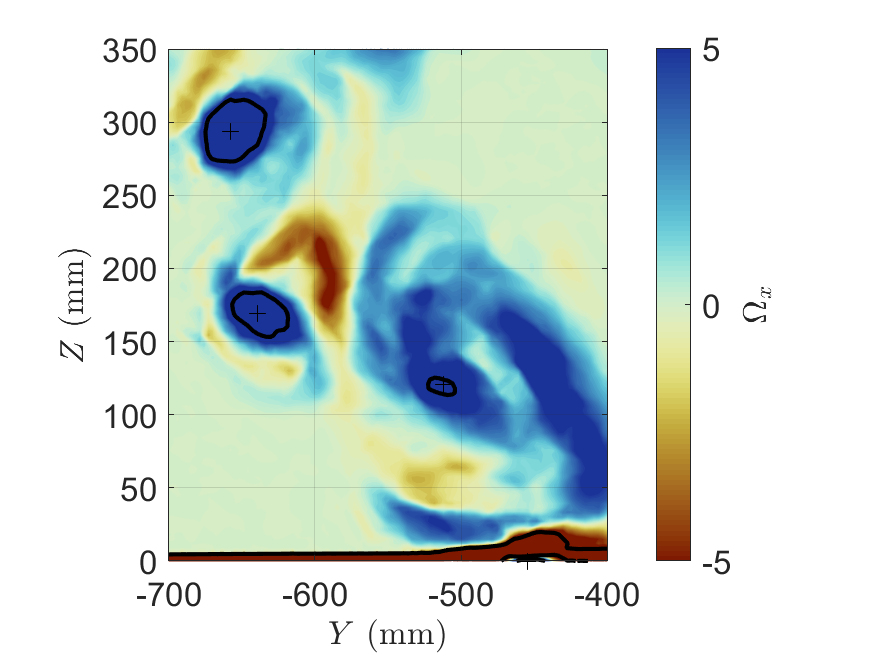}};
    \node at (-1,1.4) {\footnotesize\textbf{B}};
    \node at (-1,0.2) {\footnotesize\textbf{C}};
    \node at (-0.3,0.7) {\footnotesize\textbf{E}};
    \node at (0.8,0.6) {\footnotesize\textbf{A/D\textsubscript{in}}};
    \node at (0.1,-1) {\footnotesize \textbf{D\textsubscript{out}}};
\end{tikzpicture}
        \caption{LES - Plane 4}
        \label{fig:Omnum_Plane4}
    \end{subfigure}    
    \begin{subfigure}[b]{0.29\textwidth}
        \centering
    \begin{tikzpicture}
    \node[inner sep=0pt] (img) at (0,0)
    {\includegraphics[trim=0cm 0cm 3.8cm 0, clip, width=\linewidth]{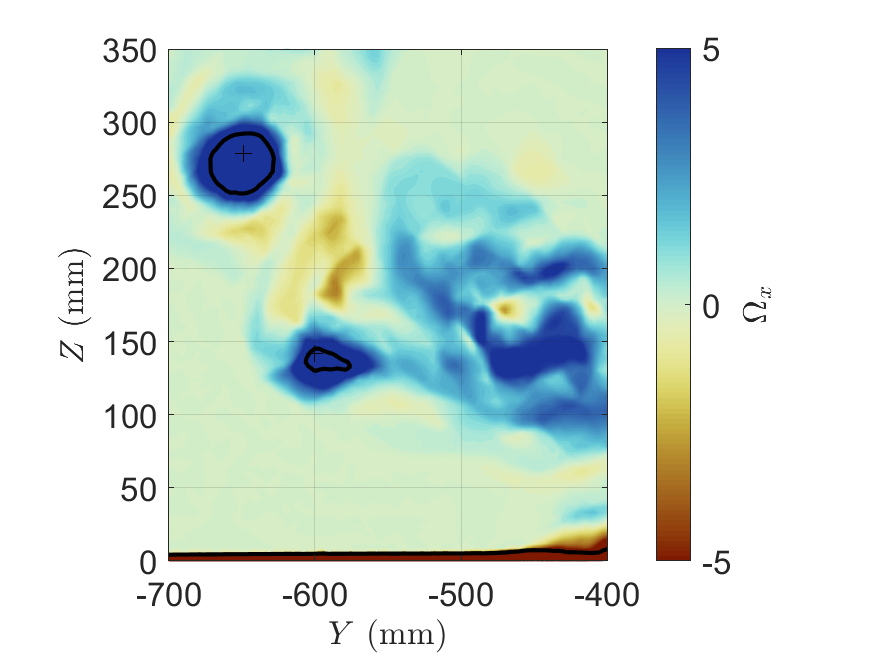}};
    \node at (-1,1.3) {\footnotesize\textbf{B}};
    \node at (-0.5,-0.2) {\footnotesize\textbf{C}};
    \node at (0,0.7) {\footnotesize\textbf{E}};
    \node at (1,0.7) {\footnotesize\textbf{A/\textbf{D\textsubscript{in}}}};
\end{tikzpicture}
        \caption{LES - Plane 5}
        \label{fig:Omnum_Plane5}
    \end{subfigure}
     \begin{subfigure}[b]{0.08\textwidth}
        \centering
\raisebox{0.25\height}{\includegraphics[trim=11cm 0cm 0cm 0, clip, width=\linewidth]{Figures/Plots_Isabella/NewFig/Plane3_Oxnum_mod.png}}
        \end{subfigure}
    \caption{Evolution of the non-dimensional streamwise vorticity at Planes 3, 4 and 5: a)-c) experiment, d)-f)  simulation.} 
    \label{fig:Om_Plane3to5}
\end{figure}

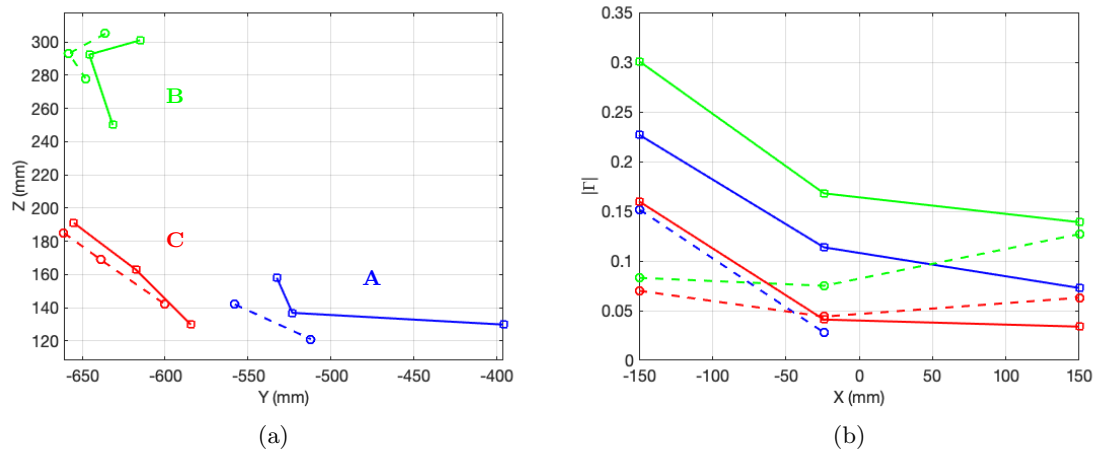
\begin{figure}[h]
   \centering
      \begin{subfigure}{0.49\textwidth}
    \begin{tikzpicture}
    \node[inner sep=0pt] (img) at (0,0)
    { \includegraphics[width=\linewidth]{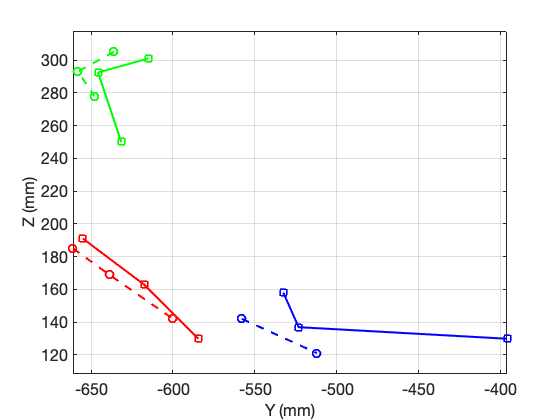}};
    \node at (-1.3,1.3) {\footnotesize\textcolor{green}{\textbf{B}}};
    \node at (1.3,-1.1) {\footnotesize\textcolor{blue}{\textbf{A}}};
    \node at (-1.3,-0.6) {\footnotesize\textcolor{red}{\textbf{C}}};
      \end{tikzpicture}
      \caption{ }
      \end{subfigure}
      \begin{subfigure}{0.49\textwidth}
\includegraphics[width=\linewidth]{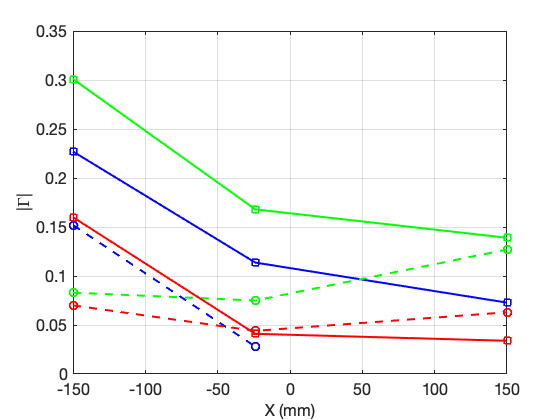}
      \caption{ }
      \label{fig:IFW_vort_circ}
      \end{subfigure}
      \caption{a) location and b) circulation of the three main vortices in the simulation (dashed line with rounded marker) compared to the experiment (continuous line with squared marker): blue vortex A, green vortex B, red vortex C.}
   \label{fig:IFW_vort_loc_circ}
\end{figure}

\subsection{Mean flow comparison}
\label{sec:main_flow}
In this section, the high-fidelity simulation results are validated against the experiments at the downstream planes defined in section \ref{sec:coordinate system}, which correspond to planes 3 to 5 from \cite{Buscariolo2022} with an extended domain in the vertical direction, to Z=350 mm, to include the flow around the endplate vortex B. 

\subsubsection{Mean vorticity}
\hfill

Starting with the streamwise component of the averaged vorticity, the differences between the two datasets are discussed in terms of three categories: (i) streamwise offset in vortex development, (ii) spatial diffusion and apparent resolution, and (iii) the secondary vorticity layer at the ground.

Figure \ref{fig:Om_Plane3to5} compares the averaged vorticity computed from the results of the simulation and the experiment at the three streamwise planes. The vortices are clearly visible, although in the simulation they appear stronger and wider, slightly less diffuse.\\
Figure \ref{fig:IFW_vort_circ}, reports the location of vortex core defined as the point with maximum vorticity and the circulation calculated in a contour region corresponding 80\% the maximum vorticity following \cite{CERRETELLI_WILLIAMSON_2003}. The latter corresponds to the regions identified by the black or blue line in figure \ref{fig:Om_Plane3to5}. Figure \ref{fig:IFW_vort_loc_circ} demonstrates that the vortices follow the same path in space in both the simulation and experiment although there is a small spanwise shift of about 20 mm in their core location. That could be due to small differences in alignment or a difference in the developing stage of the structures. The circulation calculated in figure \ref{fig:IFW_vort_circ} is generally higher in the simulation, reflecting the observation that the vortices appear generally stronger and wider compared to the experiment. The vortex B, and partially C, seem to increase the circulation as they develop downstream. Nevertheless this increment is minor and likely associated with the uncertainty on the method used to define the boundaries for calculating the circulation.\\ 

\begin{figure}[ht!]
    \centering
    \begin{subfigure}[b]{0.29\textwidth}
        \centering
    \begin{tikzpicture}
    \node[inner sep=0pt] (img) at (0,0)
    {\includegraphics[trim=0cm 0cm 3.8cm 0, clip, width=\linewidth]{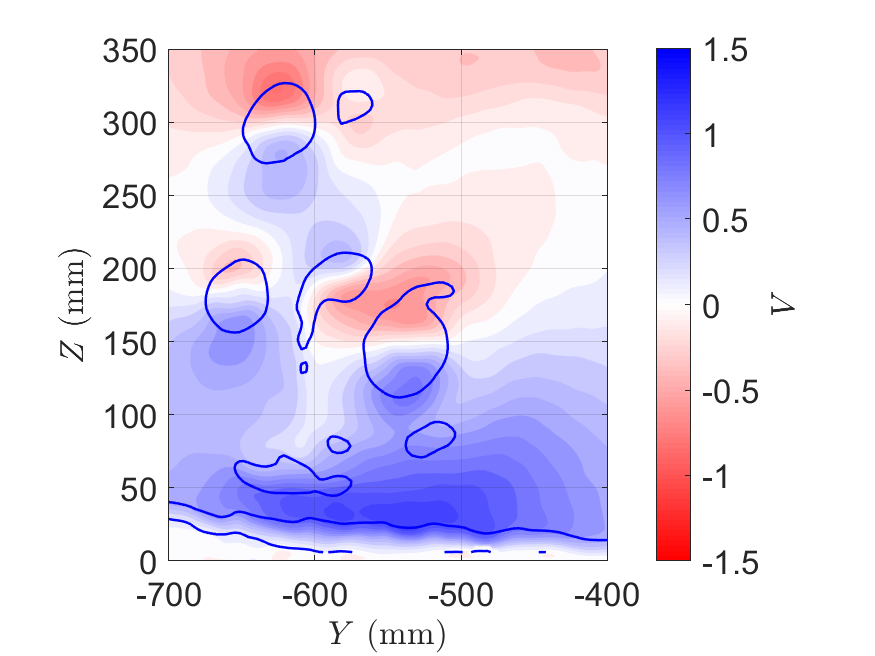}};
\node at (-0.8,1.3) {\footnotesize\textbf{B}};
    \node at (-0.8,0.6) {\footnotesize \textbf{C}};
    \node at (-0.1,0.6) {\footnotesize \textbf{E}};
    \node at (0.1,-0.1) {\footnotesize \textbf{A}};
    \node at (-0.8,-1) {\footnotesize \textbf{D\textsubscript{out}}};
    \node at (1.2,-0.8) {\footnotesize \textbf{D\textsubscript{in}}};
\end{tikzpicture}
        \caption{ PTV - Plane 3}
        \label{fig:Vexp_Plane3}
    \end{subfigure}
    \begin{subfigure}[b]{0.29\textwidth}
        \centering
    \begin{tikzpicture}
    \node[inner sep=0pt] (img) at (0,0)
    {\includegraphics[trim=0cm 0cm 3.8cm 0, clip, width=\linewidth]{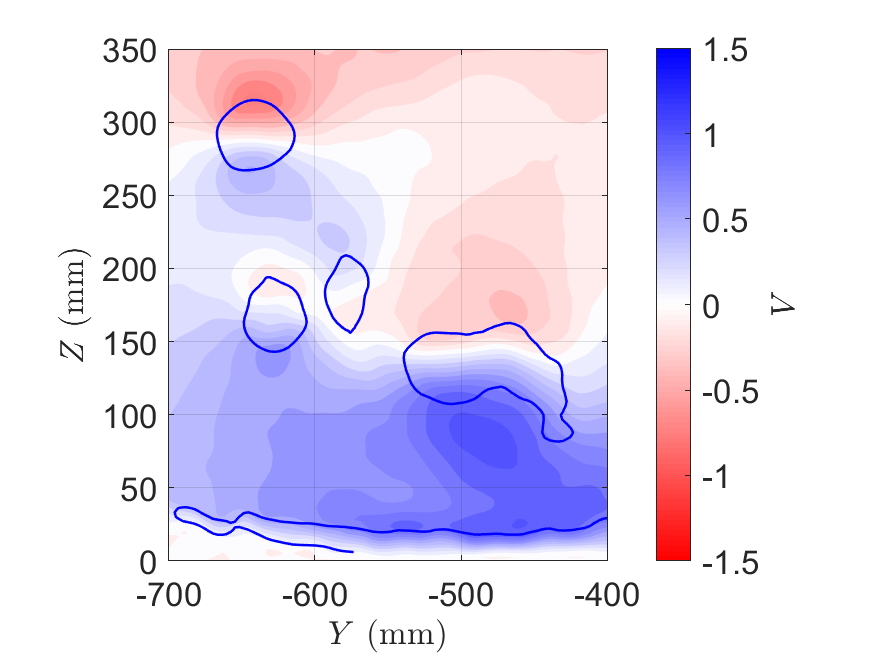}};
    \node at (-0.9,1.4) {\footnotesize \textbf{B}};
    \node at (-0.8,0.2) {\footnotesize \textbf{C}};
    \node at (-0.1,0.6) {\footnotesize \textbf{E}};
    \node at (0.8,0.2) {\footnotesize \textbf{A/\textbf{D\textsubscript{in}}}};
    \end{tikzpicture}
        \caption{ PTV - Plane 4 }
        \label{fig:Vexp_Plane4}
    \end{subfigure}    
    \begin{subfigure}[b]{0.29\textwidth}
        \centering
    \begin{tikzpicture}
    \node[inner sep=0pt] (img) at (0,0)
    {\includegraphics[trim=0cm 0cm 3.8cm 0, clip, width=\linewidth]{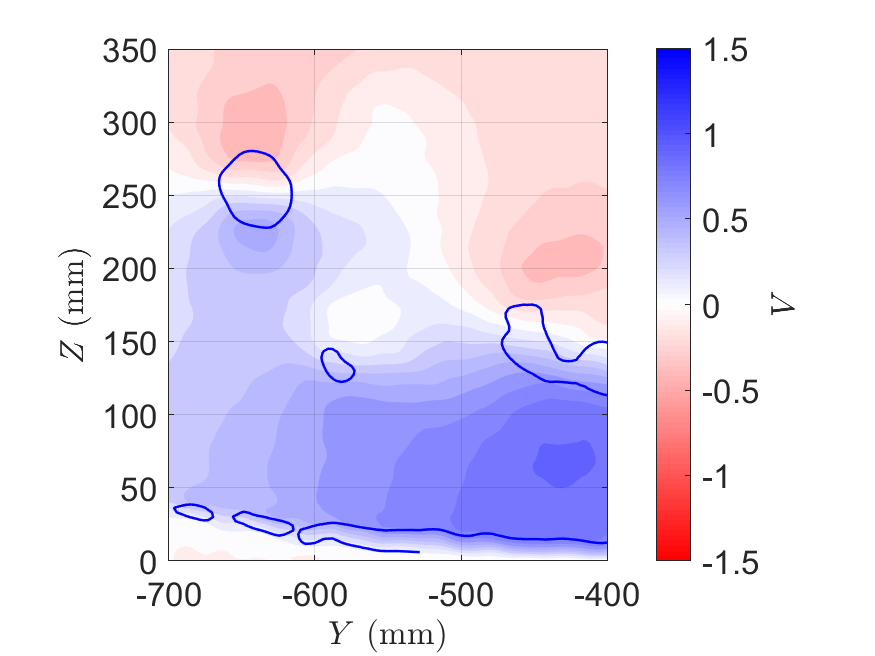}};
    \node at (-0.9,1.1) {\footnotesize \textbf{B}};
    \node at (-0.3,-0.3) {\footnotesize \textbf{C}};
    \node at (0,0.3) {\footnotesize \textbf{E}};
    \node at (1.2,0.3) {\footnotesize \textbf{A/\textbf{D\textsubscript{in}}}};
\end{tikzpicture}
        \caption{ PTV - Plane 5 }
        \label{fig:Vdiff_Plane5}
    \end{subfigure}
     \begin{subfigure}[b]{0.08\textwidth}
        \centering
\raisebox{0.25\height}{\includegraphics[trim=11cm 0cm 0cm 0, clip, width=\linewidth]{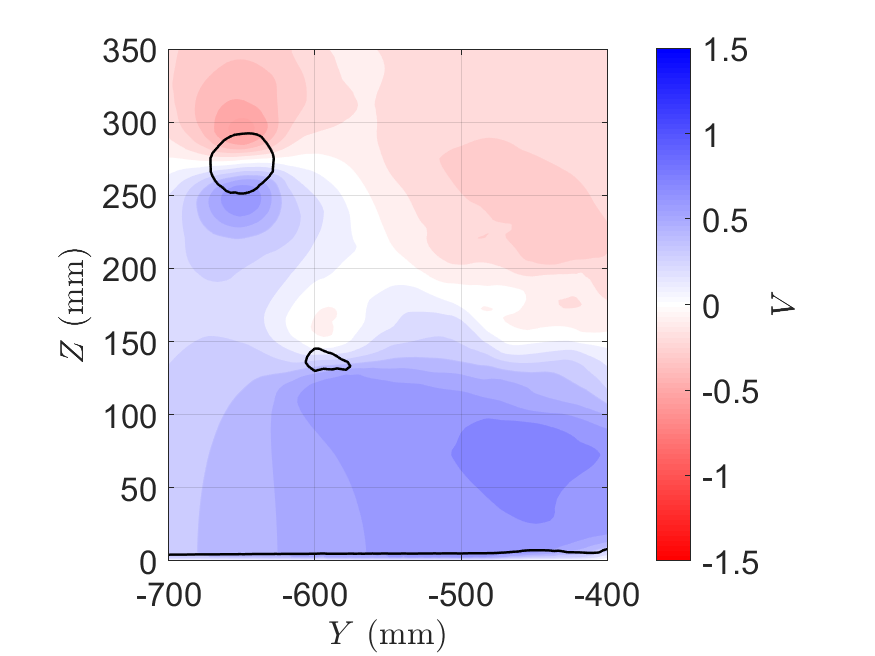}}
        \end{subfigure}
    \begin{subfigure}[b]{0.29\textwidth}
        \centering
    \begin{tikzpicture}
    \node[inner sep=0pt] (img) at (0,0)
    {\includegraphics[trim=0cm 0cm 3.8cm 0, clip, width=\linewidth]{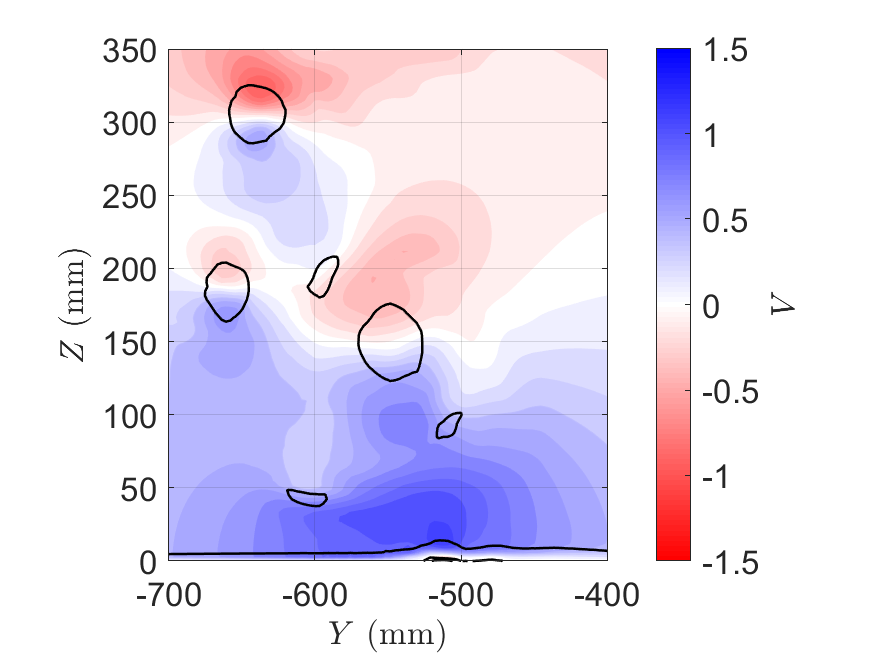}};
    \node at (-0.9,1.3) {\footnotesize \textbf{B}};
    \node at (-1,0.5) {\footnotesize \textbf{C}};
    \node at (-0.3,0.6) {\footnotesize \textbf{E}};
    \node at (0,-0.1) {\footnotesize \textbf{A}};
    \node at (-0.5,-1.3) {\footnotesize \textbf{D\textsubscript{out}}};
    \node at (1.4,-0.6) {\footnotesize \textbf{D\textsubscript{in}}};
\end{tikzpicture}
        \caption{ LES - Plane 3 }
        \label{fig:Vnum_Plane3}
    \end{subfigure}
    \begin{subfigure}[b]{0.29\textwidth}
        \centering
    \begin{tikzpicture}
    \node[inner sep=0pt] (img) at (0,0)
    {\includegraphics[trim=0cm 0cm 3.8cm 0, clip, width=\linewidth]{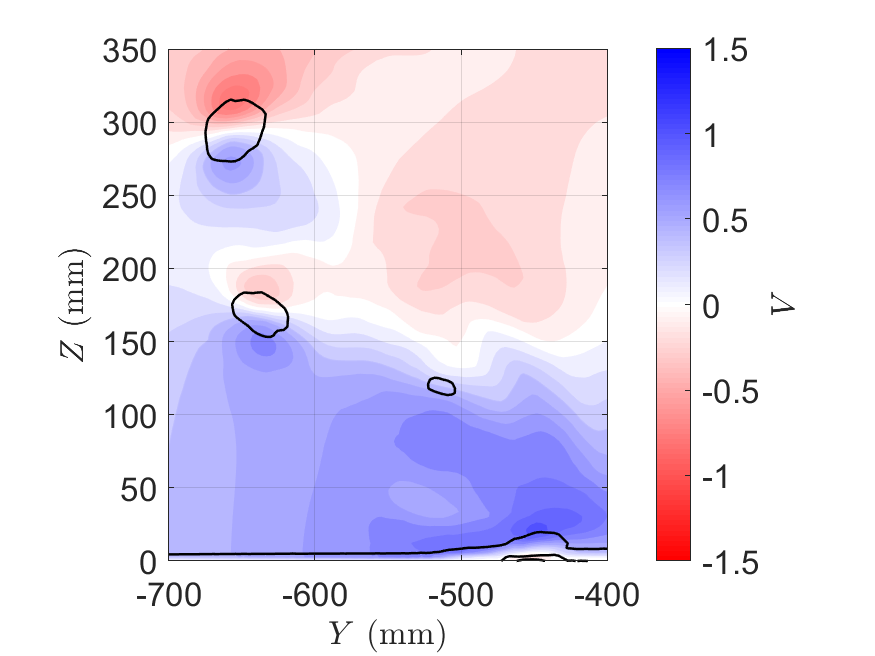}};
    \node at (-1,1.4) {\footnotesize\textbf{B}};
    \node at (-1,0.2) {\footnotesize\textbf{C}};
    \node at (-0.3,0.7) {\footnotesize\textbf{E}};
    \node at (0.8,0.6) {\footnotesize\textbf{A/D\textsubscript{in}}};
    \node at (0.1,-1) {\footnotesize \textbf{D\textsubscript{out}}};
\end{tikzpicture}
        \caption{ LES - Plane 4 }
        \label{fig:Vnum_Plane4}
    \end{subfigure}    
    \begin{subfigure}[b]{0.29\textwidth}
        \centering
   \begin{tikzpicture}
    \node[inner sep=0pt] (img) at (0,0)
    {\includegraphics[trim=0cm 0cm 3.8cm 0, clip, width=\linewidth]{Figures/Plots_Isabella/NewFig/Plane5_Vnum.png}};
    \node at (-1,1.3) {\footnotesize\textbf{B}};
    \node at (-0.5,-0.2) {\footnotesize\textbf{C}};
    \node at (0,0.7) {\footnotesize\textbf{E}};
    \node at (1,0.7) {\footnotesize\textbf{A/\textbf{D\textsubscript{in}}}};
\end{tikzpicture}
        \caption{ LES - Plane 5 }
        \label{fig:Vnum_Plane5}
    \end{subfigure}
     \begin{subfigure}[b]{0.08\textwidth}
        \centering
\raisebox{0.25\height}{\includegraphics[trim=11cm 0cm 0cm 0, clip, width=\linewidth]{Figures/Plots_Isabella/NewFig/Plane5_Vnum.png}}
        \end{subfigure}
    \begin{subfigure}[b]{0.29\textwidth}
        \centering
    \begin{tikzpicture}
    \node[inner sep=0pt] (img) at (0,0)
    {\includegraphics[trim=0cm 0cm 3.8cm 0, clip, width=\linewidth]{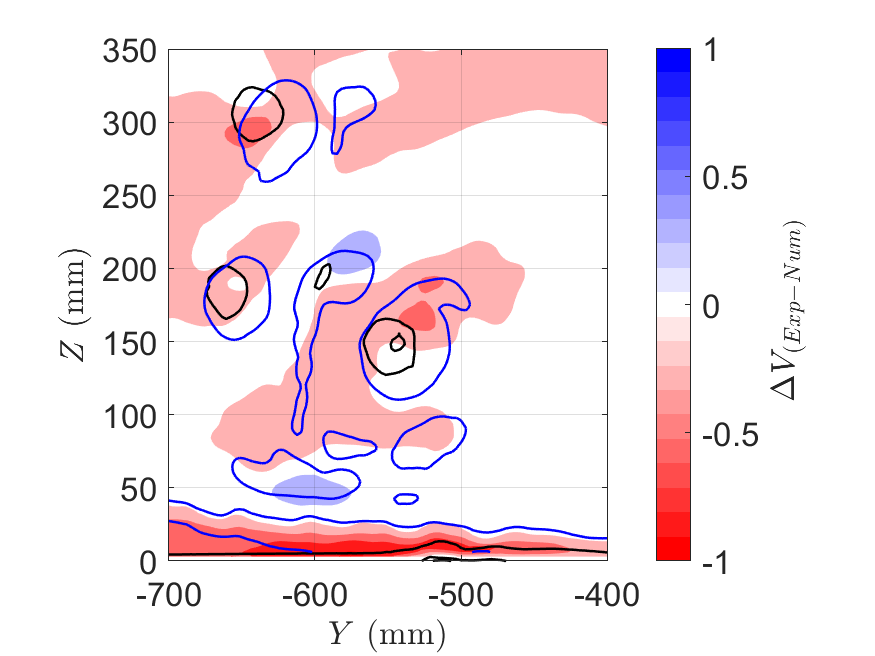}};
\end{tikzpicture}
        \caption{ Diff - Plane 3 }
        \label{fig:Vdiff_Plane3}
    \end{subfigure}
    \begin{subfigure}[b]{0.29\textwidth}
        \centering
    \begin{tikzpicture}
    \node[inner sep=0pt] (img) at (0,0)
    {\includegraphics[trim=0cm 0cm 3.8cm 0, clip, width=\linewidth]{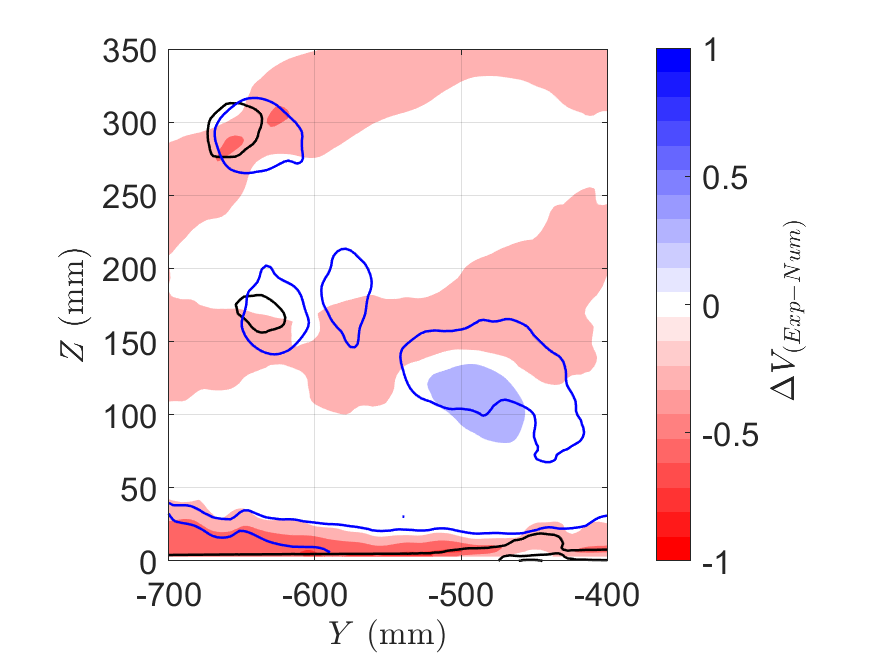}};
\end{tikzpicture}
        \caption{ Diff - Plane 4 }
        \label{fig:Vdiff_Plane4}
    \end{subfigure}    
    \begin{subfigure}[b]{0.29\textwidth}
        \centering
\includegraphics[trim=0cm 0cm 3.8cm 0, clip, width=\linewidth]{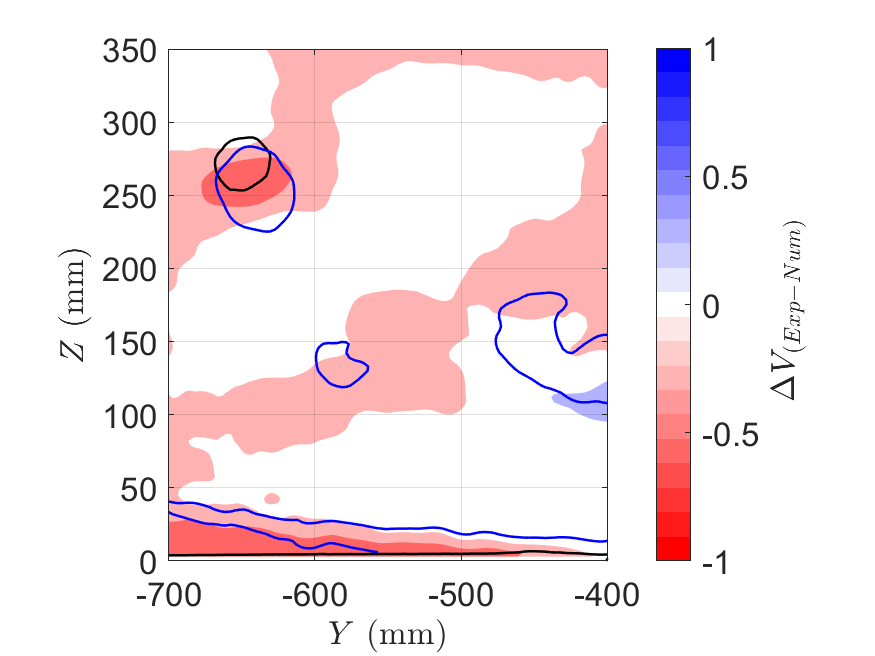}
        \caption{ Diff - Plane 5 }
        \label{fig:Vdiff_Plane5}
    \end{subfigure}
     \begin{subfigure}[b]{0.08\textwidth}
        \centering
\raisebox{0.25\height}{\includegraphics[trim=11cm 0cm 0cm 0, clip, width=\linewidth]{Figures/Plots_Isabella/NewFig/Plane5_Vdiff.png}}
        \end{subfigure}
    \caption{Non-dimensional spanwise velocity component at Plane 3, 4 and 5: a)-c) results from experiment, d)-f) results from simulation, g)-i) difference. The contour of vorticity is superimposed to the plots: blue for experiment and black for simulation.} 
    \label{fig:V_Plane3to5}
\end{figure}

In plane 3, figures \ref{fig:Omexp_Plane3} and \ref{fig:Omnum_Plane3}, all six vortices are clearly evident and at a very similar state of evolution. A secondary negative vorticity patch is also noticeable next to the endplate vortex (B) in both experiment and simulation, which is convected downstream. Looking at the downstream development of the vortices as a whole, an offset between the simulation and the experiment can be noticed. For instance, the interaction vortex (E) is oriented somehow more vertically in the simulation at plane 3, which recalls more the shape in plane 4. In addition, the downstream development could also be confirmed by the angle between the main vortex and the footplate vortex in the orbiting process. In both figure \ref{fig:Omnum_Plane3} and  \ref{fig:Omnum_Plane4} the angle formed by the line connecting the centre of vorticity is larger than the one in the results from the experiment figures \ref{fig:Omexp_Plane3} and \ref{fig:Omexp_Plane4}.\\ 

The most substantive discrepancy between the two datasets is the boundary layer vorticity at the ground, a region poorly resolved by prior PIV measurements \cite{Buscariolo2022} and therefore a specific target of the present volumetric PTV dataset. That is the region of strong negative vorticity close to the ground which is a consequence of having a system of vortices operating in ground effect \cite{Puel2000interaction}. The region of secondary vorticity is thin lying close to the ground in the simulation, while in the experiment it is thicker and lifted off the ground in the outbound regions. This discrepancy is present in all planes considered. Nevertheless, it is interesting to notice that the simulation manages to capture the secondary vorticity peeling off the ground around Y=-500 mm in figure \ref{fig:Omnum_Plane3} and around Y=-450 mm in figure \ref{fig:Omnum_Plane4}. That eject of secondary vorticity, which is in agreement with literature of vortices in ground effect \cite{Puel2000interaction}, is more clearly observed in other locations in the experiment, figures \ref{fig:Vort4} and \ref{fig:Vort5}. This would need a dedicated sensitivity study on the ability of the simulation to resolve the near-ground flow and a validation of the PTV results in that location. A secondary vortex of positive vorticity does not quite form (figure \ref{fig:Omnum_Plane4}).

In Plane 4, figures \ref{fig:Omexp_Plane4} and \ref{fig:Omnum_Plane4}, the main difference is the outward foot plate vortex (D\textsubscript{out}) which is almost fully dissipated in the experiment while it has still quite a lot of energy in the simulation. 
\begin{figure}[ht!]
    \centering
    \begin{subfigure}[b]{0.29\textwidth}
        \centering
    \begin{tikzpicture}
    \node[inner sep=0pt] (img) at (0,0)
    {\includegraphics[trim=0cm 0cm 3.8cm 0, clip,width=\linewidth]{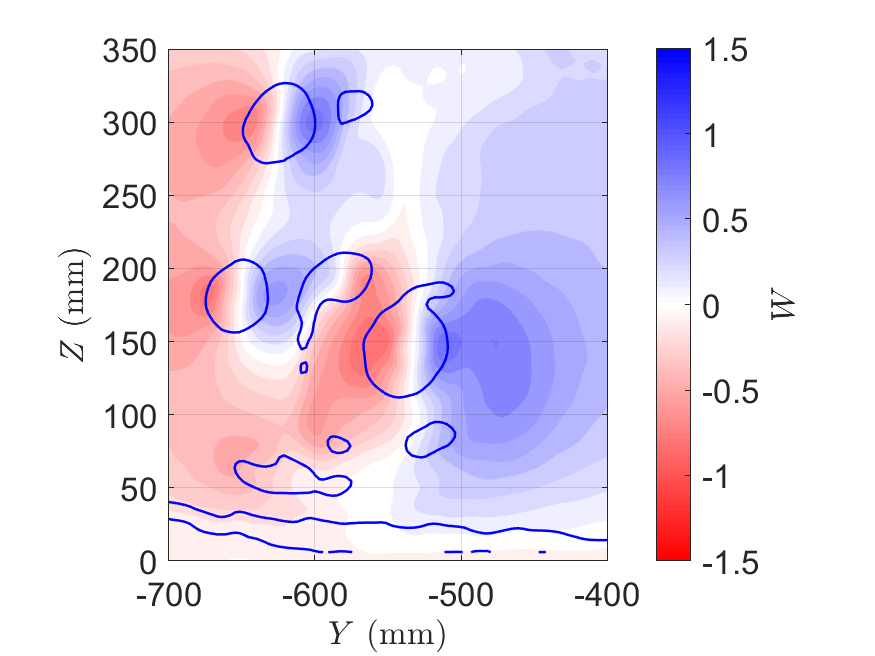}};
\node at (-0.8,1.3) {\footnotesize\textbf{B}};
    \node at (-0.8,0.6) {\footnotesize \textbf{C}};
    \node at (-0.1,0.6) {\footnotesize \textbf{E}};
    \node at (0.1,-0.1) {\footnotesize \textbf{A}};
    \node at (-0.8,-1) {\footnotesize \textbf{D\textsubscript{out}}};
    \node at (1.2,-0.8) {\footnotesize \textbf{D\textsubscript{in}}};
\end{tikzpicture}
        \caption{ PTV - Plane 3}
        \label{fig:Wexp_Plane3}
    \end{subfigure}
    \begin{subfigure}[b]{0.29\textwidth}
        \centering
    \begin{tikzpicture}
    \node[inner sep=0pt] (img) at (0,0)
    {\includegraphics[trim=0cm 0cm 3.8cm 0, clip,width=\linewidth]{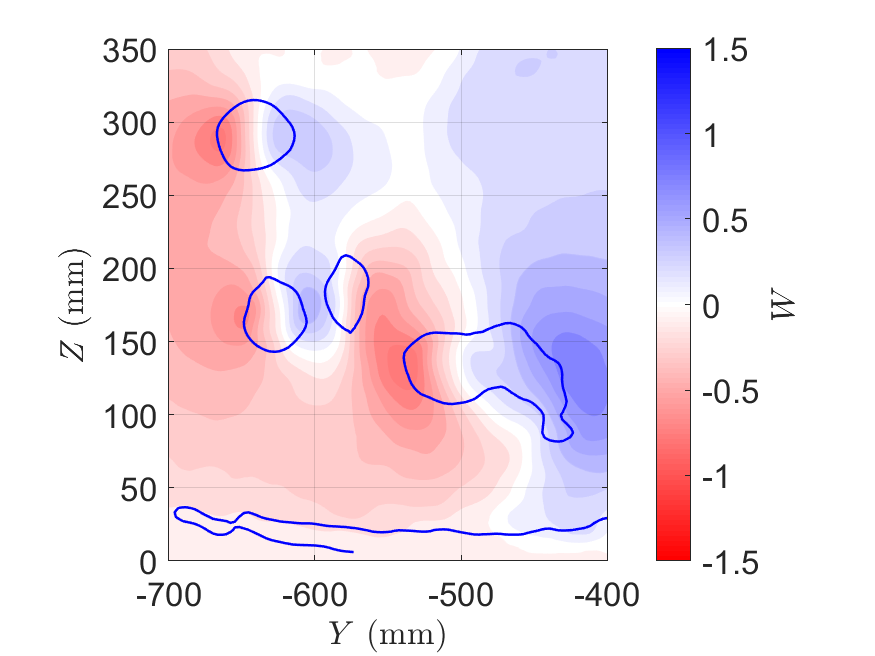}};
    \node at (-0.9,1.4) {\footnotesize \textbf{B}};
    \node at (-0.8,0.2) {\footnotesize \textbf{C}};
    \node at (-0.1,0.6) {\footnotesize \textbf{E}};
    \node at (0.8,0.2) {\footnotesize \textbf{A/\textbf{D\textsubscript{in}}}};
    \end{tikzpicture}
        \caption{ PTV - Plane 4 }
        \label{fig:Wexp_Plane4}
    \end{subfigure}    
    \begin{subfigure}[b]{0.29\textwidth}
        \centering
    \begin{tikzpicture}
    \node[inner sep=0pt] (img) at (0,0)
    {\includegraphics[trim=0cm 0cm 3.8cm 0, clip,width=\linewidth]{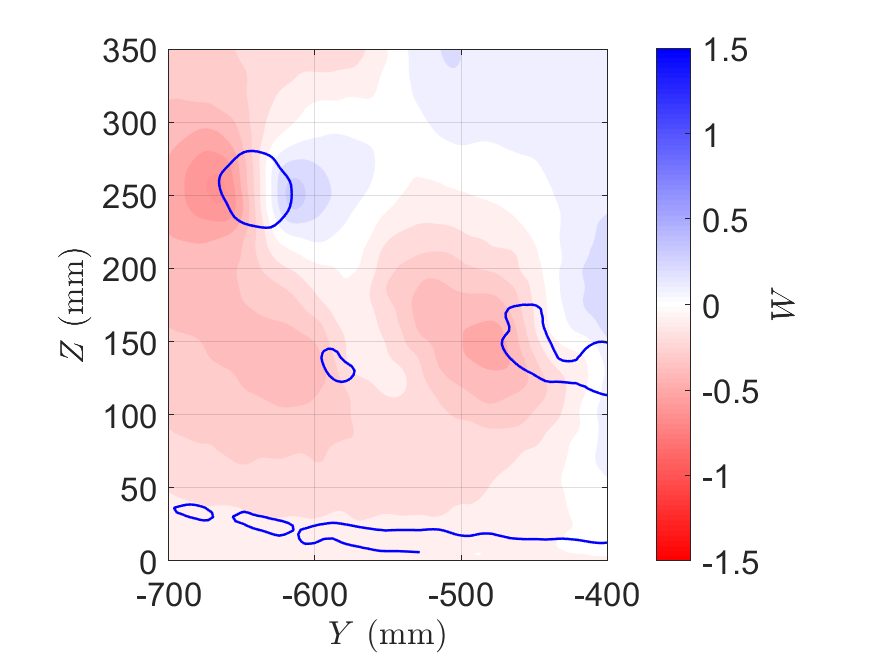}};
    \node at (-0.9,1.1) {\footnotesize \textbf{B}};
    \node at (-0.3,-0.3) {\footnotesize \textbf{C}};
    \node at (0,0.3) {\footnotesize \textbf{E}};
    \node at (1.2,0.3) {\footnotesize \textbf{A/\textbf{D\textsubscript{in}}}};
\end{tikzpicture}
        \caption{ PTV - Plane 5 }
        \label{fig:Wdiff_Plane5_exp}
    \end{subfigure}
             \begin{subfigure}[b]{0.08\textwidth}
        \centering
\raisebox{0.25\height}{\includegraphics[trim=11cm 0cm 0cm 0, clip, width=\linewidth]{Figures/Plots_Isabella/NewFig/Plane5_Wexp.png}}
        \end{subfigure}
    \begin{subfigure}[b]{0.29\textwidth}
        \centering
    \begin{tikzpicture}
    \node[inner sep=0pt] (img) at (0,0)
    {\includegraphics[trim=0cm 0cm 3.8cm 0, clip, width=\linewidth]{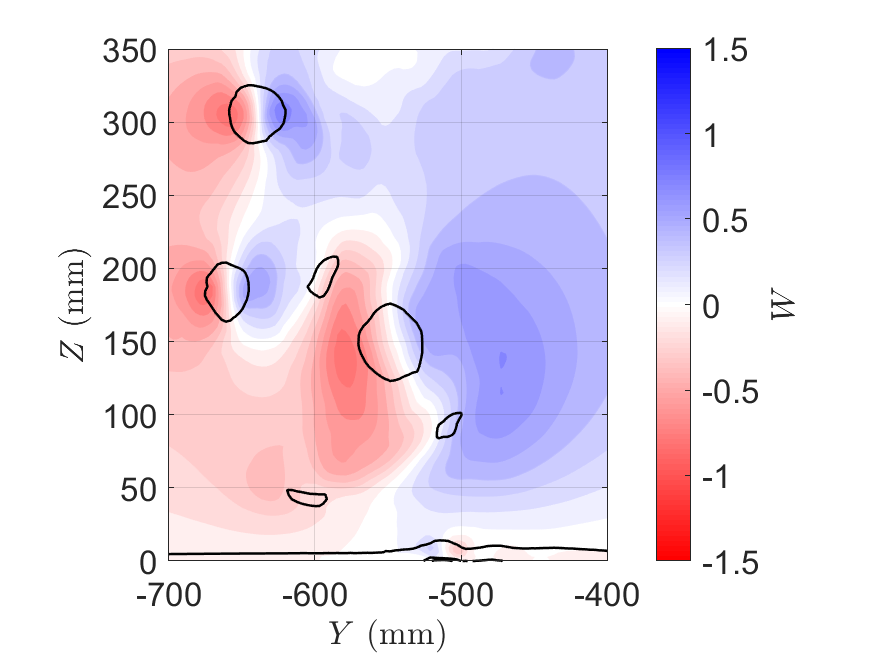}};
    \node at (-0.9,1.3) {\footnotesize \textbf{B}};
    \node at (-1,0.5) {\footnotesize \textbf{C}};
    \node at (-0.3,0.6) {\footnotesize \textbf{E}};
    \node at (0,-0.1) {\footnotesize \textbf{A}};
    \node at (-0.5,-1.3) {\footnotesize \textbf{D\textsubscript{out}}};
    \node at (1.4,-0.6) {\footnotesize \textbf{D\textsubscript{in}}};
\end{tikzpicture}
        \caption{ LES - Plane 3 }
        \label{fig:Wnum_Plane3}
    \end{subfigure}
    \begin{subfigure}[b]{0.29\textwidth}
        \centering
    \begin{tikzpicture}
    \node[inner sep=0pt] (img) at (0,0)
    {\includegraphics[trim=0cm 0cm 3.8cm 0, clip, width=\linewidth]{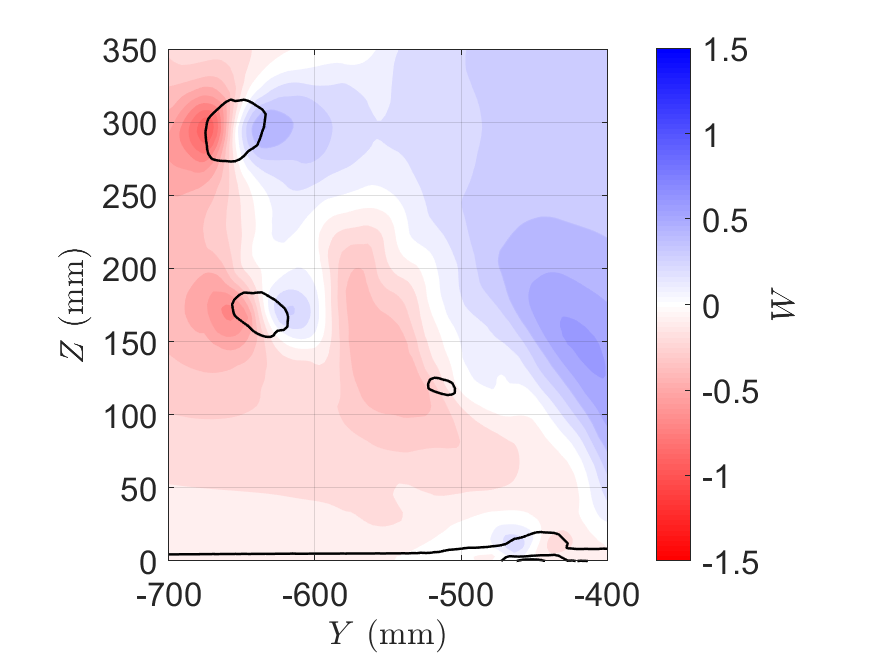}};
    \node at (-1,1.4) {\footnotesize\textbf{B}};
    \node at (-1,0.2) {\footnotesize\textbf{C}};
    \node at (-0.3,0.7) {\footnotesize\textbf{E}};
    \node at (0.8,0.6) {\footnotesize\textbf{A/D\textsubscript{in}}};
    \node at (0.1,-1) {\footnotesize \textbf{D\textsubscript{out}}};
\end{tikzpicture}
        \caption{ LES - Plane 4 }
        \label{fig:Wnum_Plane4}
    \end{subfigure}    
    \begin{subfigure}[b]{0.29\textwidth}
        \centering
   \begin{tikzpicture}
    \node[inner sep=0pt] (img) at (0,0)
    {\includegraphics[trim=0cm 0cm 3.8cm 0, clip,width=\linewidth]{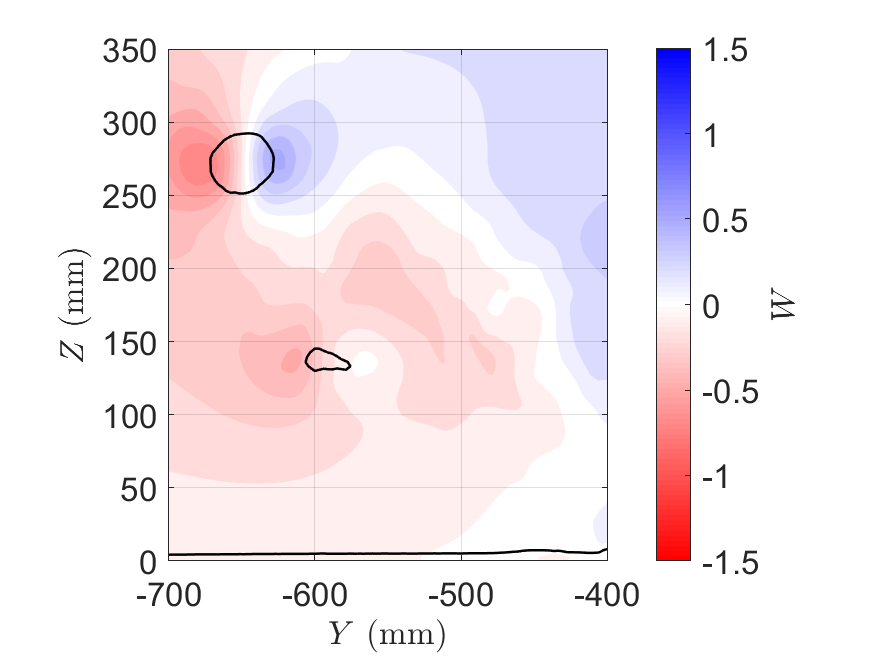}};
    \node at (-1,1.3) {\footnotesize\textbf{B}};
    \node at (-0.5,-0.2) {\footnotesize\textbf{C}};
    \node at (0,0.7) {\footnotesize\textbf{E}};
    \node at (1,0.7) {\footnotesize\textbf{A/\textbf{D\textsubscript{in}}}};
\end{tikzpicture}
        \caption{ LES - Plane 5 }
        \label{fig:Wnum_Plane5}
    \end{subfigure}
             \begin{subfigure}[b]{0.08\textwidth}
        \centering
\raisebox{0.25\height}{\includegraphics[trim=11cm 0cm 0cm 0, clip, width=\linewidth]{Figures/Plots_Isabella/NewFig/Plane5_Wnum.png}}
        \end{subfigure}
    \begin{subfigure}[b]{0.29\textwidth}
        \centering
    \begin{tikzpicture}
    \node[inner sep=0pt] (img) at (0,0)
    {\includegraphics[trim=0cm 0cm 3.8cm 0, clip,width=\linewidth]{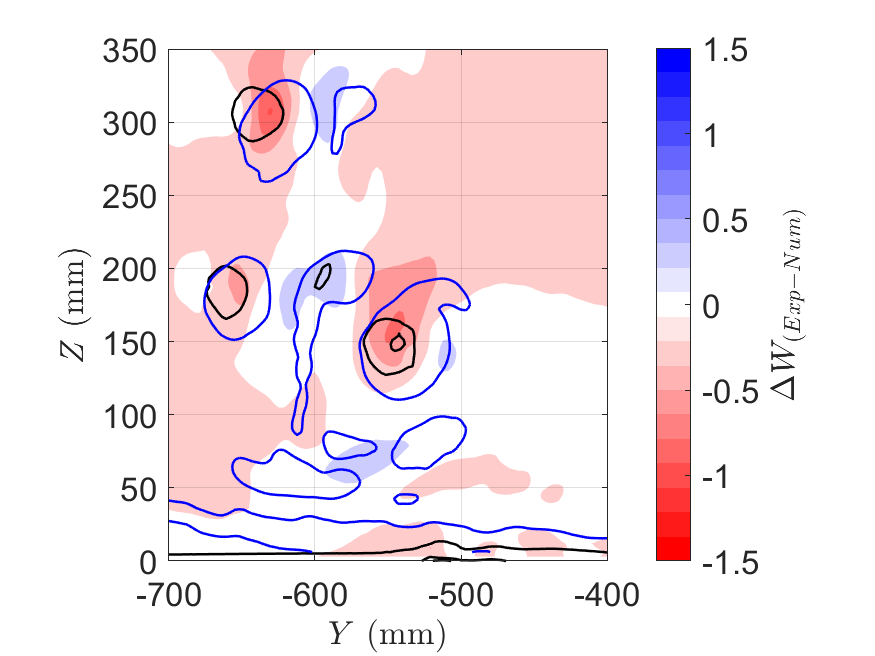}};
\end{tikzpicture}
        \caption{ Diff - Plane 3 }
        \label{fig:Wdiff_Plane3}
    \end{subfigure}
    \begin{subfigure}[b]{0.29\textwidth}
        \centering
    \begin{tikzpicture}
    \node[inner sep=0pt] (img) at (0,0)
    {\includegraphics[trim=0cm 0cm 3.8cm 0, clip,width=\linewidth]{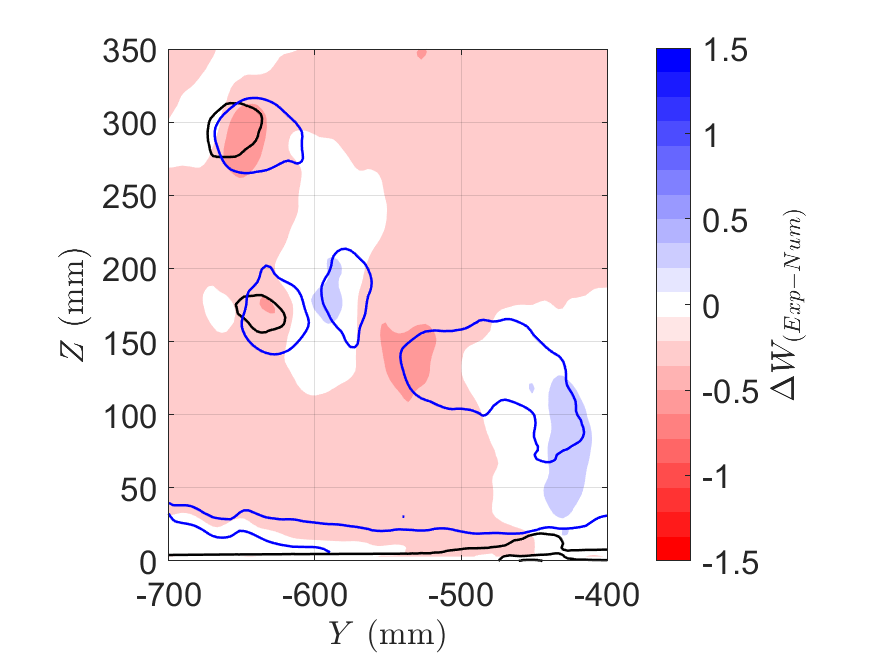}};
\end{tikzpicture}
        \caption{ Diff - Plane 4 }
        \label{fig:Wdiff_Plane4}
    \end{subfigure}    
    \begin{subfigure}[b]{0.29\textwidth}
        \centering
\includegraphics[trim=0cm 0cm 3.8cm 0, clip,width=\linewidth]{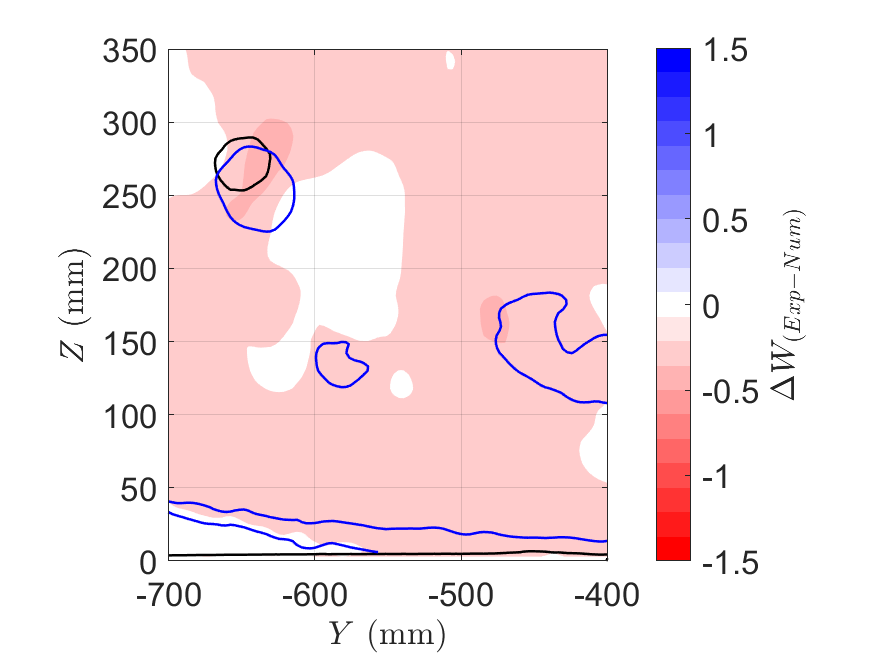};
        \caption{ Diff - Plane 5 }
        \label{fig:Wdiff_Plane5}
    \end{subfigure}
         \begin{subfigure}[b]{0.08\textwidth}
        \centering
\raisebox{0.25\height}{\includegraphics[trim=11cm 0cm 0cm 0, clip, width=\linewidth]{Figures/Plots_Isabella/NewFig/Plane5_Wdiff.png}}
        \end{subfigure}
    \caption{ Non-dimensional wall-normal velocity component at Plane 3, 4 and 5: a)-c) results from experiment, d)-f) results from simulation, g)-i) difference. The contour of vorticity is superimposed to the plots: blue for experiment and black for simulation.} 
    \label{fig:W_Plane3to5}
\end{figure}
The main vortex (A) and the inboard footplate vortex (D\textsubscript{in}) appear larger in the simulation. The canard vortex in \ref{fig:Omnum_Plane4} has a more elliptical shape in the simulation which could be in agreement with the idea of a small offset in vortex development stage between experiment and simulation.\\
Finally, in Plane 5, figure \ref{fig:Omexp_Plane5} and \ref{fig:Omnum_Plane5}, the results are again very similar with the remaining vorticities being stronger in the in the simulation than in the experiment, and the main vortex and footplate vortex are again more diffused in space.
Despite detailed discrepancies the results are well align in the description of the overall flow development. 

\subsubsection{Mean velocity}
\begin{figure}[ht!]
    \centering
    \begin{subfigure}[b]{0.29\linewidth}
        \centering
    \begin{tikzpicture}
    \node[inner sep=0pt] (img) at (0,0)
    {\includegraphics[trim=0cm 0cm 3.8cm 0, clip,width=\linewidth]{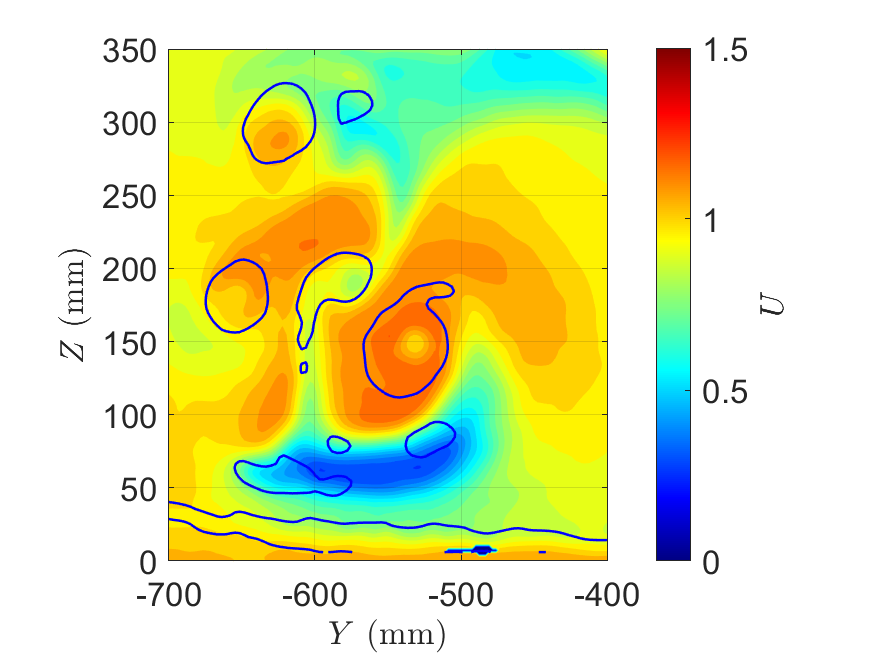}};
\node at (-0.8,1.3) {\footnotesize\textbf{B}};
    \node at (-0.8,0.6) {\footnotesize \textbf{C}};
    \node at (-0.1,0.6) {\footnotesize \textbf{E}};
    \node at (0.1,-0.1) {\footnotesize \textbf{A}};
    \node at (-0.8,-1) {\footnotesize \textbf{D\textsubscript{out}}};
    \node at (1.2,-0.8) {\footnotesize \textbf{D\textsubscript{in}}};
\end{tikzpicture}
        \caption{ PTV - Plane 3}
        \label{fig:Uexp_Plane3}
    \end{subfigure}
    \begin{subfigure}[b]{0.29\linewidth}
        \centering
    \begin{tikzpicture}
    \node[inner sep=0pt] (img) at (0,0)
    {\includegraphics[trim=0cm 0cm 3.8cm 0, clip,width=\linewidth]{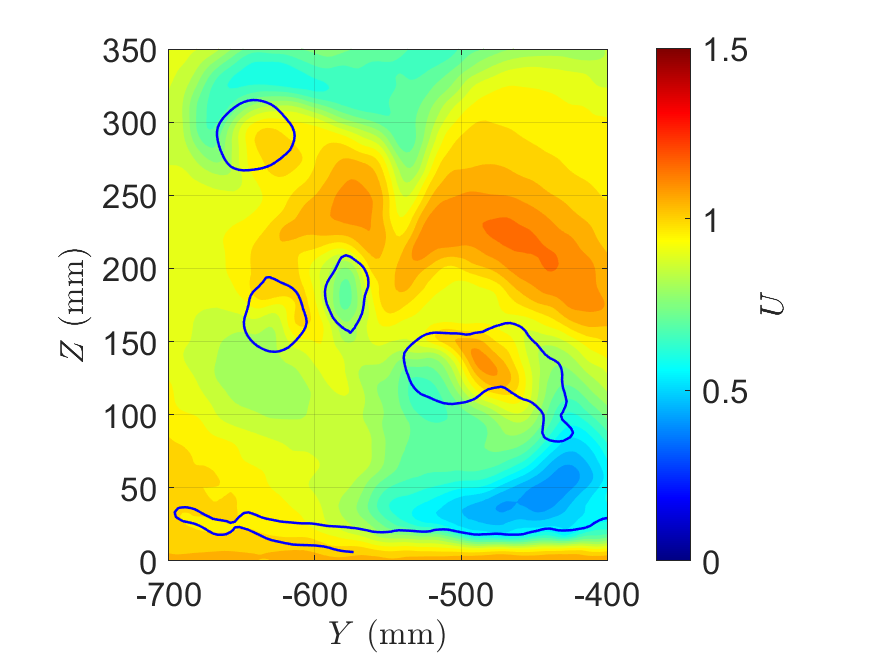}};
    \node at (-0.9,1.4) {\footnotesize \textbf{B}};
    \node at (-0.8,0.2) {\footnotesize \textbf{C}};
    \node at (-0.1,0.6) {\footnotesize \textbf{E}};
    \node at (0.8,0.2) {\footnotesize \textbf{A/\textbf{D\textsubscript{in}}}};
    \end{tikzpicture}
        \caption{ PTV - Plane 4 }
        \label{fig:Uexp_Plane4}
    \end{subfigure}    
    \begin{subfigure}[b]{0.29\linewidth}
        \centering
    \begin{tikzpicture}
    \node[inner sep=0pt] (img) at (0,0)
    {\includegraphics[trim=0cm 0cm 3.8cm 0, clip,width=\linewidth]{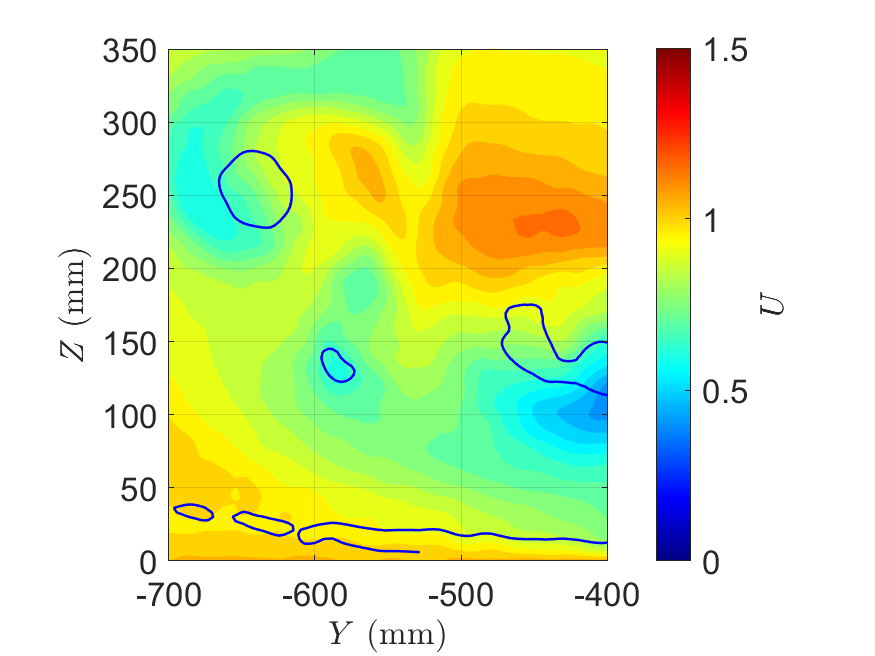}};
    \node at (-0.9,1.1) {\footnotesize \textbf{B}};
    \node at (-0.3,-0.3) {\footnotesize \textbf{C}};
    \node at (0,0.3) {\footnotesize \textbf{E}};
    \node at (1.2,0.3) {\footnotesize \textbf{A/\textbf{D\textsubscript{in}}}};
\end{tikzpicture}
        \caption{ PTV - Plane 5 }
        \label{fig:Udiff_Plane5}
    \end{subfigure}
             \begin{subfigure}[b]{0.08\textwidth}
        \centering
\raisebox{0.25\height}{\includegraphics[trim=11cm 0cm 0cm 0, clip, width=\linewidth]{Figures/Plots_Isabella/NewFig/Plane5_Uexp.png}}
        \end{subfigure}
    \begin{subfigure}[b]{0.29\linewidth}
        \centering
    \begin{tikzpicture}
    \node[inner sep=0pt] (img) at (0,0)
    {\includegraphics[trim=0cm 0cm 3.8cm 0, clip,width=\linewidth]{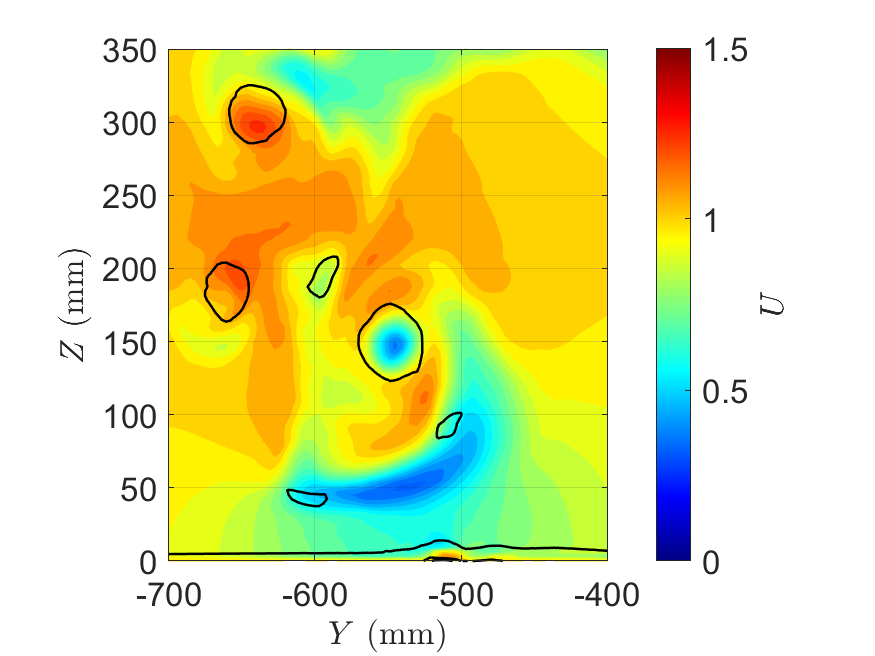}};
    \node at (-0.9,1.3) {\footnotesize \textbf{B}};
    \node at (-1,0.5) {\footnotesize \textbf{C}};
    \node at (-0.3,0.6) {\footnotesize \textbf{E}};
    \node at (0,-0.1) {\footnotesize \textbf{A}};
    \node at (-0.5,-1.3) {\footnotesize \textbf{D\textsubscript{out}}};
    \node at (1.4,-0.6) {\footnotesize \textbf{D\textsubscript{in}}};
\end{tikzpicture}
        \caption{ LES - Plane 3 }
        \label{fig:Unum_Plane3}
    \end{subfigure}
    \begin{subfigure}[b]{0.29\textwidth}
        \centering
    \begin{tikzpicture}
    \node[inner sep=0pt] (img) at (0,0)
    {\includegraphics[trim=0cm 0cm 3.8cm 0, clip,width=\linewidth]{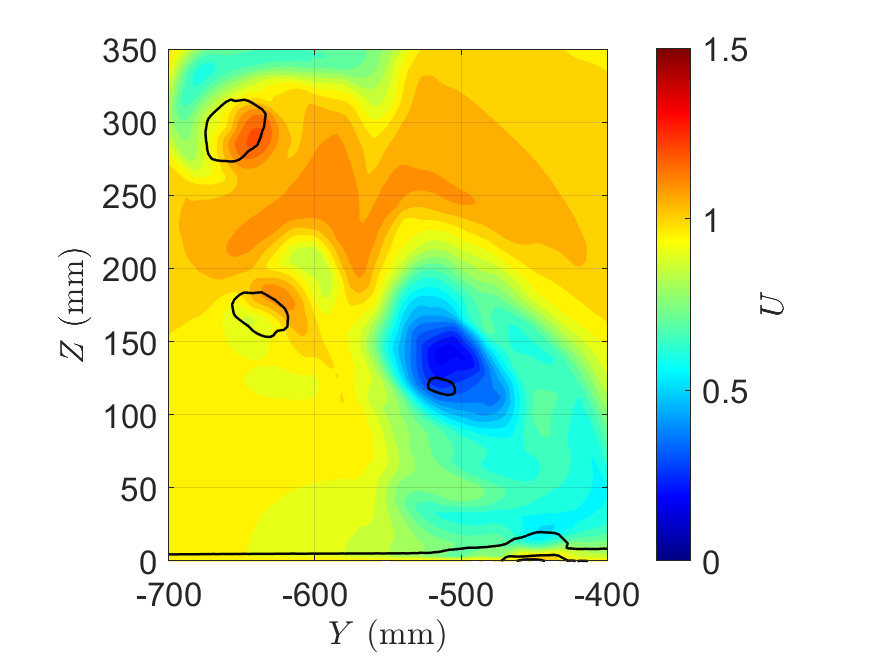}};
    \node at (-1,1.4) {\footnotesize\textbf{B}};
    \node at (-1,0.2) {\footnotesize\textbf{C}};
    \node at (-0.3,0.7) {\footnotesize\textbf{E}};
    \node at (0.8,0.6) {\footnotesize\textbf{A/D\textsubscript{in}}};
    \node at (0.1,-1) {\footnotesize \textbf{D\textsubscript{out}}};
\end{tikzpicture}
        \caption{ LES - Plane 4 }
        \label{fig:Unum_Plane4}
    \end{subfigure}    
    \begin{subfigure}[b]{0.29\textwidth}
        \centering
   \begin{tikzpicture}
    \node[inner sep=0pt] (img) at (0,0)
    {\includegraphics[trim=0cm 0cm 3.8cm 0, clip,width=\linewidth]{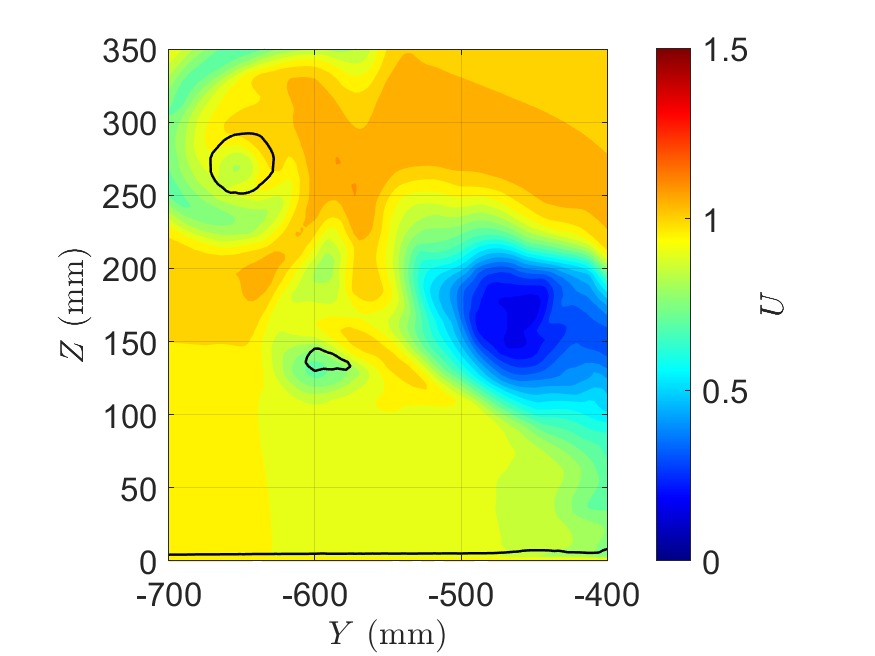}};
    \node at (-1,1.3) {\footnotesize\textbf{B}};
    \node at (-0.5,-0.2) {\footnotesize\textbf{C}};
    \node at (0,0.7) {\footnotesize\textbf{E}};
    \node at (1,0.7) {\footnotesize\textbf{A/\textbf{D\textsubscript{in}}}};
\end{tikzpicture}
        \caption{ LES - Plane 5 }
        \label{fig:Unum_Plane5}
    \end{subfigure}
             \begin{subfigure}[b]{0.08\textwidth}
        \centering
\raisebox{0.25\height}{\includegraphics[trim=11cm 0cm 0cm 0, clip, width=\linewidth]{Figures/Plots_Isabella/NewFig/Plane5_Unum.png}}
        \end{subfigure}
    \begin{subfigure}[b]{0.29\textwidth}
        \centering
    \begin{tikzpicture}
    \node[inner sep=0pt] (img) at (0,0)
    {\includegraphics[trim=0cm 0cm 3.8cm 0, clip,width=\linewidth]{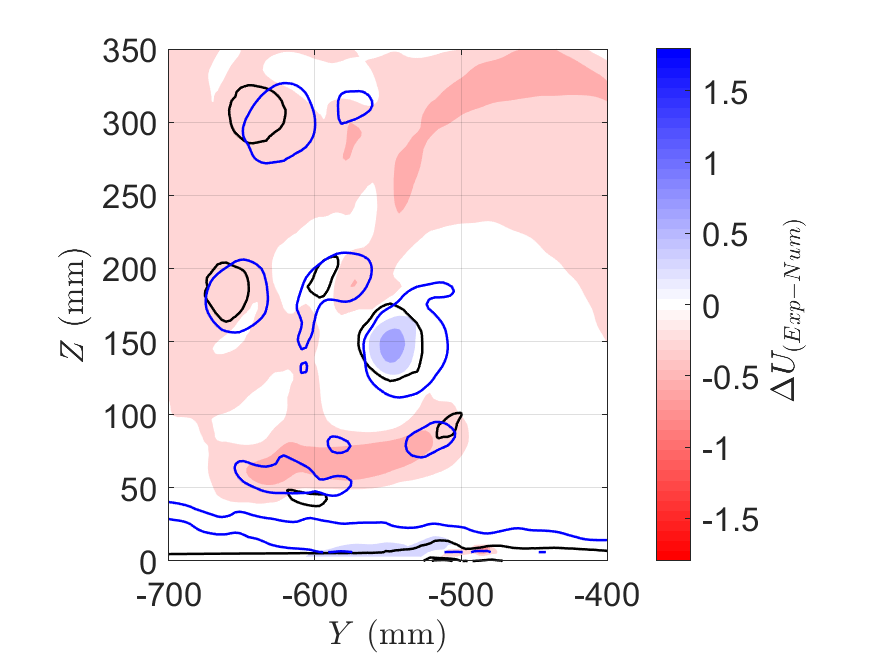}};
\end{tikzpicture}
        \caption{ Diff - Plane 3 }
        \label{fig:Udiff_Plane3}
    \end{subfigure}
    \begin{subfigure}[b]{0.29\textwidth}
        \centering
    \begin{tikzpicture}
    \node[inner sep=0pt] (img) at (0,0)
    {\includegraphics[trim=0cm 0cm 3.8cm 0, clip,width=\linewidth]{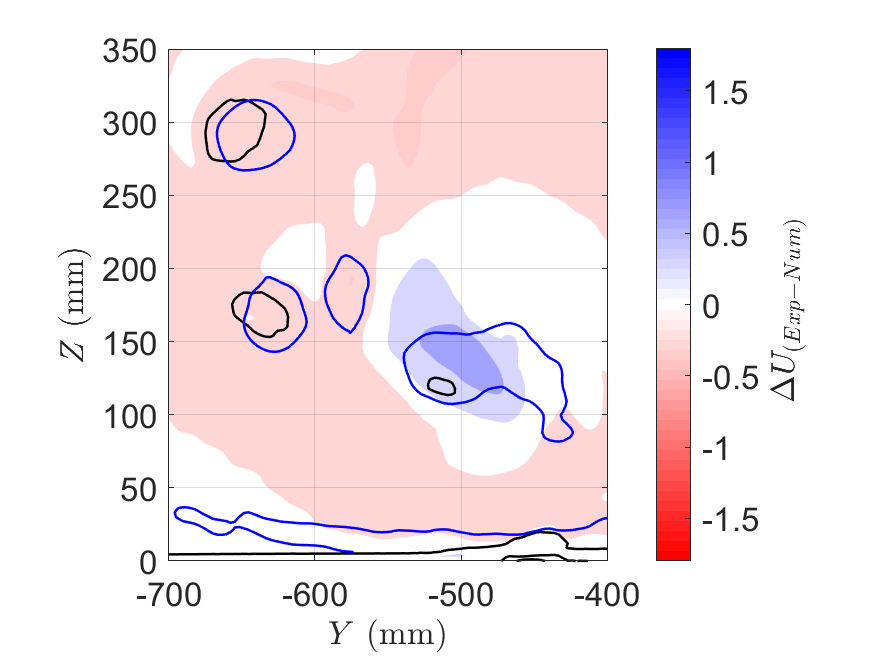}};
\end{tikzpicture}
        \caption{ Diff - Plane 4 }
        \label{fig:Udiff_Plane4}
    \end{subfigure}    
    \begin{subfigure}[b]{0.29\textwidth}
        \centering
\includegraphics[trim=0cm 0cm 3.8cm 0, clip,width=\linewidth]{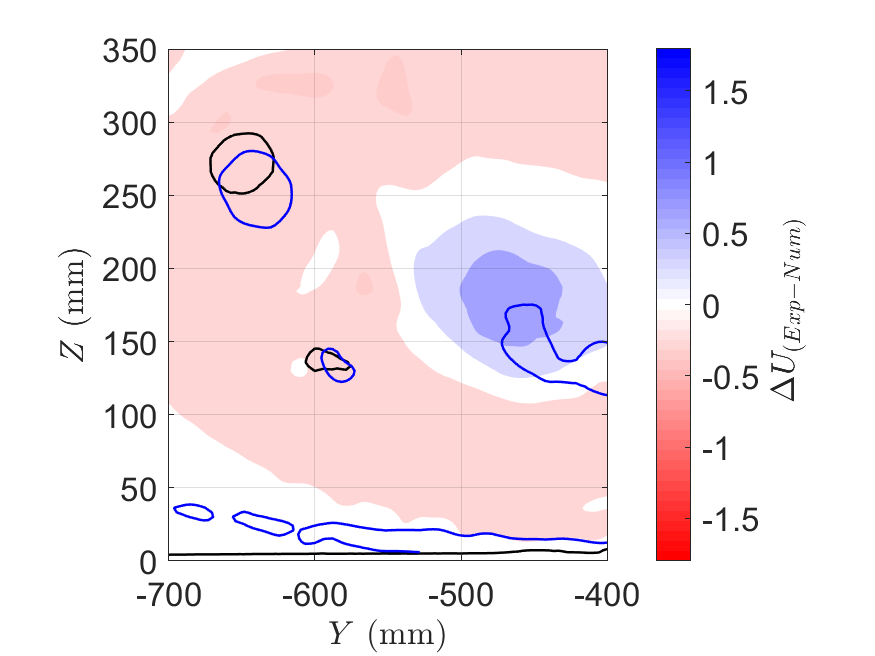}
        \caption{ Diff - Plane 5 }
        \label{fig:Udiff_Plane5}
    \end{subfigure}
             \begin{subfigure}[b]{0.08\textwidth}
        \centering
\raisebox{0.25\height}{\includegraphics[trim=11cm 0cm 0cm 0, clip, width=\linewidth]{Figures/Plots_Isabella/NewFig/Plane5_Udiff.png}}
        \end{subfigure}
    \caption{Non-dimensional streamwise velocity component at Plane 3, 4 and 5: a)-c) results from experiment, d)-f) results from simulation, g)-i) difference. The contour of vorticity is superimposed to the plots: blue for experiment and black for simulation.} 
    \label{fig:U_Plane3to5}
\end{figure}

 \hspace{1em}\\
Figures \ref{fig:V_Plane3to5} to \ref{fig:U_Plane3to5} show the evolution of the three velocity components across the same streamwise planes. To help interpret the data, the isocontour of the  streamwise vorticity is superimposed on the plots, blue is used for the experimental results, and black for the simulation. In addition, a point-to-point subtraction between the experimental results and the simulation is reported for a quantitative comparison at each domain location.
The agreement is generally good, as demonstrated by the average and maximum velocity difference within the domain reported in table \ref{tab:Plane3_vel_diff}. Nevertheless, there are some differences that are worth comment.

\begin{table}[h]
    \centering
    \begin{tabular}{|c|c|c|c|c|c|c|}
    \hline
    \rule{0pt}{3ex} &  $\overline{|\Delta U|}$ & $|\Delta U|_{max}$ &$\overline{|\Delta V|}$ & $|\Delta V|_{max}$&  $\overline{|\Delta W|}$ & $|\Delta W|_{max}$\\
    \hline
     Plane 3 &  0.11 & 0.77& 0.12 &  0.97  & 0.11 & 0.94 \\
    \hline
     Plane 4 &  0.13 & 0.86  & 0.11 & 0.71  & 0.12 &  0.58\\
    \hline
     Plane 5 &  0.11 & 0.77 & 0.12 & 0.96 &0.11 &  0.93\\
     \hline
     \end{tabular}
    \caption{Average and maximum difference of the three velocity components between experiment and simulation results at Plane 3 over the considered domain.}
    \label{tab:Plane3_vel_diff}
\end{table}

Analysing figures \ref{fig:V_Plane3to5} and \ref{fig:W_Plane3to5}, both spanwise and wall-normal velocity components are zero in the core of the vortices and change sign around them in agreement with the sense of rotation. Since all vortices, except for the interaction vortex E, rotate in the same direction the flow appears almost to be divided in two well-defined regions of flow moving respectively up or down and left or right. These regions are located at about the same place for both simulation and experiments. Close to the ground, around the location of the footplate vortex, a large region of strong positive spanwise velocity V is present, with absolute values comparable to the streamwise freestream velocity. That indicates that the flow, highly three-dimensional, is pushed inwards, towards the centre of the car. The intensity of the spanwise velocity, in this region of strong in-wash, is similar in the experiment and in the simulation, as also noticeable in the point-by-point difference (figure \ref{fig:Vdiff_Plane3}). This is an improvement from the simulations of O'Sullivan et al. \cite{OSullivan2025} which seem to over-predict the inwash when compared to the experiment. 
Nevertheless, the layer closest to the ground where the spanwise velocity goes to zero, in agreement with the boundary condition, is thicker in the experiment compared to the simulation particularly in the more outboard region of the domain. This difference, which leads to the larger values in the point-by-point plots (figure \ref{fig:Vdiff_Plane3} to \ref{fig:Vdiff_Plane5}) is the reason behind the differences in the near-ground vorticity described in the previous section.

The largest discrepancy between the simulation and the experiment is in the development of the streamwise velocity component. Although the averaged point-to-point difference remains relatively low when averaged across the entire Field of View (FoV), the region around the main vortex and the footplate ($-600$ mm $\le Y \leq -500$ mm) appears dissimilar (figures \ref{fig:Uexp_Plane3}, \ref{fig:Uexp_Plane4} and \ref{fig:Unum_Plane3}, \ref{fig:Unum_Plane4}). In particular, the region of low-speed, almost stagnant, flow within the core of vortex A observed in the simulation is not present in the experimental results or, more precisely, the velocity gradient across the vortex is very small. On the other hand, the low-speed region around the foot plate vortex is much wider in the experiment at plane 3, figure \ref{fig:Uexp_Plane3}. This discrepancy persists in the downstream planes, as shown in figures \ref{fig:Uexp_Plane3}, \ref{fig:Uexp_Plane4}, \ref{fig:Unum_Plane3}, \ref{fig:Unum_Plane4}. Unfortunately, comparison with the previous experiment is not possible because Buscariolo et al. \cite{Buscariolo2022} reported only the spanwise and streamwise velocity components. A possible reason for this difference may be in the flow conditions which were perturbed by the LNA in the experiment, but further tests will be needed to verify this hypothesis.

\subsection{Standard deviation}
\begin{figure}[ht!]
    \centering
    \begin{subfigure}[b]{0.32\textwidth}
\centering
    \begin{tikzpicture}
    \node[inner sep=0pt] (img) at (0,0)
    {\includegraphics[width=\linewidth]{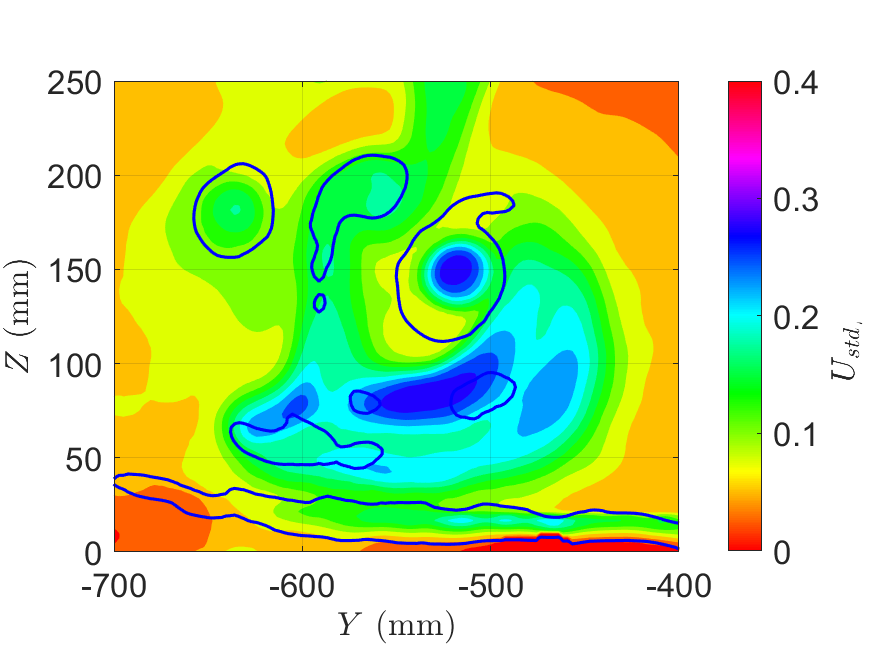}};
    \node at (-1.2,1.2) {\small \textbf{C}};
    \node at (-0.5,1.2) {\small \textbf{E}};
    \node at (-0.5,0.4) {\small \textbf{A}};
    \node at (-1.2,-0.8) {\footnotesize \textbf{D\textsubscript{out}}};
    \node at (0.5,-0.7) {\footnotesize \textbf{D\textsubscript{in}}};
\end{tikzpicture}
\caption{PTV - Plane 3}
    \label{fig:Ustdexp_Plane3}
    \end{subfigure} 
    \begin{subfigure}[b]{0.32\textwidth}
        \centering
    \begin{tikzpicture}
    \node[inner sep=0pt] (img) at (0,0)
    {\includegraphics[width=\linewidth]{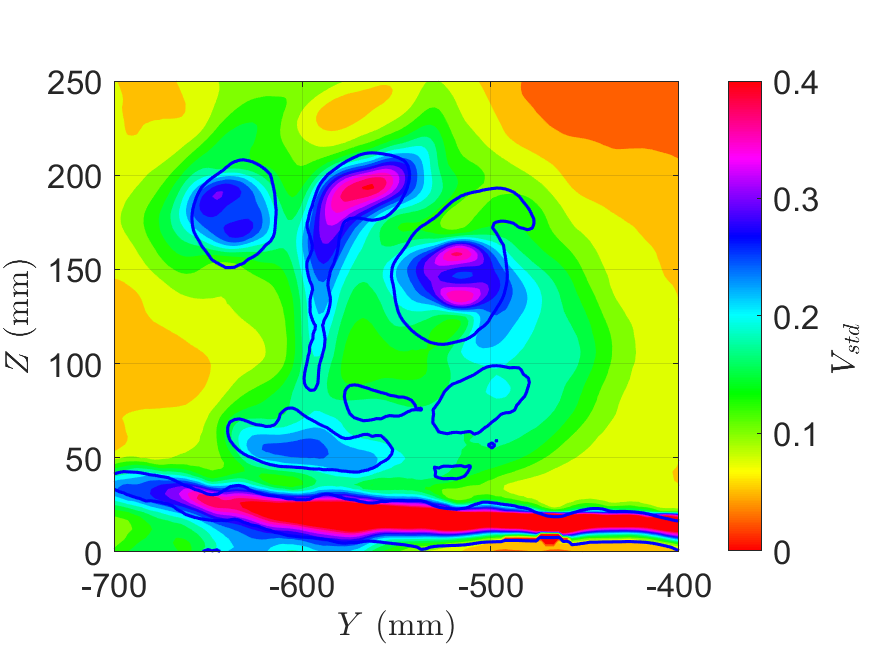}};
    \node at (-1.2,1.2) {\small \textbf{C}};
    \node at (-0.5,1.2) {\small \textbf{E}};
    \node at (-0.5,0.4) {\small \textbf{A}};
    \node at (-1.2,-0.8) {\footnotesize \textbf{D\textsubscript{out}}};
    \node at (0.5,-0.7) {\footnotesize \textbf{D\textsubscript{in}}};
\end{tikzpicture}
        \caption{PTV - Plane 3}
        \label{fig:Vstdexp_Plane3}
    \end{subfigure}  
    \begin{subfigure}[b]{0.32\textwidth}
        \centering
    \begin{tikzpicture}
    \node[inner sep=0pt] (img) at (0,0)
    {\includegraphics[width=\linewidth]{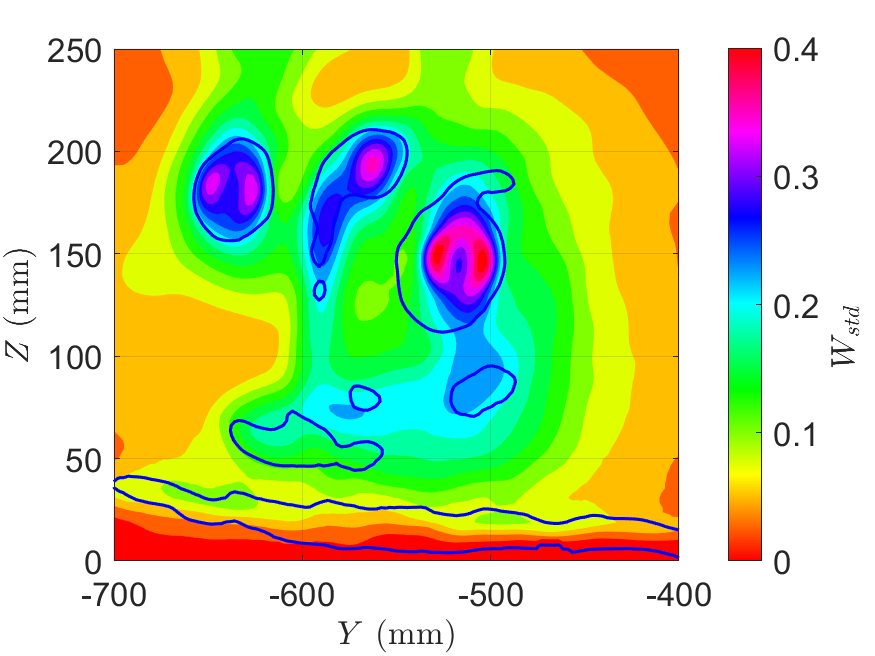}};
    \node at (-1.2,1.2) {\small \textbf{C}};
    \node at (-0.5,1.2) {\small \textbf{E}};
    \node at (-0.5,0.4) {\small \textbf{A}};
    \node at (-1.2,-0.8) {\footnotesize \textbf{D\textsubscript{out}}};
    \node at (0.5,-0.7) {\footnotesize \textbf{D\textsubscript{in}}};
\end{tikzpicture}
        \caption{PTV - Plane 3}
        \label{fig:Wstdexp_Plane3}
    \end{subfigure}
    \begin{subfigure}[b]{0.32\textwidth}
        \centering
    \begin{tikzpicture}
    \node[inner sep=0pt] (img) at (0,0)
    {\includegraphics[width=\linewidth]{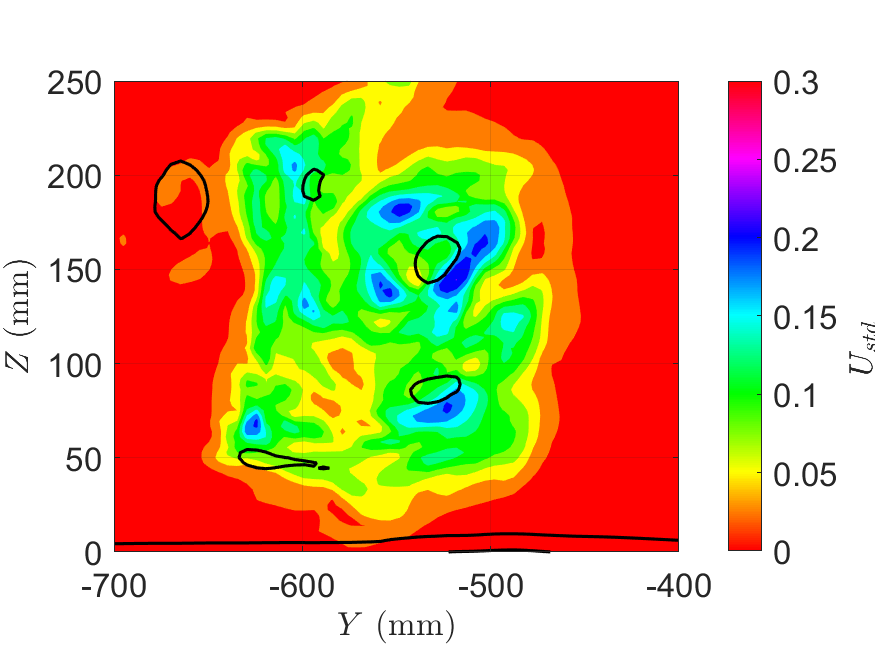}};
    \node at (-1.3,1) {\small \textcolor{black}{\textbf{C}}};
    \node at (-0.5,1.1) {\small \textcolor{black}{\textbf{E}}};
    \node at (-0.3,0.4) {\small \textcolor{black}{\textbf{A}}};
    \node at (-1,-0.9) {\footnotesize \textcolor{black}{\textbf{D\textsubscript{out}}}};
    \node at (0.3,-0.7) {\footnotesize \textcolor{black}{\textbf{D\textsubscript{in}}}};
\end{tikzpicture}
        \caption{LES - Plane 3}
        \label{fig:Ustdnum_Plane3}
    \end{subfigure}
    \begin{subfigure}[b]{0.32\textwidth}
        \centering
    \begin{tikzpicture}
    \node[inner sep=0pt] (img) at (0,0)
    {\includegraphics[width=\linewidth]{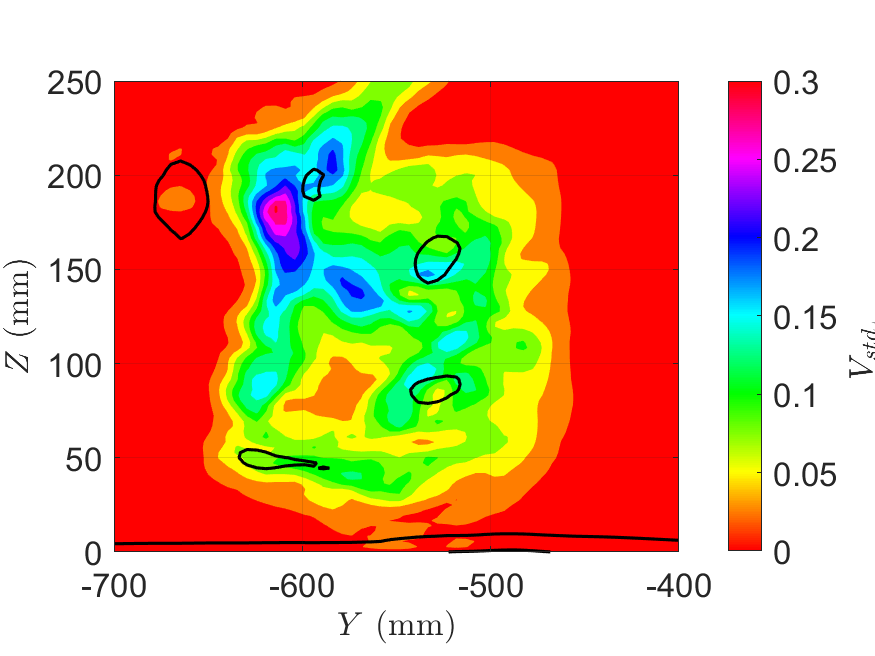}};
    \node at (-1.3,1) {\small \textcolor{black}{\textbf{C}}};
    \node at (-0.5,1.1) {\small \textcolor{black}{\textbf{E}}};
    \node at (-0.3,0.4) {\small \textcolor{black}{\textbf{A}}};
    \node at (-1,-0.9) {\footnotesize \textcolor{black}{\textbf{D\textsubscript{out}}}};
    \node at (0.3,-0.7) {\footnotesize \textcolor{black}{\textbf{D\textsubscript{in}}}};
\end{tikzpicture}
        \caption{LES - Plane 3}
        \label{fig:Vstdnum_Plane3}
    \end{subfigure}    
    \begin{subfigure}[b]{0.32\textwidth}
        \centering
\begin{tikzpicture}
    \node[inner sep=0pt] (img) at (0,0)
    {\includegraphics[width=\linewidth]{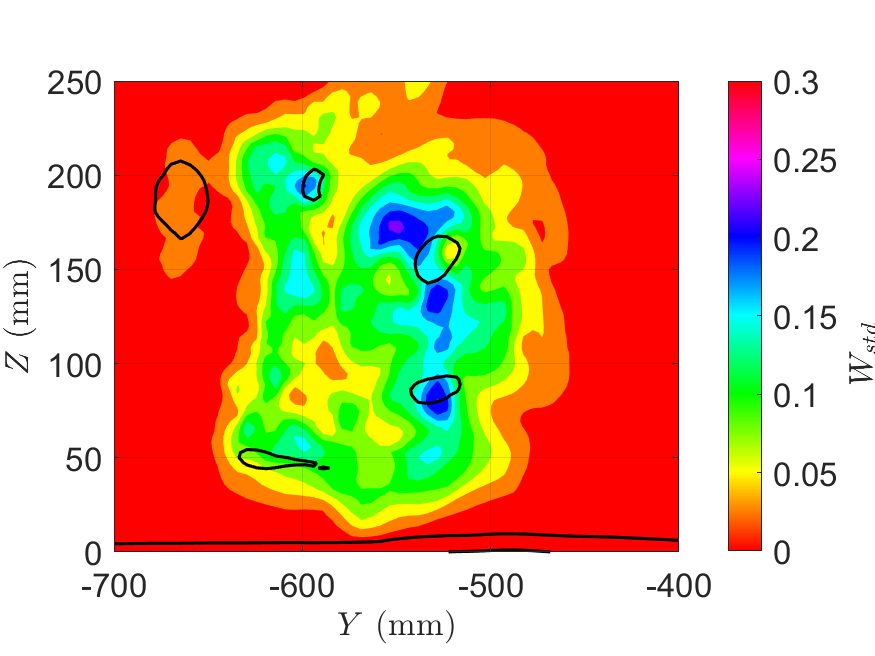}};
    \node at (-1.3,1) {\small \textcolor{black}{\textbf{C}}};
    \node at (-0.5,1.1) {\small \textcolor{black}{\textbf{E}}};
    \node at (-0.3,0.4) {\small \textcolor{black}{\textbf{A}}};
    \node at (-1,-0.9) {\footnotesize \textcolor{black}{\textbf{D\textsubscript{out}}}};
    \node at (0.3,-0.7) {\footnotesize \textcolor{black}{\textbf{D\textsubscript{in}}}};
\end{tikzpicture}
        \caption{LES - Plane 3}
        \label{fig:Wstdnum_Plane3}
    \end{subfigure}
    \caption{ Comparison between simulation and experiments: standard deviation of the streamwise, spanwise and the normal velocity components at $X=-150$ mm (Plane 3).} 
    \label{fig:STD_Plane3}
\end{figure}
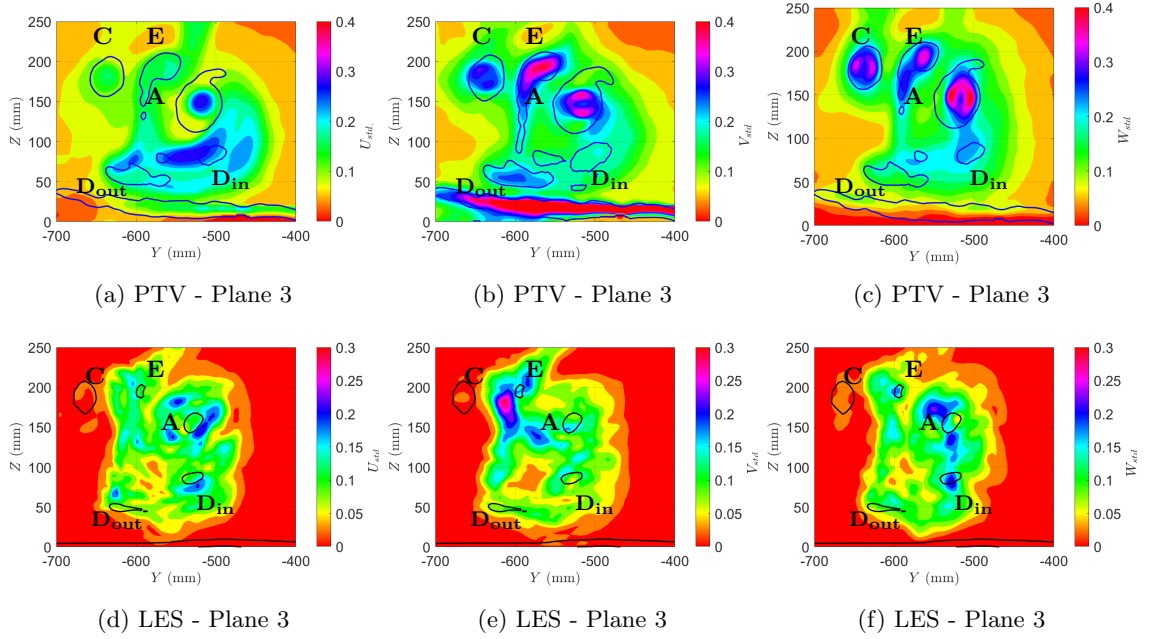

Figure \ref{fig:STD_Plane3} shows the standard deviation for all three velocity components ($U_{std},V_{std}, W_{std}$), defined as:
\begin{equation}
    U_{std}=\sqrt{\sum(u-U)^2/N}
\end{equation}
\noindent where $N$ is the number of samples. The results from the simulation are compared to the experiment for the plane 3. Generally, the level of fluctuations is higher in the experiment since the general level of turbulence, which was tried to be matched by the simulation, is already higher in the incoming flow partially due to the tunnel itself and mostly to the presence of the LNAs.

The standard deviation appears more disperse and homogeneous in the simulation with respect to the experiment. 
For the simulation results, figure \ref{fig:Ustdnum_Plane3} to \ref{fig:Wstdnum_Plane3}, the level of fluctuations is of a similar order for all three velocity components, with the largest fluctuations always located around the vortices. This is in contrast to the experimental results, figure \ref{fig:Ustdexp_Plane3} to \ref{fig:Wstdexp_Plane3}, where the region of higher fluctuations, particularly for the streamwise and spanwise components, are aligned with the core of the vortices. That could be due to vortex meandering and in plane interaction between structure, an analysis on time-resolved data, which will be a follow up to this work, may clarify this observation. It is interesting to notice the high fluctuations appear also in the location of vortex E which orbits around A and C. 
Finally, the most interesting result is the higher fluctuation regions in the proximity of the ground in figure \ref{fig:Vstdexp_Plane3}. Its development in the 3D flow demonstrates that it is an effect associated with the strong in-wash close to the moving ground and to the secondary vorticity region. A similar region of high fluctuations is not observed in the results from the simulation.\\

\section{Discussion}
\label{sec:discussion}
One of the new findings of this work is the formation of a counter-rotating vortex, named the \textit{interaction vortex} (E). Its role in the downstream development of the flow has already been discussed, but it is worth analysing its origin further. 

\begin{figure}[ht!]
    \centering
\begin{subfigure}[b]{0.45\textwidth}
        \centering
    \begin{tikzpicture}
    \node[inner sep=0pt] (img) at (0,0)
    {\includegraphics[width=\linewidth]{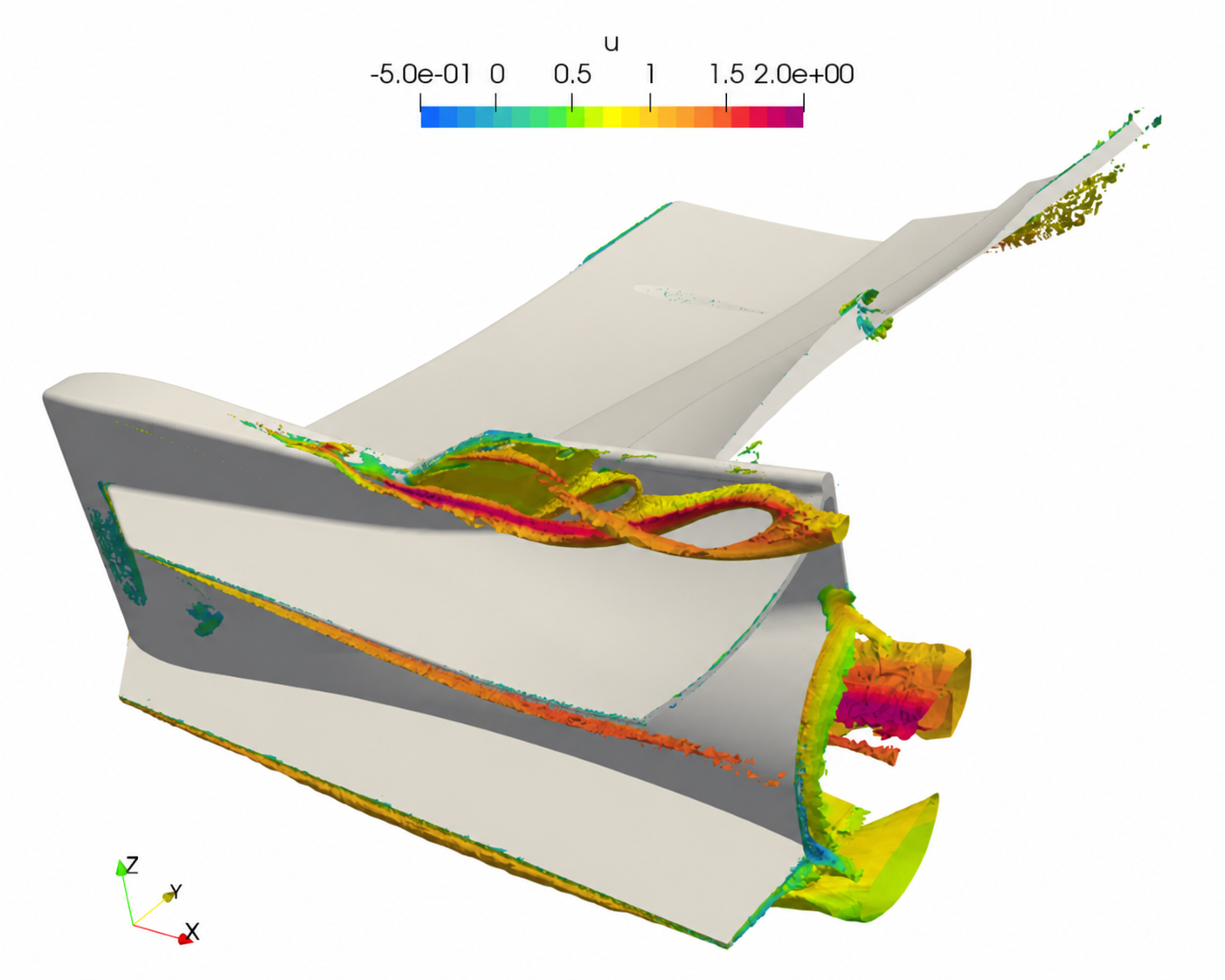}};
    \node at (-1,0.4) {\small\textbf{B}};
    \node at (-1,-0.8) {\small\textbf{C}};
    \node at (2.2,-1) {\small\textbf{A}};
    \node at (2,-2) {\small \textbf{$\textrm{D}$}};
  \draw[->, thick] (1.9,0) -- (1.3,-1);
\end{tikzpicture}
        \caption{ }
        \label{fig:Vort_formation1}
    \end{subfigure}
\begin{subfigure}[b]{0.45\textwidth}
        \centering
    \begin{tikzpicture}
    \node[inner sep=0pt] (img) at (0,0)
    {\includegraphics[width=0.9\linewidth]{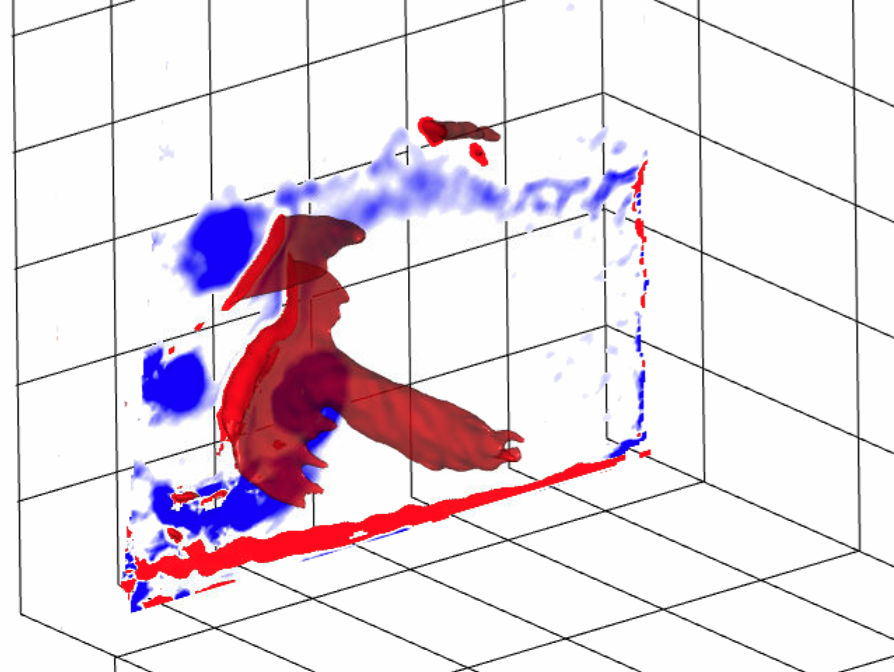}};
    \node at (-1.5,1.2) {\small\textbf{B}};
    \node at (-1.8,0.1) {\small\textbf{C}};
    \node at (0,-0.2)  {\small\textbf{E}};
    \node at (-0.7,-1.2) {\small \textbf{$\textrm{D}$}};
  \draw[->, thick] (-0.5,0.4) -- (-1.2,-0.3);
  \draw[->, thick] (1.1,-0.7) -- (1.5,-0.9) node[pos=1, right] {\tiny X};
  \draw[->, thick] (1.1,-0.7) -- (1.1,-0.3) node[pos=1, left] {\tiny Z};
  \draw[->, thick] (1.1,-0.7) -- (1.5,-0.6) node[pos=2, left] {\tiny Y};

  \def\barwidth{0.4cm}
  \def\barheight{4cm}
\begin{scope}[shift={($(img.east)+(0.5cm,-0.5*\barheight)$)}]
  \shade[bottom color=blue, top color=white]
    (0, 0) rectangle (\barwidth,\barheight/2);
  \shade[bottom color=white, top color=red]
    (0,\barheight/2) rectangle (\barwidth,\barheight);
  \draw (0,0) rectangle (\barwidth,\barheight);
  \foreach \val/\frac in {-10/0, -5/0.25, 0/0.5, 5/0.75, 10/1} {
    \draw (\barwidth,\frac*\barheight) -- ++(0.12,0);
    \node[anchor=west] at (\barwidth+0.15,\frac*\barheight) {$\val$};
  }
  \node[rotate=90, anchor=north] at (\barwidth*2,\barheight/2) {$\Omega_x$};
  \end{scope}
\end{tikzpicture}
        \caption{ }
        \label{fig:Vort_formation2}
    \end{subfigure}
   \caption{Formation of the interaction vortex (E). a) 3D View from the rear for the IFW - Iso-contour of zero total pressure, coloured by the streamwise velocity component \cite{Liosi2026}. b) Experimental results, streamwise vorticity at $X= - 226$ mm and isocontour of negative vorticity downstream: the vortex sheet (indicated by the arrow) rolls up into the interaction vortex as it separates from the trailing edge of the endplate.}
    \label{fig:Vort_formation}
\end{figure}
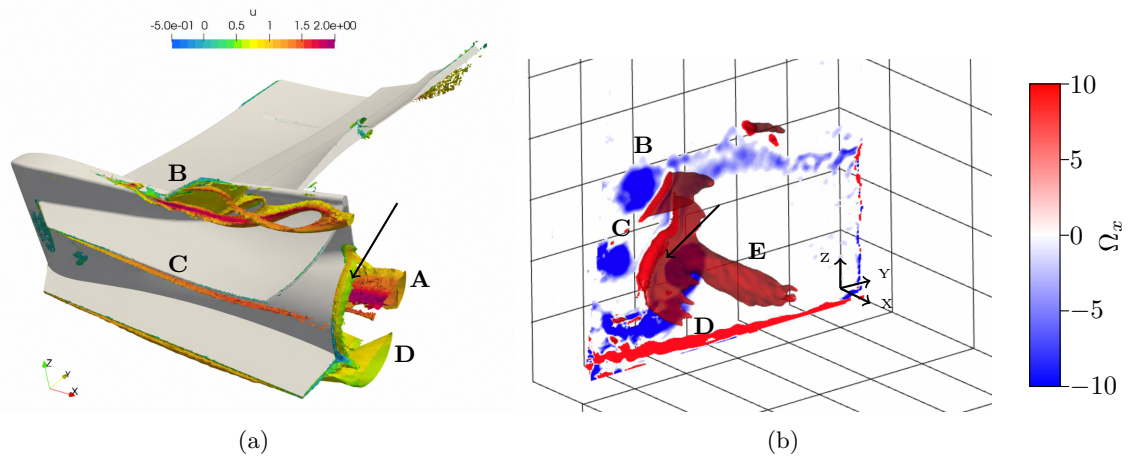

Figure \ref{fig:Vort_formation1}, obtained from the simulation, shows a shear layer that forms on the inner curved surface of the endplate and separates from the surface at the trailing edge. Similarly, figure \ref{fig:Vort_formation2} shows a region of distributed negative vorticity behind the trailing edge of the endplate, as depicted by the experiment, in the same location (black arrow in figure). This vortex sheet eventually rolls up forming the interaction vortex with the clear, rounded coherent structure observed only from about 50 mm downstream.

The vorticity sheet of (E), negative at the surface of the endplate, could originate from the action of the main vortex (A) on the inner surface of the endplate. In fact, the main vortex is formed at the leading edge of the endplate and travels very close to the surface, which is effectively a stationary wall; thereby generating secondary flows. 
To further understand the origin of this structure, one can analyse the vorticity transport equation in the streamwise direction as derived from the Reynolds average equations \cite{bradshaw1987turbulent}:
\begin{equation}
    \resizebox{\textwidth}{!}{$
    \textcolor{Cyan}{\underbrace{\color{black} U \frac{\partial \Omega_x}{\partial x}+V \frac{\partial \Omega_x}{\partial y}+W\frac{\partial \Omega_x} {\partial z}}_\textrm{Advection}}= \textcolor{red}{\underbrace{\color{black}\Omega_x \frac{\partial U}{\partial x}+\Omega_y \frac{\partial U}{\partial y}+\Omega_z \frac{\partial U}{\partial z}}_\textrm{Stretching/Tilting}}+\textcolor{blue}{\underbrace{\color{black}\Big(\frac{\partial^2}{\partial y^2}-\frac{\partial^2}{\partial z^2}\Big)(-\overline{vw})}_\textrm{Shear stress}}+\textcolor{green}{\underbrace{\color{black}\frac{\partial^2}{\partial y \partial z}(\overline{v^2}-\overline{w^2})}_\textrm{Normal stress}}+\textcolor{orange}{\underbrace{\color{black} \frac{1}{Re}\nabla^2\Omega_x}_\textrm{Viscosity}}$}
    \label{eq:vort_trans}
\end{equation}
\begin{figure}[ht!]
    \centering
    \begin{subfigure}[b]{0.49\textwidth}
\centering
    \begin{tikzpicture}
    \node[inner sep=0pt] (img) at (0,0)
    {\includegraphics[trim=0cm 0cm 1.2cm 0, clip,width=\linewidth]{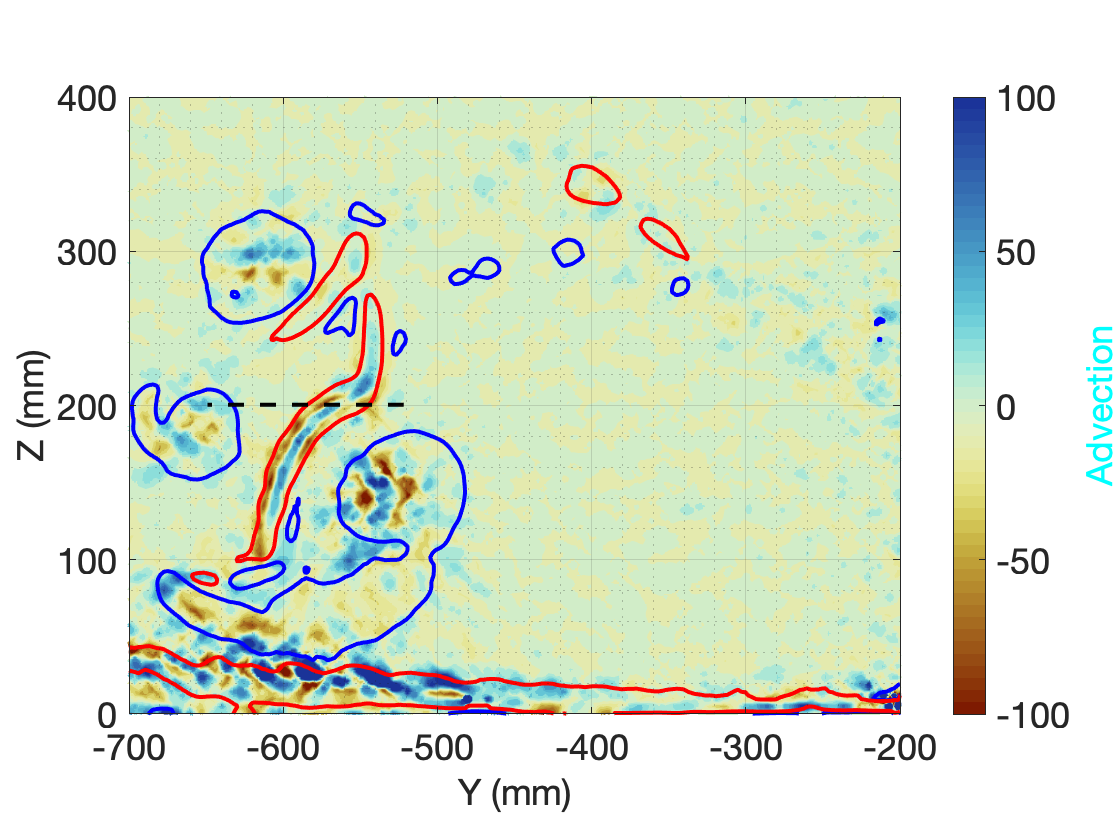}}; 
    \node at (-2.5,0.4) {\small \textbf{B}};
    \node at (-1.6,1.5) {\small \textbf{C}};
    \node at (-0.9,0.5) {\small \textbf{E}};
    \node at (-0.3,-0.3) {\small \textbf{A}};
    \node at (-2.5,-0.8) {\footnotesize \textbf{D\textsubscript{out}}};
    \node at (-0.8,-1.5) {\footnotesize \textbf{D\textsubscript{in}}};
\end{tikzpicture}
\caption{ \textcolor{cyan}{Advection}}
    \label{fig:Advection}
    \end{subfigure} 
    \begin{subfigure}[b]{0.49\textwidth}
        \centering
    \begin{tikzpicture}
    \node[inner sep=0pt] (img) at (0,0)
    {\includegraphics[trim=0cm 0cm 1.2cm 0, clip,width=\linewidth]{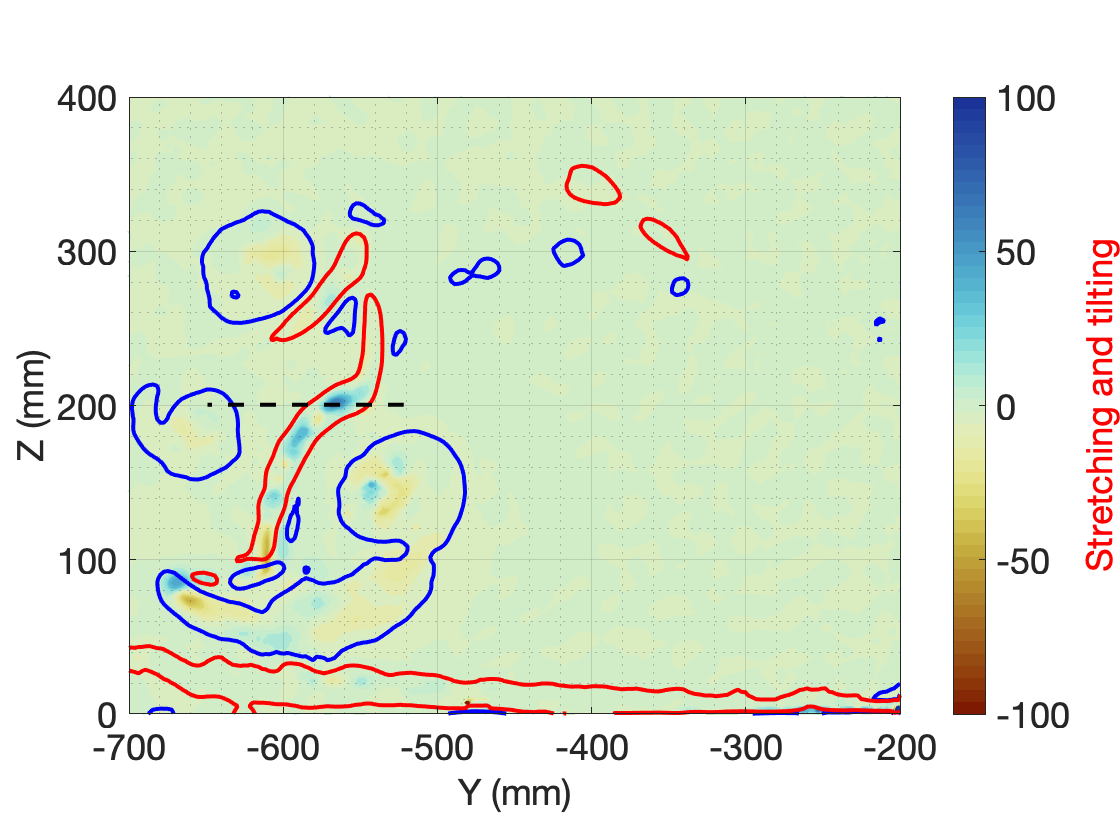}}; 
    \node at (-2.5,0.4) {\small \textbf{B}};
    \node at (-1.6,1.5) {\small \textbf{C}};
    \node at (-0.9,0.5) {\small \textbf{E}};
    \node at (-0.3,-0.3) {\small \textbf{A}};
    \node at (-2.5,-0.8) {\footnotesize \textbf{D\textsubscript{out}}};
    \node at (-0.8,-1.5) {\footnotesize \textbf{D\textsubscript{in}}};
\end{tikzpicture}
        \caption{\textcolor{red}{Stretching and tilting} }
        \label{fig:Stretching_tilting}
    \end{subfigure}  
    \begin{subfigure}[b]{0.49\textwidth}
        \centering
    \begin{tikzpicture}
    \node[inner sep=0pt] (img) at (0,0)
    {\includegraphics[trim=0cm 0cm 1.2cm 0, clip,width=\linewidth]{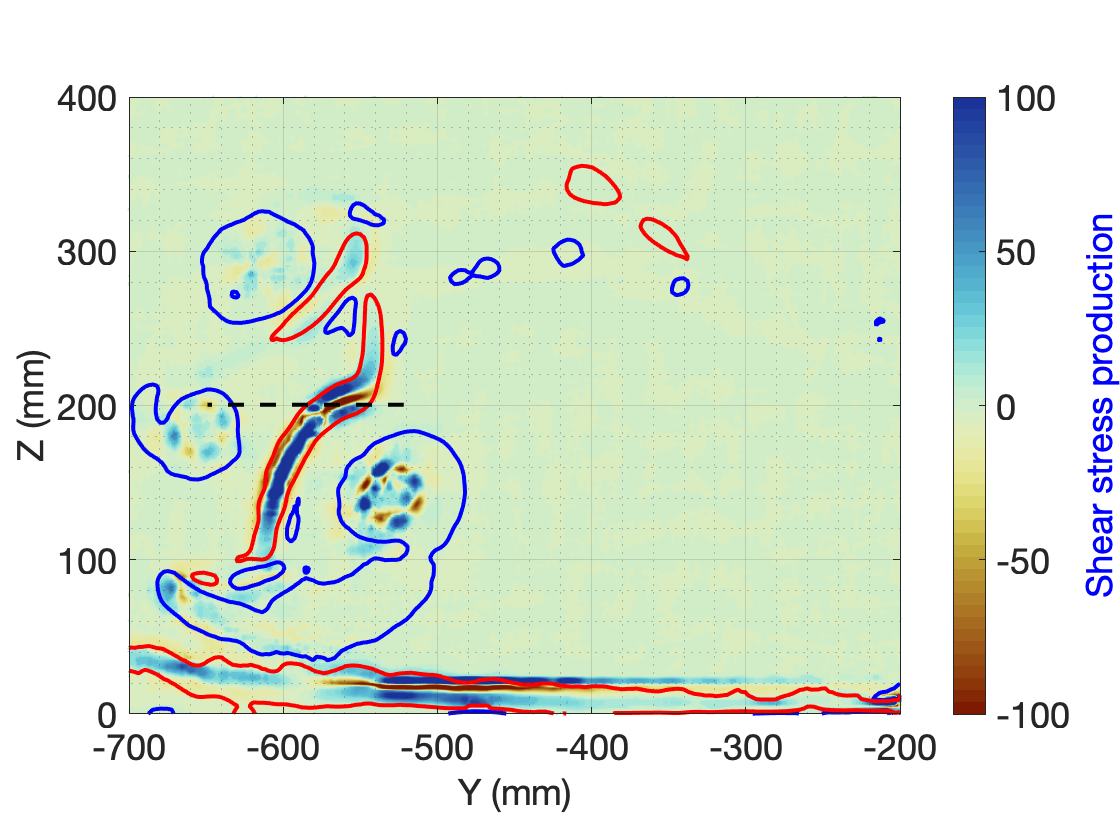}}; 
    \node at (-2.5,0.4) {\small \textbf{B}};
    \node at (-1.6,1.5) {\small \textbf{C}};
    \node at (-0.9,0.5) {\small \textbf{E}};
    \node at (-0.3,-0.3) {\small \textbf{A}};
    \node at (-2.5,-0.8) {\footnotesize \textbf{D\textsubscript{out}}};
    \node at (-0.8,-1.5) {\footnotesize \textbf{D\textsubscript{in}}};
\end{tikzpicture}
        \caption{ \textcolor{blue}{Shear stress production}}
        \label{fig:Shear_stress_prod}
    \end{subfigure}
    \begin{subfigure}[b]{0.49\textwidth}
        \centering
    \begin{tikzpicture}
    \node[inner sep=0pt] (img) at (0,0)
    {\includegraphics[trim=0cm 0cm 1.2cm 0, clip,width=\linewidth]{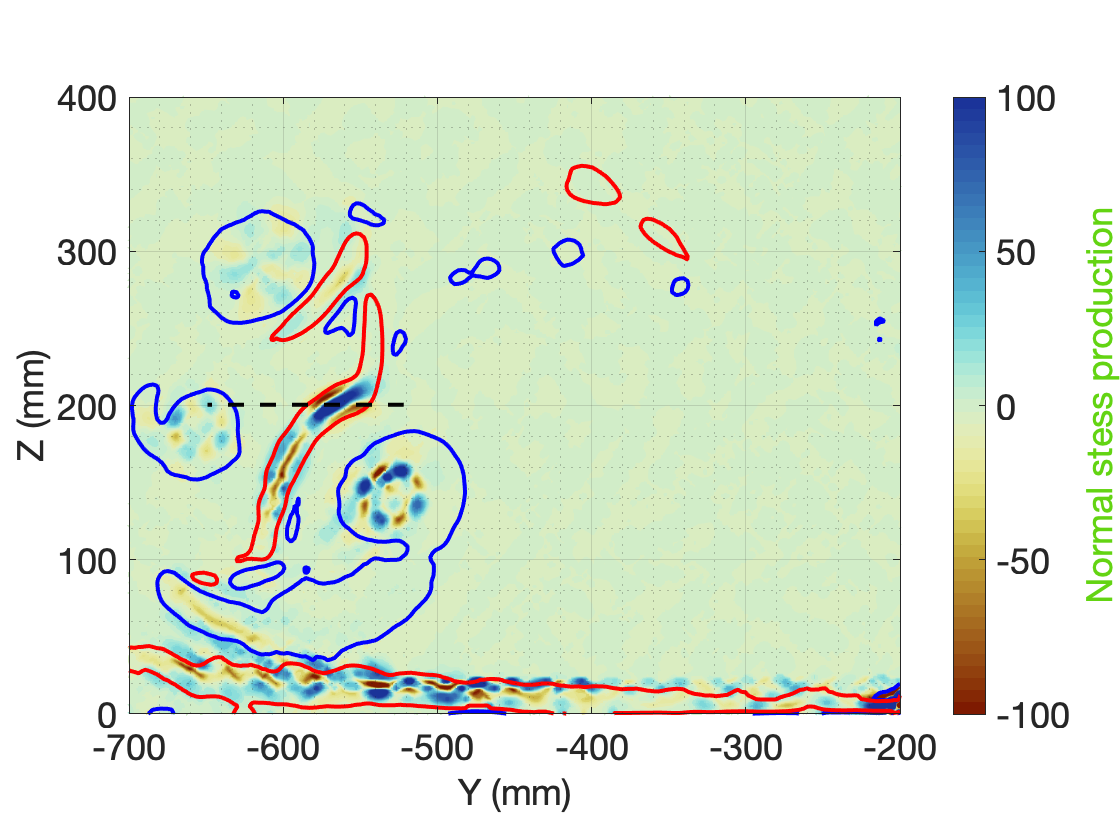}}; 
    \node at (-2.5,0.4) {\small \textbf{B}};
    \node at (-1.6,1.5) {\small \textbf{C}};
    \node at (-0.9,0.5) {\small \textbf{E}};
    \node at (-0.3,-0.3) {\small \textbf{A}};
    \node at (-2.5,-0.8) {\footnotesize \textbf{D\textsubscript{out}}};
    \node at (-0.8,-1.5) {\footnotesize \textbf{D\textsubscript{in}}};
\end{tikzpicture}
        \caption{\textcolor{green}{Normal stress production} }
        \label{fig:normal_stress_prod}
    \end{subfigure}
    \begin{subfigure}[b]{0.49\textwidth}
        \centering
    \begin{tikzpicture}
    \node[inner sep=0pt] (img) at (0,0)
    {\includegraphics[trim=0cm 0cm 1.6cm 0, clip,width=\linewidth]{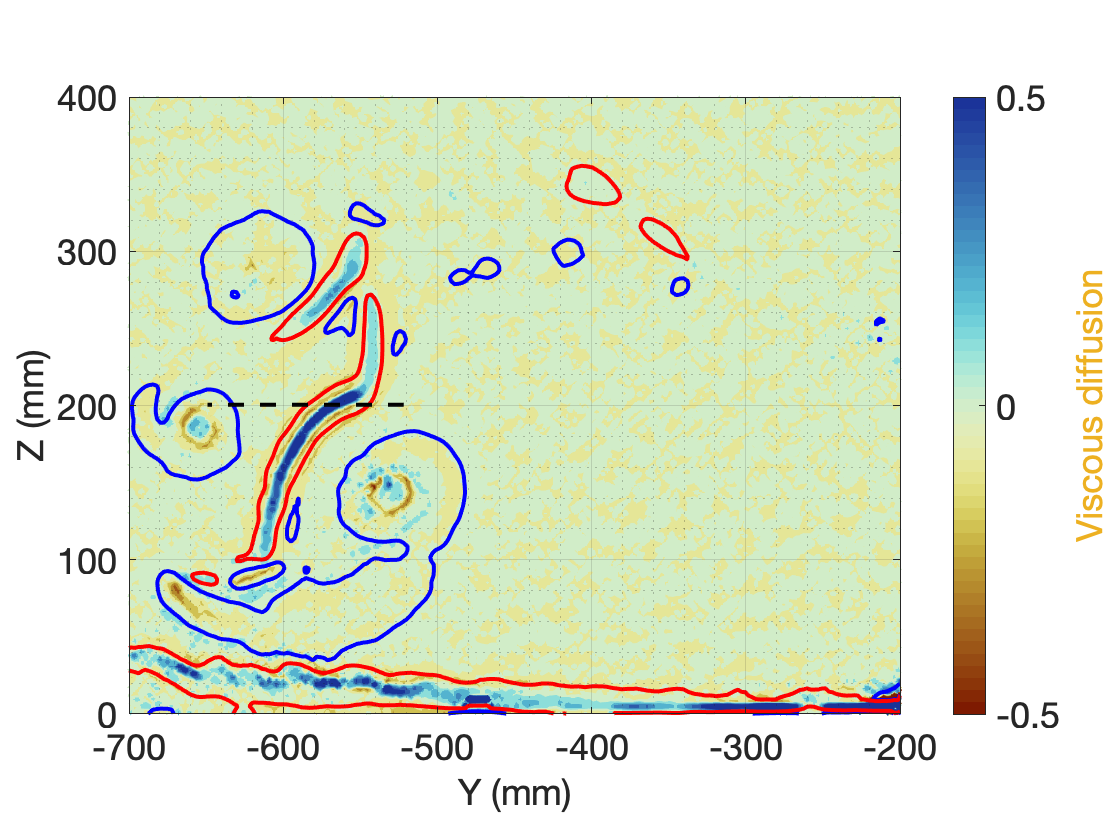}};    
    \node at (-2.5,0.4) {\small \textbf{B}};
    \node at (-1.6,1.5) {\small \textbf{C}};
    \node at (-0.9,0.5) {\small \textbf{E}};
    \node at (-0.3,-0.3) {\small \textbf{A}};
    \node at (-2.5,-0.8) {\footnotesize \textbf{D\textsubscript{out}}};
    \node at (-0.8,-1.5) {\footnotesize \textbf{D\textsubscript{in}}};
\end{tikzpicture}
        \caption{\textcolor{orange}{Viscous diffusion} }
        \label{fig:Viscous}
    \end{subfigure}    
    \begin{subfigure}[b]{0.49\textwidth}
        \centering
\begin{tikzpicture}
    \node[inner sep=0pt] (img) at (0,0)
    {\includegraphics[width=\linewidth]{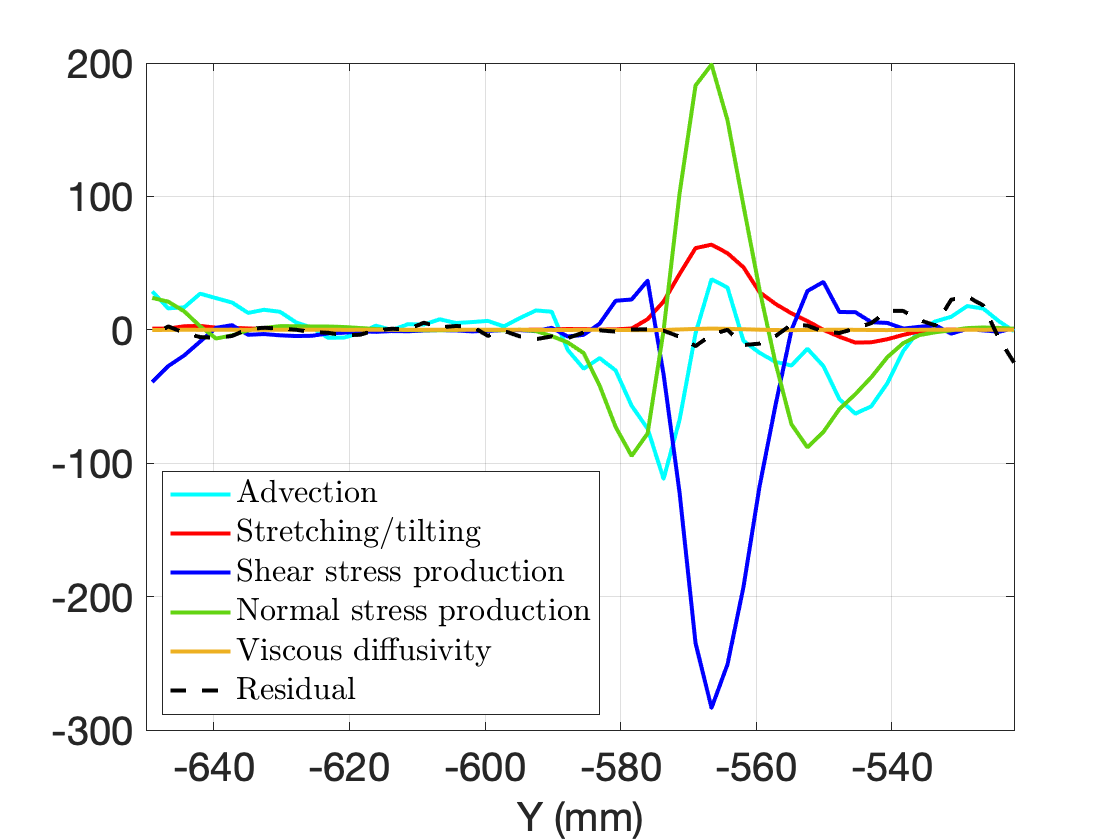}};
\end{tikzpicture}
        \caption{ }
        \label{fig:Eq_terms_Y}
    \end{subfigure}
    \caption{a) to e) individual terms of the vorticity transport equation \ref{eq:vort_trans} calculated from the PTV data at $X=200$ mm ($X^*=0.8$) with isocontour of vorticity overlaid on top. f) linear plot along the spanwise direction at $Z=200$ mm.} 
    \label{fig:vorticity_transport}
\end{figure}
Two mechanisms of vorticity formation can be identified in a streamwise vorticity dominated turbulent flow \cite{bradshaw1987turbulent}:
1) Skew-induced: generated by a vorticity in the orthogonal directions that is tilted by a spanwise or a wall-normal shear layer, as described by the second and third term on the right-hand side of \ref{eq:vort_trans}(Prandtl's first kind of secondary flow);
2) Stress-induced: due to the Reynolds stresses gradients either for a shear stresses curvature or for the normal stresses anisotropy, described by the fourth and fifth term on the right-hand side (Prandtl's second kind of secondary flow).
\begin{figure}[ht!]
    \centering
\begin{subfigure}[b]{0.49\textwidth}
\centering
    \includegraphics[width=\textwidth]{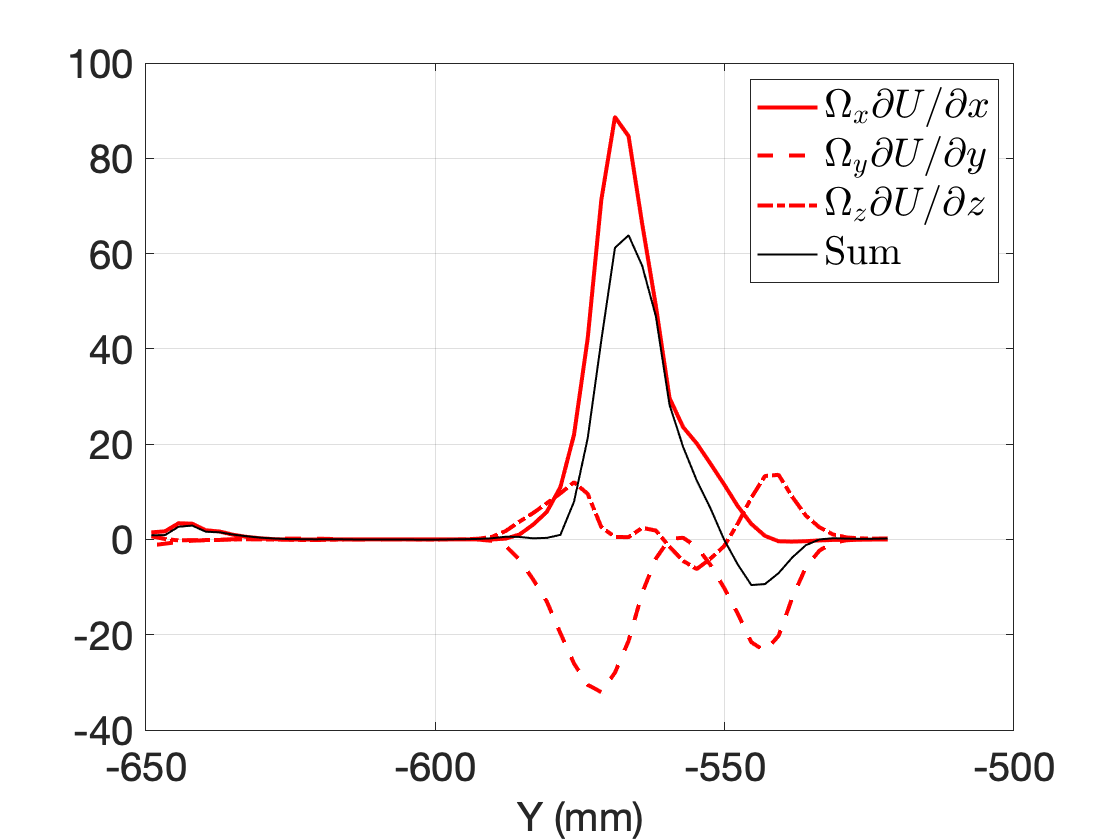}
    \caption{}
    \end{subfigure}
\begin{subfigure}[b]{0.49\textwidth}
\centering
    \includegraphics[width=\textwidth]{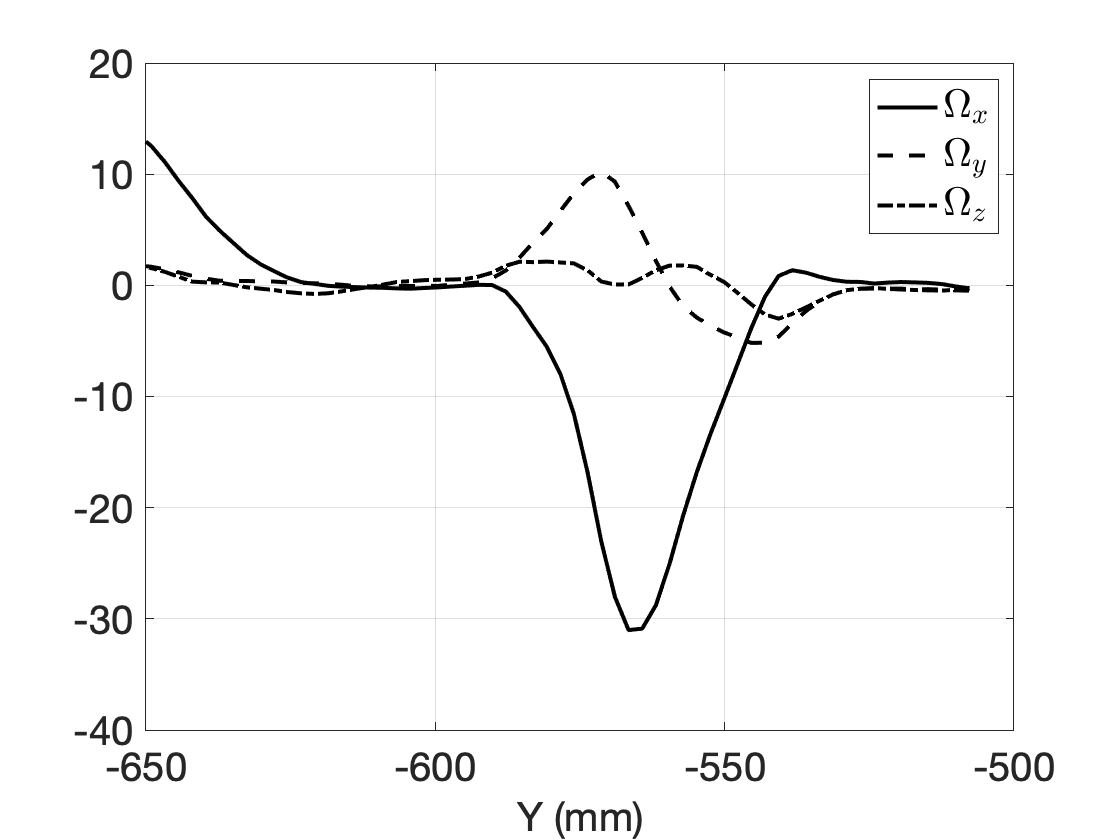}
    \caption{}
    \label{fig:vorticity_comp}
    \end{subfigure}
    \caption{a) Stretching and tilting and b) vorticity component along the spanwise direction at $Z=200$ mm.} 
    \label{fig:stetch_tilt_terms}
\end{figure}
\begin{figure}[ht!]
    \centering
    \begin{tikzpicture}
    \node[inner sep=0pt] (img) at (0,0)
    {\includegraphics[width=0.49\linewidth]{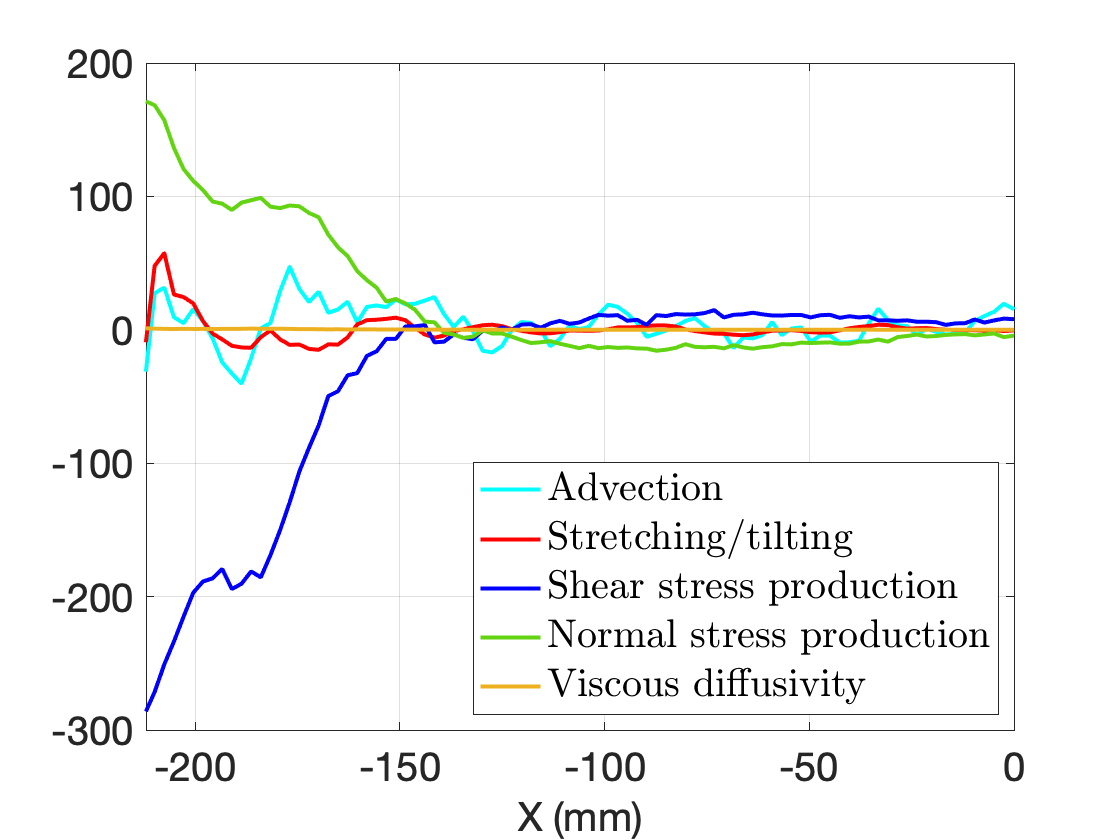}};
    \node[inner sep=0pt] (img) at (1.8,3.8){\fbox{\includegraphics[width=0.2\linewidth]{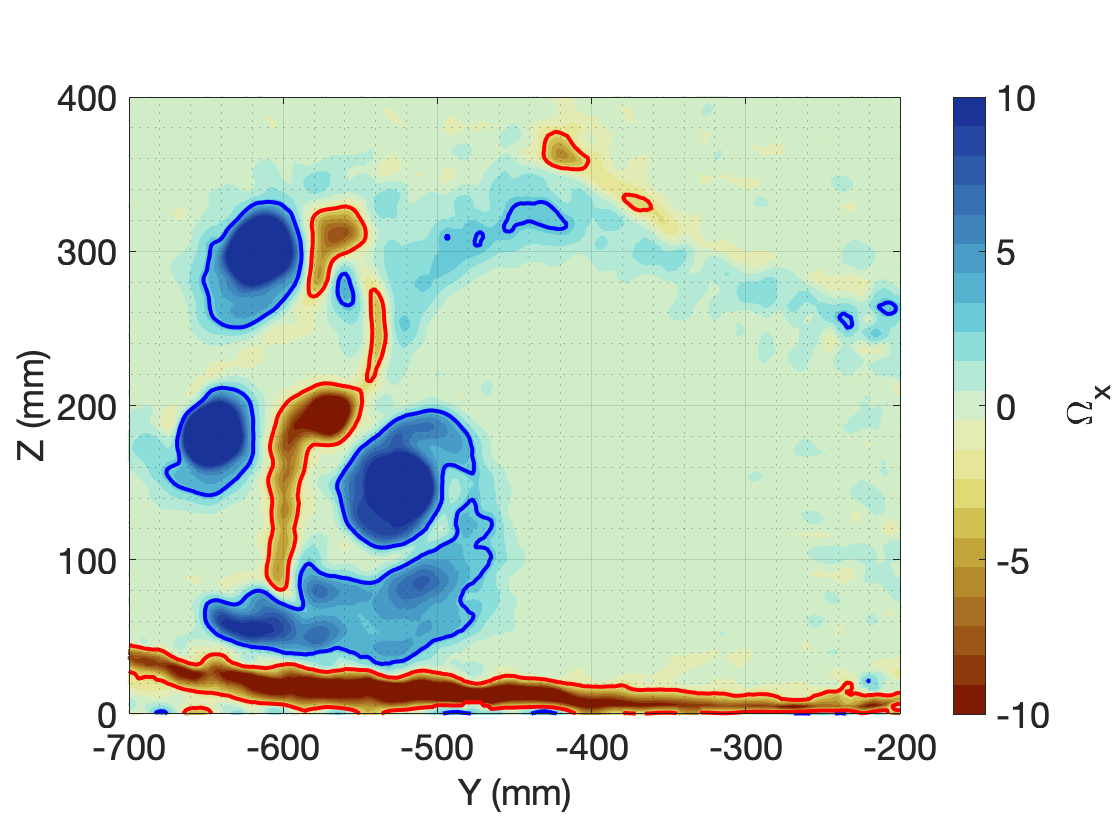}}};
    \draw[->, thick] (-1.1,0.8) -- (1,2.5);
    \node[inner sep=0pt] (img) at (-1.8,3.8){\fbox{\includegraphics[width=0.2\linewidth]{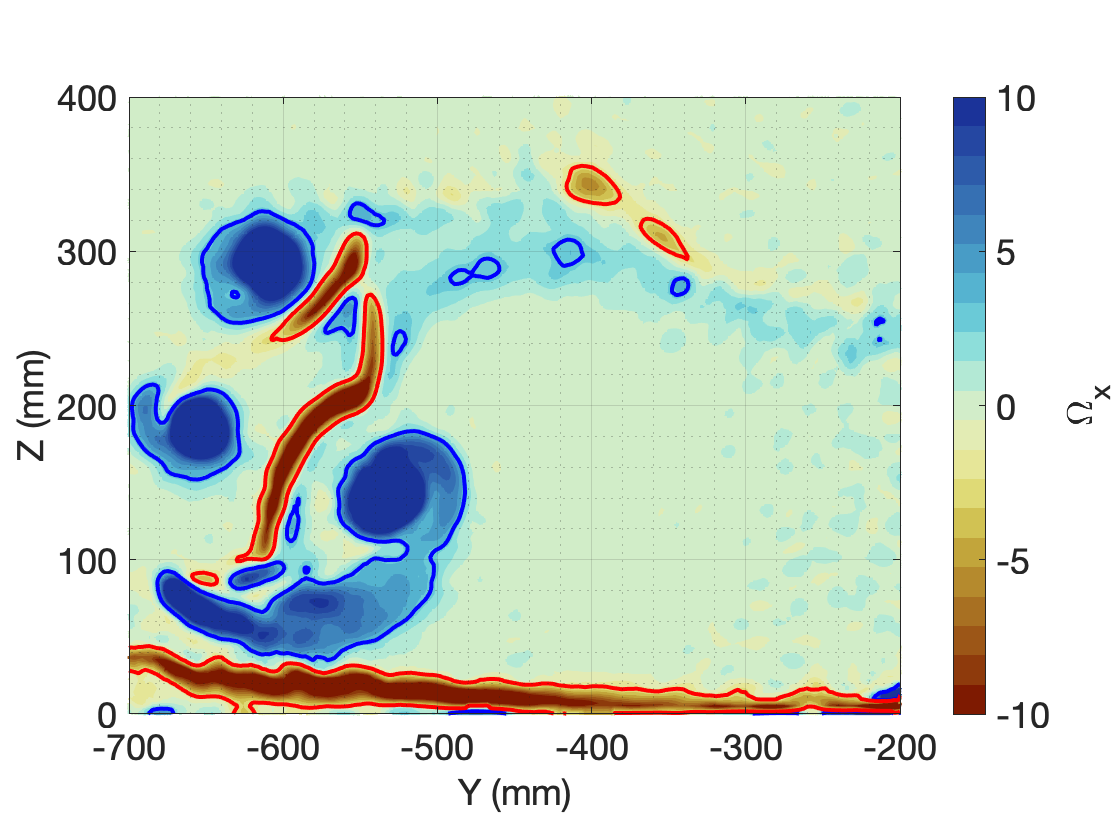}}};
    \draw[->, thick] (-2.4,1.5) -- (-2,2.5);
    \end{tikzpicture}
    \caption{Vorticity transport equation at the core of the interaction vortex along the streamline direction. The contour plots on the top of the figure show the streamwise vorticity at $X=-200$ mm and $X=-150$ mm to demonstrate how the Reynolds stresses production terms are higher when the vortex is in the formation process, whilst comparable to the other terms when the vortex is formed.}
    \label{fig:all_terms_X}
\end{figure}

\begin{figure}[ht!]
    \centering
\centering
    \begin{tikzpicture}
    \node[inner sep=0pt] (img) at (0,0)
    {\includegraphics[trim=0cm 0cm 0cm 0, clip, width=0.49\textwidth]{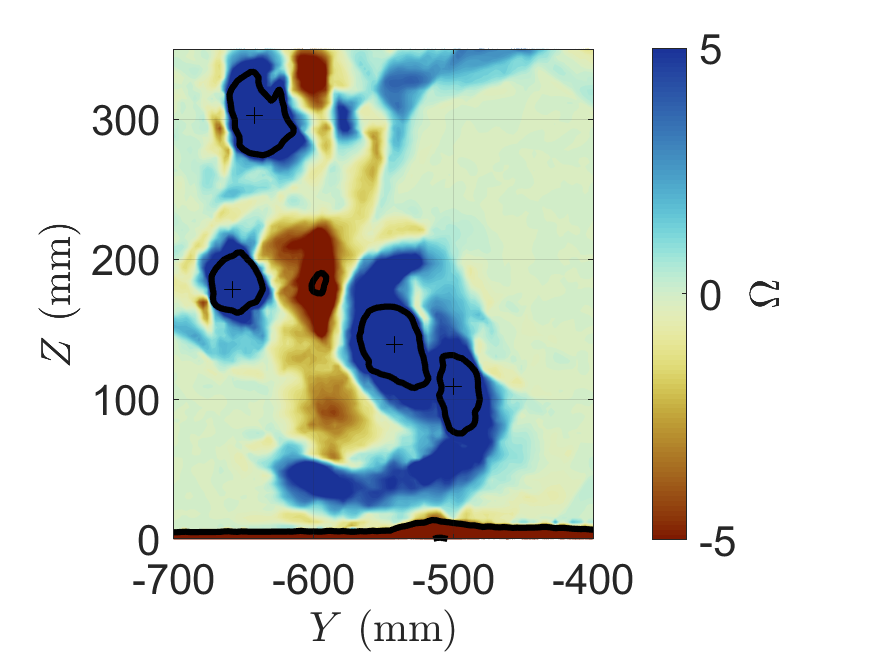}};
    \node at (-2,1.6) {\footnotesize\textbf{B}};
    \node at (-2.1,0.9) {\footnotesize \textbf{C}};
    \node at (-1.3,1.1) {\footnotesize \textbf{E}};
    \node at (0.1,0.2) {\footnotesize \textbf{A}};
    \node at (-0.8,-1) {\footnotesize \textbf{D\textsubscript{out}}};
    \node at (1,-0.8) {\footnotesize \textbf{D\textsubscript{in}}};
\end{tikzpicture}
\caption{LES results at $Re_c=208045$ for Plane 3.}
    \label{fig:Omexp_Plane3_Re220k}
\end{figure}

Figure \ref{fig:vorticity_transport} compares the individual terms of the equation \ref{eq:vort_trans} at the plane $X$ = -200 mm ($X^*=0.8$), a location at which the interaction vortex has not formed yet. To better quantify the relative magnitude of the terms, figure \ref{fig:Eq_terms_Y} shows a linear plot across the interaction vortex in the spanwise direction at $Z$ = 200 mm (corresponding to the black dashed line on the contour plots). The main term appears to be the shear stress curvature with a contribution of opposite sign given by the normal stress anisotropy and a relevant contribution of the stretching and tilting term. It can be concluded that the mechanism is shear stress driven. To further confirm this finding, the stretching and tilting terms are decomposed in figure \ref{fig:stetch_tilt_terms}. The main contribution is from the streamwise vorticity being stretched in the streamwise direction:
$\Omega_x \frac{\partial U}{\partial x}>\Omega_z \frac{\partial U}{\partial z}>\Omega_y \frac{\partial U}{\partial y}$. Therefore the contribution of skew-induced vorticity is negligible. For completeness, the magnitudes of individual vorticity components is compared in figure \ref{fig:vorticity_comp}.\\
From this analysis it can be concluded that the production of vorticity that leads to the roll up of the interaction vortex is a mechanism of the second type driven by the shear stresses generated at the wall of the curved vertical surface of the endplate with the streamwise vorticity being stretched as it progresses downstream. 
Finally, further evidence is given by the five terms of the vorticity transport equation \ref{eq:vort_trans} calculated in the core of the interaction vortex and plotted along the streamwise direction at all measured locations (figure \ref{fig:all_terms_X}). This confirms that the Reynolds stresses are the terms that contribute the most when the vortex is not formed yet, to then equalise the magnitude of the other terms once the coherent structure appears evident at $X=-150$ mm.

There remain two open questions regarding the interaction vortex: why the structure was not observed in previous works, and whether it still persists at higher Reynolds number.
The answer to the first question lies in the focus of previous study. Buscariolo et al. \cite{Buscariolo2022} and O'Sullivan \cite{Sullivan2023} only reported either isocontour of vorticity of the same sign or contour plots of the wall-normal and spanwise velocity from which it is difficult to discern velocity gradients, making it hard to even attempt to identify hints of the interaction vortex. The only evidence of the interaction vortex in previous works is contained in Pegrum's PhD thesis \cite{Pegrum2006}, where contour plots of vorticity, centred around zero, show the opposite sign vorticity sheet between the main and the canard vortices, similar to figures \ref{fig:Vort1sim} and \ref{fig:Vort1}, at X=-300 mm, essentially at the trailing edge of the IFW, for $U_\infty=12.5$ m/s and $U_\infty=25$ m/s and for different ride heights with/without the rolling road (see figure 4.5 and 4.8 and 4.9 in \cite{Pegrum2006}). Unfortunately, no information of the downstream development and interaction vortex formation is available either because of a lack of data or because the three-dimensional plots show isocontour of vorticity of the same sign hiding structures rotating in opposite directions. Regarding the second question, Pegrum's results \cite{Pegrum2006} demonstrate topology agreement at all Reynolds numbers tested, but show no evidence of interaction vortex.
To address this question, figure \ref{fig:Omexp_Plane3_Re220k} shows the results from the Nektar++ simulation at $Re=208045$ using the same methodology discussed in this paper, which demonstrated the Reynolds number independence on the topology of the vortical wake and the existence of the interaction vortex also at higher Reynolds number.

\section{Conclusions}
\label{sec:conclusions}
In this work, volumetric PTV was employed for the first time to perform an experimental investigation of the wake behind the IFW, an F1 benchmark configuration. In particular, the STB algorithm was used to reconstruct velocity fields within a large downstream volume of the IFW, covering most of the wake. The experimental results were compared with an implicit LES simulation of the same geometry and Reynolds number, using higher-order Spectral/hp element method implemented in Nektar++.

The large measurement volume enables a more complete description of the wake development behind the IFW than previously available, improving the understanding of its vortical evolution. In particular, a previously unreported structure, herein referred to as the interaction vortex, has been identified in both the experimental and numerical datasets.

This vortex appears to play a significant role in shaping the downstream wake topology by inhibiting the merging process between the main and the canard vortex. Its origin can be traced to a vorticity sheet shed from the trailing edge of the endplate. It was demonstrated that the formation process of the interaction vortex is stress-induced corresponding to the second kind of Prandtl's secondary flow \cite{bradshaw1987turbulent}. The Reynolds stresses originate from the flow at the inner surface of the curved endplate, probably due to the interaction of the main vortex with the stationary surface. This observation shows the importance of correctly estimating the Reynolds stresses when numerically simulating vortex dominated flows. This key result provides a challenge for turbulence modelling because the Reynolds stresses (and their anisotropy) change more slowly than the mean rate of strain. The analysis of the vortex formation shows the richness of volumetric PTV measurements and the potential to unveil physical understanding.
The discussion includes an analysis on why the vortex was not observed by previous studies and results of the simulation at higher Reynolds number to demonstrate the existence of the interaction vortex at higher speeds.

The comparison between the experimental and the simulation results shows a good agreement in  capturing the overall vortex system in the wake of the IFW, including the location and the downstream evolution. Nevertheless, some differences are observed when examining the flow more detail. 

In particular, in the simulation the flow appeared decelerating in the streamwise direction at the location of the main vortex A, whereas it is accelerated in the experiment. It is hypothesised that this is due to the influence of the LNAs to the incoming flow. When comparing the evolution of the vorticity in the streamwise planes, there is a small offset  between the results from the experiment and the simulation. The major difference appears in the secondary vorticity close to the ground that is much thinner in the simulation compared to the experiment. This effect appeared to be related to the spanwise velocity component which, close to the ground, is resolved differently by the two methods. Further study will need to address this discrepancy, either validating the PTV results with more resolved data close to the ground or evaluating a sensitivity analysis on the simulation. Large differences were noticed on the standard deviation, which may be due to the statistics not being well resolved in the simulation or the meandering of the vortices not been captured well. Nevertheless, the overall agreement has improved compared to previous attempts.
These results demonstrate the ability of tomographic PTV to capture complex three-dimensional flows over a large domain using a relatively simple setup, making it well suited for industrial applications. For F1 aerodynamics, resolving the formation and evolution of the interaction vortex provides insight directly relevant to endplate design, where controlling vortex interactions in the wake influences both downforce and drag. Beyond this specific application, the accuracy and efficiency demonstrated here, combined with the reduced wind tunnel running time required, support the adoption of tomographic PTV wherever dense volumetric data is needed under similar testing constraints.
This work has not only validated the results from simulation via the first PTV experiment on the IFW, but has contributed to create a robust, industrial benchmark for volumetric flow measurements. 

Further work will focus on time-resolved analysis of the dataset, using techniques such as fine-scale reconstruction flow to further investigate the unsteady vortex dynamics. In addition, configurations of the IFW incorporating a rotating wheel will be analysed to extend the analysis towards a more realistic F1 scenario.

\newpage \section{Appendix}

\subsection{PTV convergence}
\label{sec:convergence}
\begin{figure}[!ht]
   \centering
      \begin{subfigure}{0.49\textwidth}
\includegraphics[width=\linewidth]{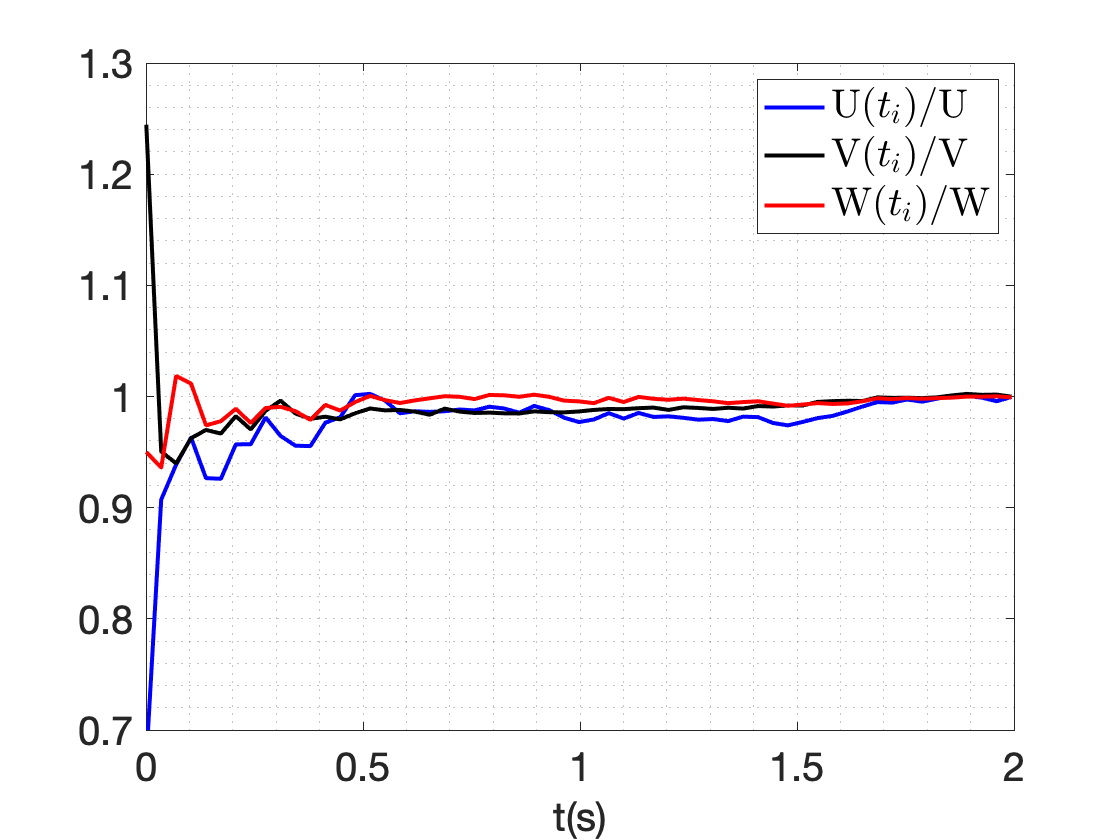}
      \caption{ }
      \end{subfigure}
      \begin{subfigure}{0.49\textwidth}
\includegraphics[width=\linewidth]{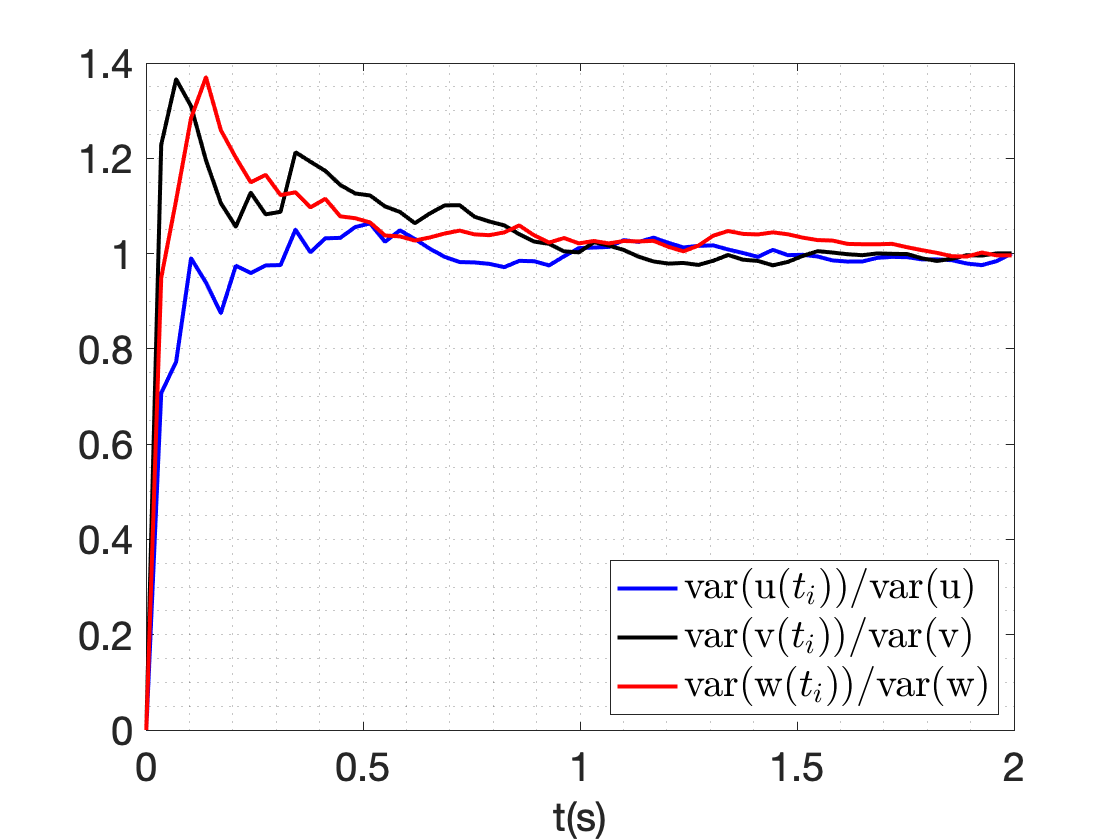}
      \caption{ }
      \end{subfigure}
      \begin{subfigure}{0.49\textwidth}
\includegraphics[width=\linewidth]{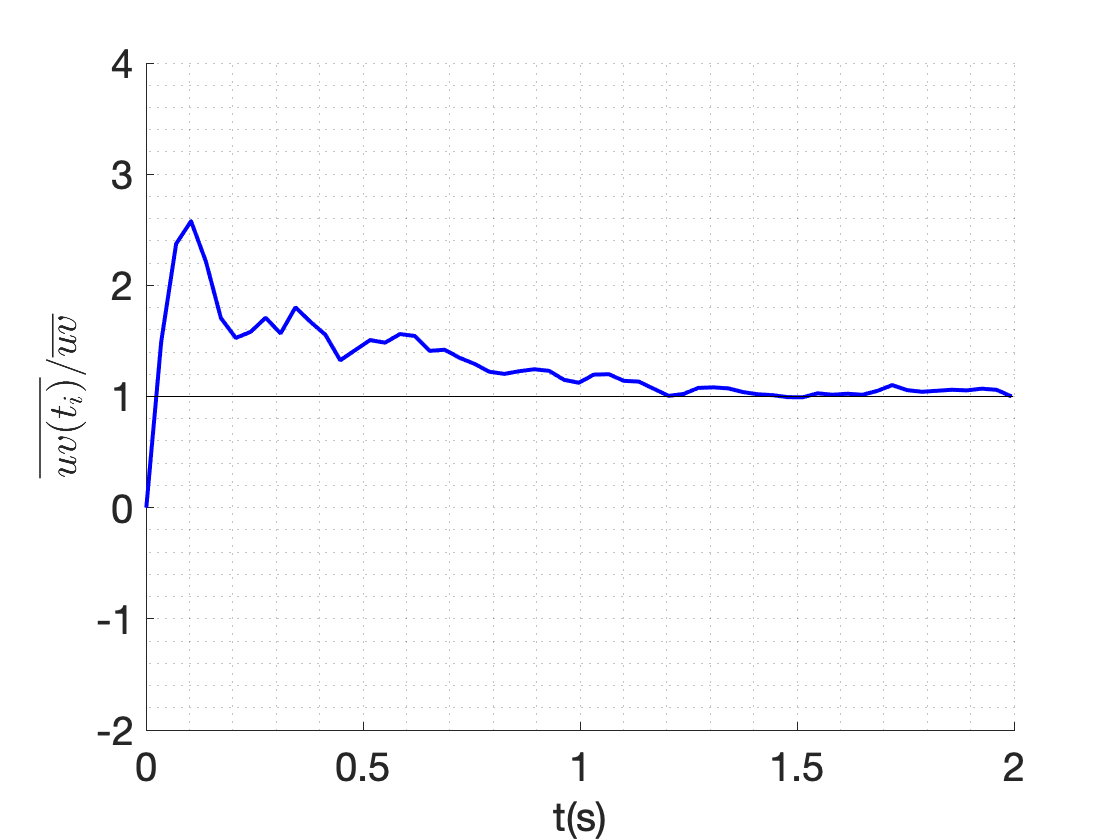}
      \caption{ }
      \end{subfigure}
      \caption{An example of data convergence at Y=-500 mm and Z=-80 mm at plane 3, corresponding to an area of high fluctuations: a) average velocity, b) variance, c) Reynold stress component $\overline{uv}$.}
   \label{fig:PTV_conv}
\end{figure}

To analyse the data in time, the fine scale reconstruction algorithm (VIC\#) has been applied to convert the particles from tracks to grid instead of the average binning method referred in the text. This algorithm applies data assimilation techniques to reconstruct the volume and is able to reconstruct the flow even in locations where at a specific instant of time particles may be missed. It is not in the interest of this paper to discuss the fine scale reconstruction methodology in details, since the paper only concern with average quantities, but the results are used here only to show a typical convergence plot, figure \ref{fig:PTV_conv}.

\subsection{Simulation: Temporal Convergence \& Data-acquisition}
The end of initial transients, or alternatively the start of time-averaging is obtained via the MSER criterion on the forces signal. According to the MSER, for a timeseries of size $N$ where each sample is expressed by $g_i$, the timestamp $d$ of the timeseries, that signifies the end of the initial transients is defined as

\begin{equation}
   d = min\left[ \frac{1}{(N-d)^2} \sum_{i=d}^{N-1}(g_i - \overline{g}_{N,d}) \right]
\end{equation}

where $\overline{g}_{N,d}$ is the truncated average of the timeseries with starting index $d$ until the end of the timeseries. This non-linear equation is solved iteratively until the timestamp that minimizes the mean square error of the truncated timeseries is found.

\begin{figure}[!ht]
   \centering
      \begin{subfigure}{0.49\textwidth}
      \includegraphics[width=\linewidth]{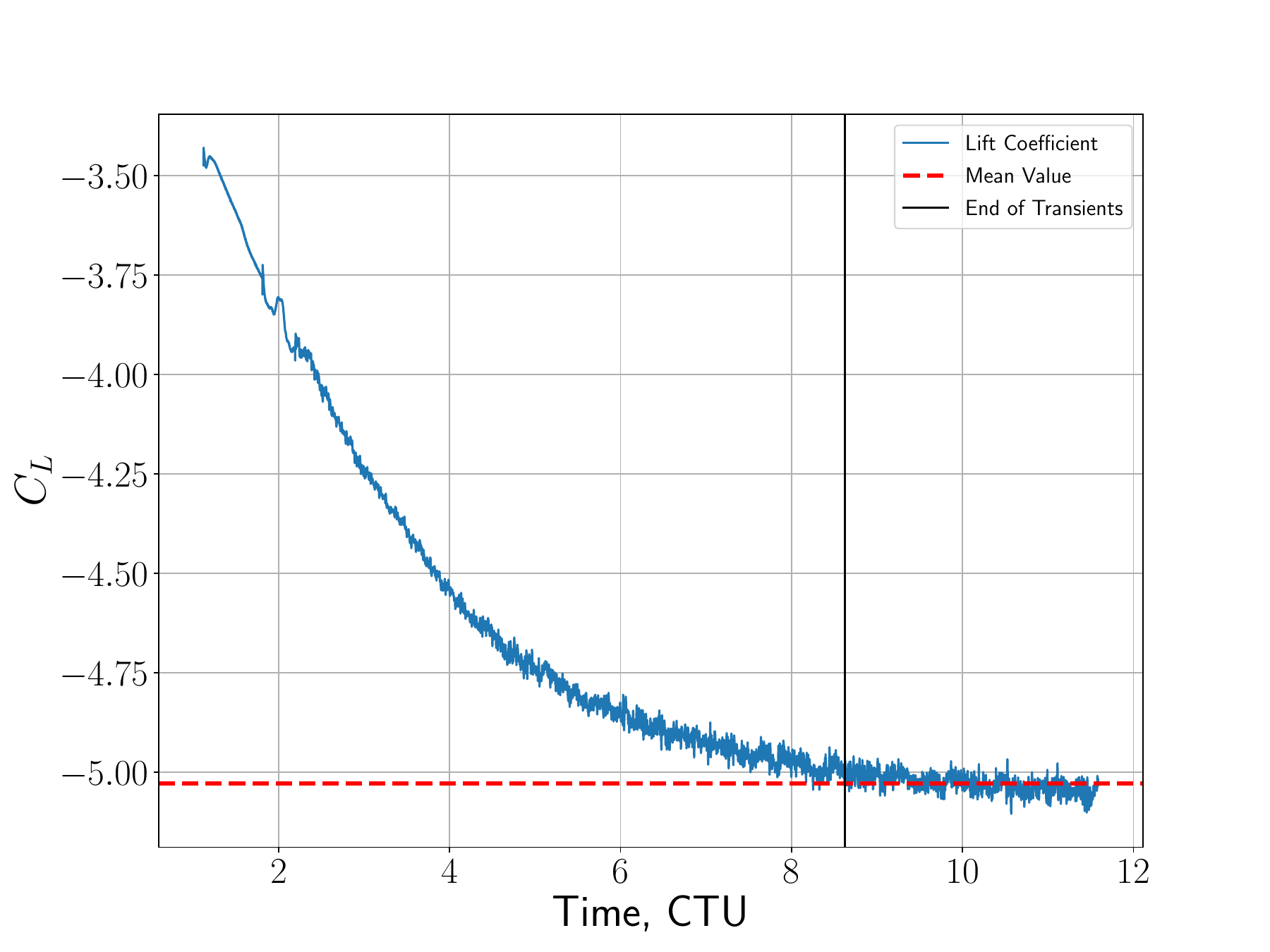}
      \caption{ }
      \end{subfigure}
      \begin{subfigure}{0.49\textwidth}
      \includegraphics[width=\linewidth]{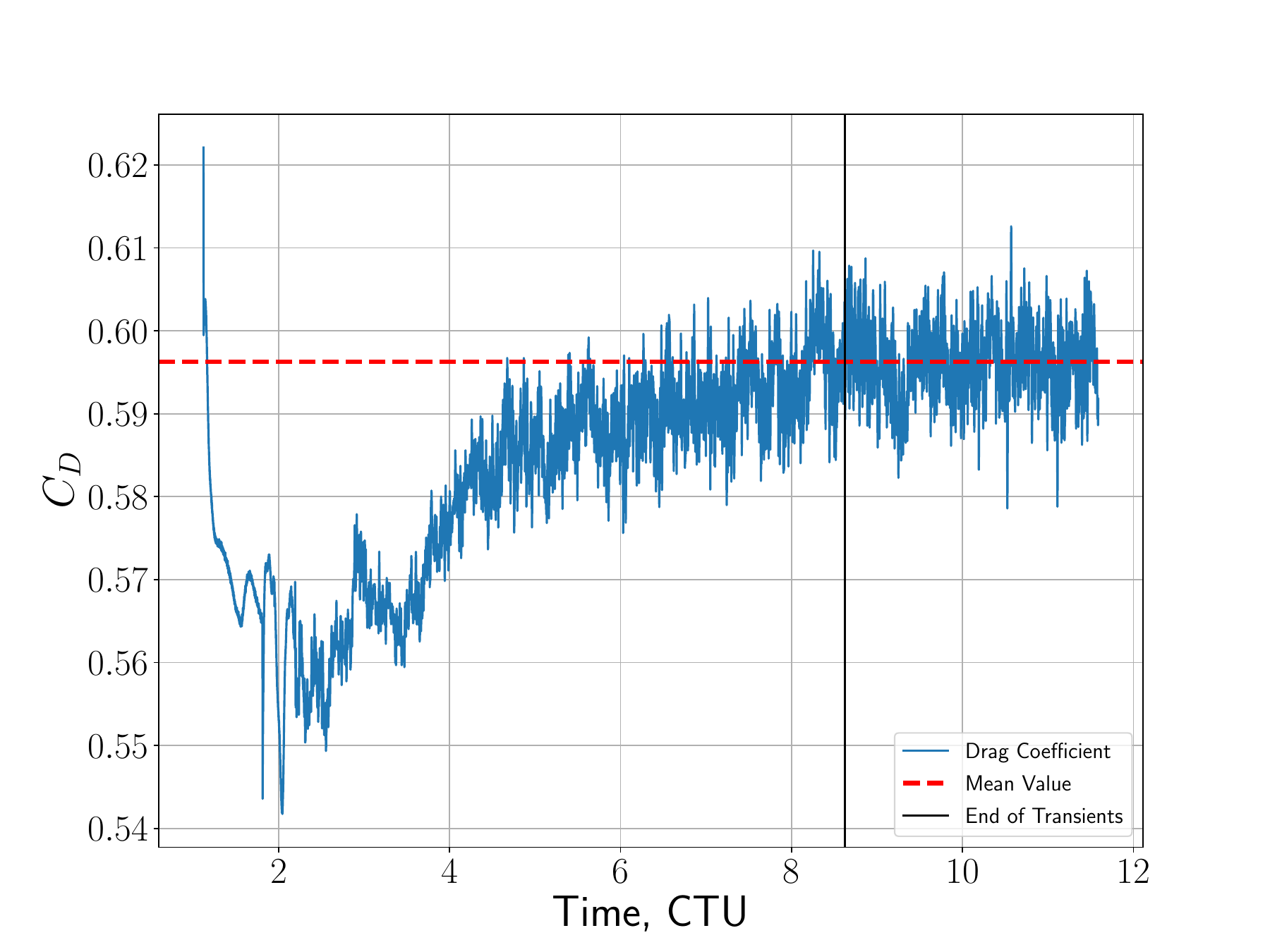}
      \caption{ }
      \end{subfigure}
      \caption{Evolution of the a) lift and b) drag coefficients over time.}
   \label{fig:forcesTraces}
\end{figure}

The temporal evolution of the lift and drag coefficients is presented in figure \ref{fig:forcesTraces}. The onset of the time-averaging window is indicated by the vertical black line in figure, as determined by the MSER criterion, which identifies the point at which the initial transient phase of the flow is complete. Both coefficients exhibit considerable changes in magnitude from the start of the simulation until the fourth CTU, owing to rapid flow development, before their rate of change decreases rapidly as they approach a converged state. In the experiments, data-acquisition begins only once the freestream velocity is fully established around the geometry. The numerical model follows the same principle, with data-acquisition and time-averaging beginning only after the initial transient has passed. Accordingly, the numerical simulation was run for 1 s longer than the experimental campaign, to eliminate the initial flow transient prior to the time-averaging stage.

The simulation was terminated once the standard deviation of the integral quantities fell below 0.5$\%$ of their corresponding mean values, yielding a total averaging duration of 3.5 CTUs. As an additional check, the peak-to-zero amplitude of both integral quantities within this window likewise remained below below 1.5$\%$ of the mean, confirming that the selected window is sufficient for statistical convergence of the global force coefficients.

\begin{figure}[!ht]
   \centering
      \begin{subfigure}{0.4\textwidth}
\includegraphics[width=\linewidth]{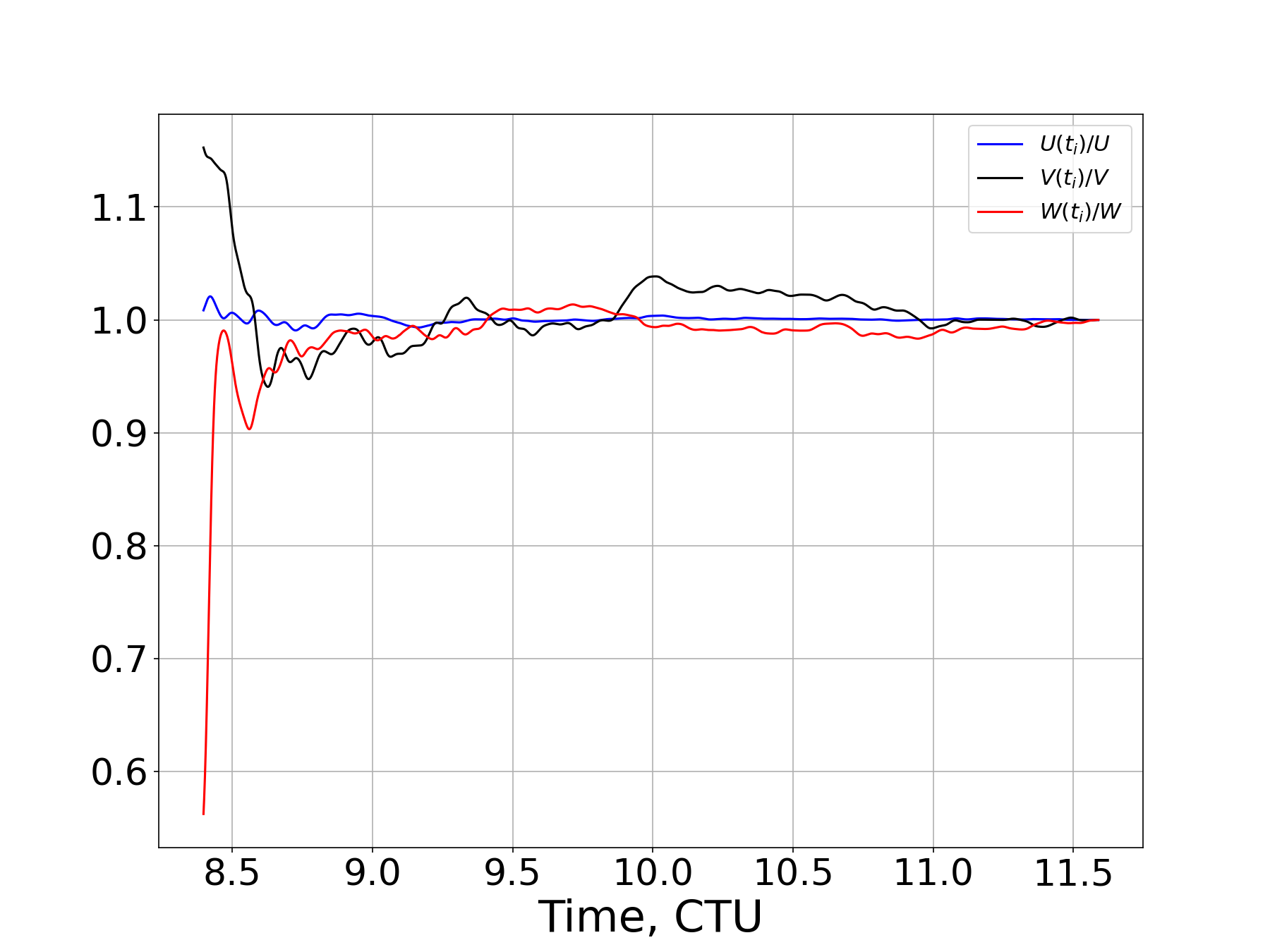}
      \caption{Normalised cumulative average for the velocity components}
      \end{subfigure}
      \begin{subfigure}{0.4\textwidth}
\includegraphics[width=\linewidth]{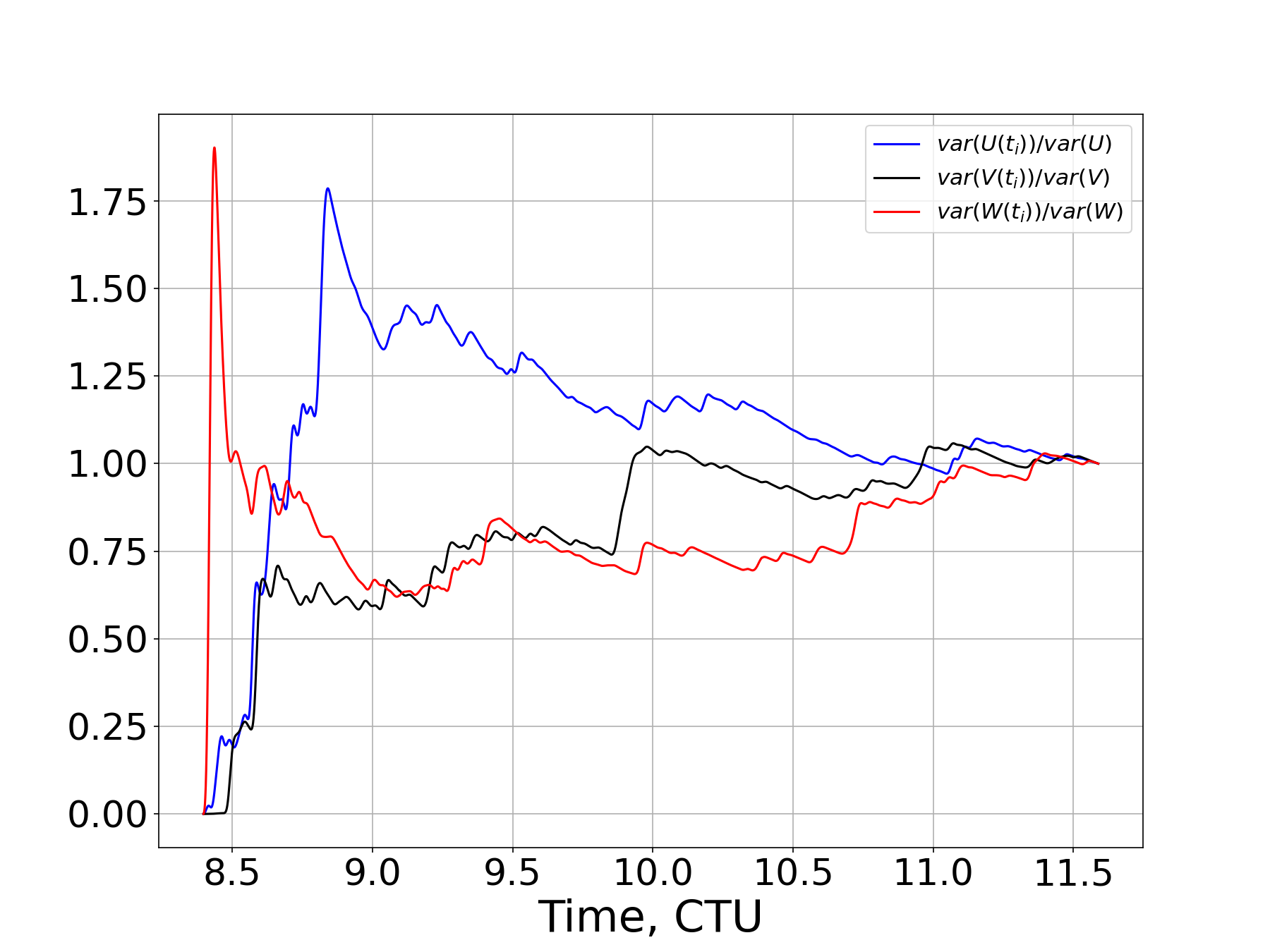}
      \caption{Normalised cumulative variance for the velocity components}
      \end{subfigure}
      \caption{The normalised statistics of the velocity components to assess convergence over time within the selected time-averaging window.}
   \label{fig:simulation_conv}
\end{figure}

The temporal convergence of the velocity components is examined for the specified start of time-averaging and duration in figure \ref{fig:simulation_conv} for a probe point in the wake of the wing. This point is located at $(X, Y, Z) = (0.294, -0.556, 0.112)$ m and the acquisition frequency for the field variables was 10000 Hz. It is representative for the convergence of the large scales behind wing, as it is found in the vicinity of the main vortex. In figure \ref{fig:simulation_conv}, the normalized cumulative average for the velocity components clearly converges rapidly, so the selected time-averaging window is sufficient. This is not the case for the normalised cumulative variance of the velocity components, as they continue to exhibit non-negligible variation until approximately 11 CTU, indicating that full statistical convergence of the second-order moments is not achieved within the present time-averaging window. The present time-averaging window was selected primarily on the basis of the integral coefficients and the mean flow field, rather than achieving statistical convergence of second-order moments, such as variance and Reynolds stresses. From an academic standpoint, this represents a limitation of the present study, from an industrial perspective, however, it reflects a pragmatic compromise between computational cost and solution accuracy.

\begin{figure}[!ht]
    \centering
    \includegraphics[width=0.4\linewidth]{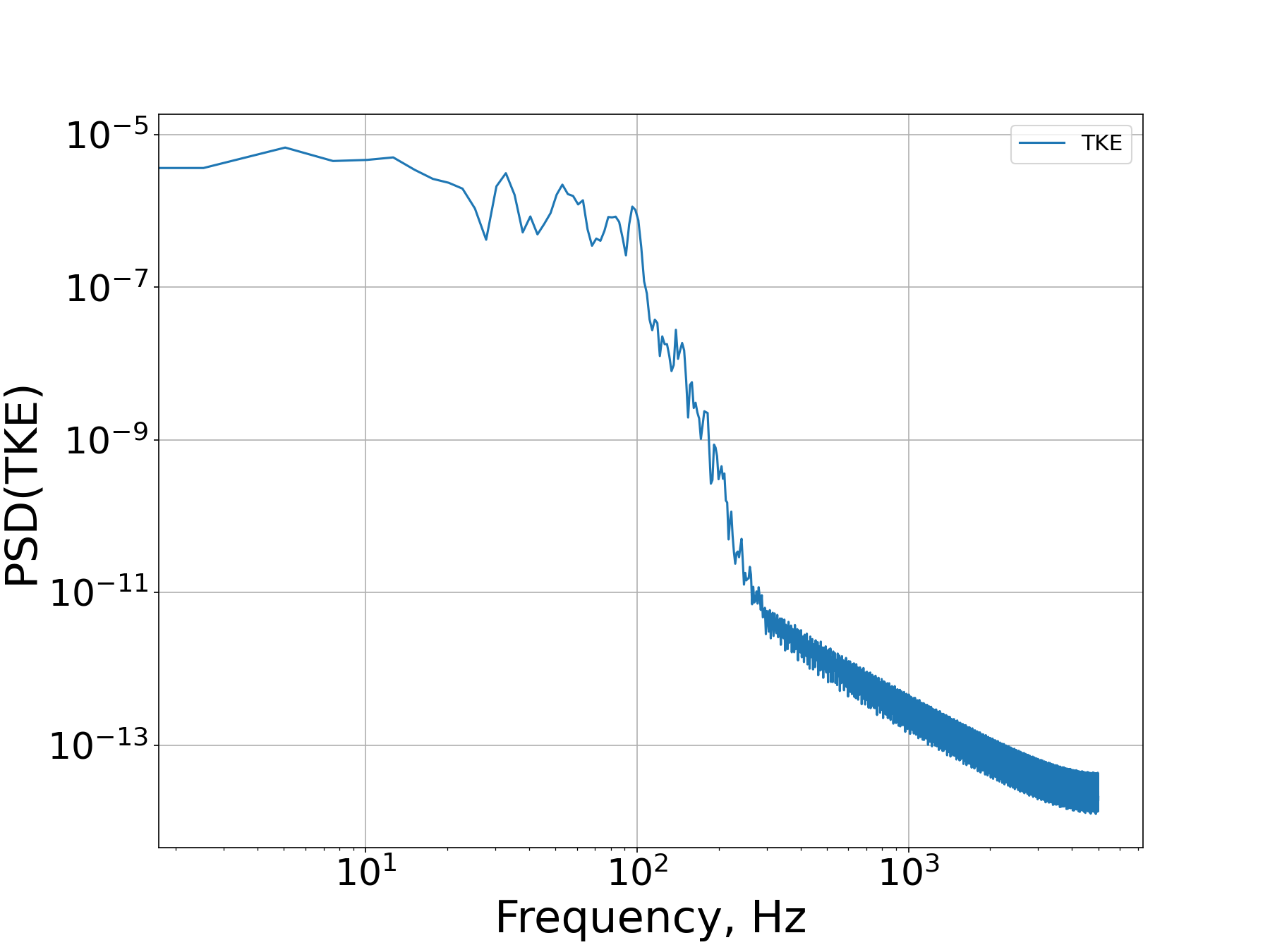}
    \caption{Power Spectral Density (PSD) of the Turbulent Kinetic Energy (TKE)}
    \label{fig:psd_tke}
\end{figure}

Preliminary examination of the turbulent kinetic energy spectrum at the same probe location in figure \ref{fig:psd_tke} shows resolved energy content across 3.5 Hz up to 300 Hz, where each characteristic frequency is resolved with a minimum of three datapoints. While the theoretical minimum resolvable frequency, based on a minimum of three cycles captured within the averaging window, is 3.5 Hz, the statistical reliability of the spectral estimate decreases near this lower limit. For this reason, we consider 5 Hz a more conservative bound for reliably resolved content. A fully converged spectral analysis was beyond the scope of the present study.

\subsection{Simulation: Aerodynamic Load}
The aerodynamic load is computed from the numerical model, assessing its convergence over time and its sensitivity with the Reynolds number, in comparison to the findings from \cite{Buscariolo2022, Liosi2024}.

\begin{table} [hbt!]
    \centering
    \begin{tabular}{cccccc}
    \hline
   Configuration       & $Re_c$     & $C_{L,mean}$ & $C_{D,mean}$ & $C_{L,std}$ & $C_{D,std}$ \\\hline
   Current Simulation  &  74896 & -5.026       & 0.596        &  0.025      & 0.004 \\
   Liosi et al.
   \cite{Liosi2024}    & 208045 & -5.676       & 0.593        &  0.010      & 0.002 \\
   O'Sullivan et al.
   \cite{Sullivan2023} & 440000 & -6.043       & 0.594        &  0.012      & 0.003 \\
    \hline
   \end{tabular}
   \caption{\label{tab:force_stat}Statistics for the Lift \& Drag Coefficients}
\end{table}

\begin{figure}[!ht]
   \centering
      \begin{subfigure}{0.4\textwidth}
      \includegraphics[width=\linewidth]{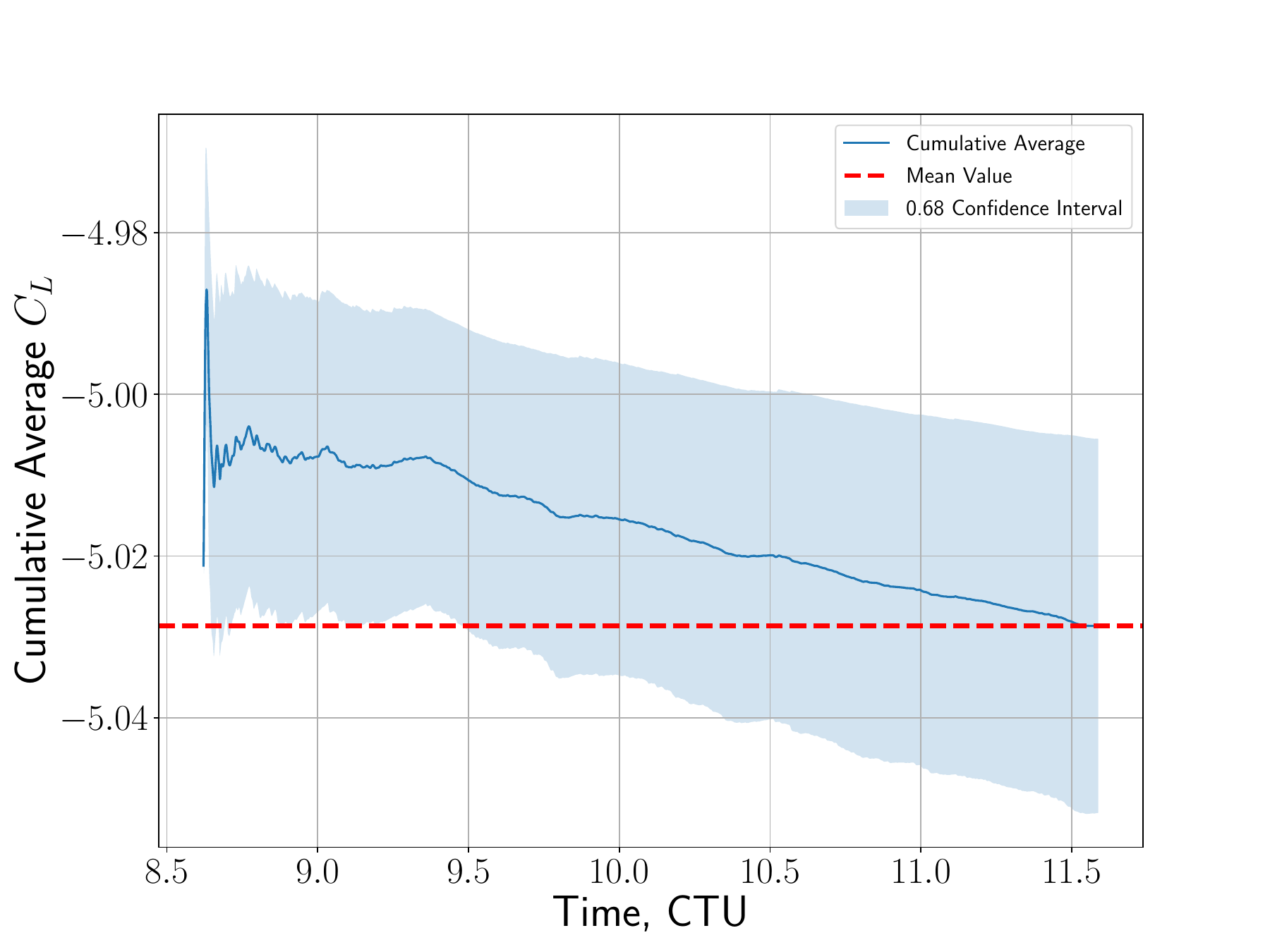}
      \end{subfigure}
      \begin{subfigure}{0.4\textwidth}
      \includegraphics[width=\linewidth]{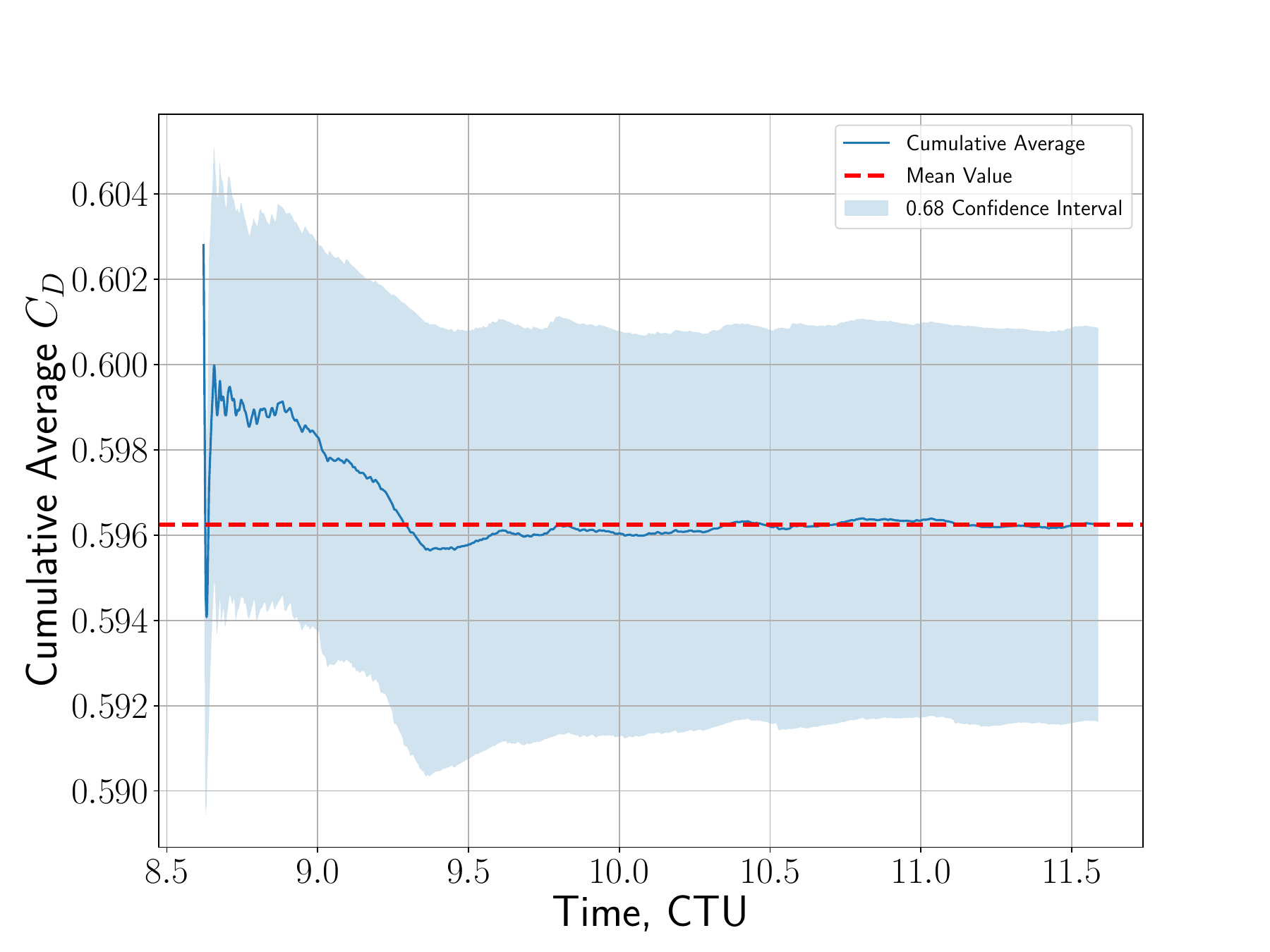}
      \end{subfigure}
      \caption{Cumulative average for the lift(left) and drag(right) coefficients after the end of initial transients}
   \label{fig:cumulativeForces}
\end{figure}

The time-averaged lift and drag coefficients of the numerical model, compared with existing literature at higher $Re$ numbers, are summarised in Table \ref{tab:force_stat}. It is evident that as the Reynolds number increases, the lift coefficient also rises, while the drag coefficient remains relatively constant at similar levels of temporal convergence. The temporal convergence of the integral values is indicated by their standard deviation, which shows strong similarity among the different numerical models. The increase in lift coefficient with higher freestream velocity is expected, as it leads to stronger suction below the wing, which strengthens the ground-effect mechanism responsible for the downforce production from the wing. Consequently, it is anticipated that the vortical system will exhibit stronger vortices at increased Re. The most dominant component of the drag coefficient is induced drag, which remains consistent across this range of $Re$ numbers.

\subsection{Uncertainty quantification for the PTV}
\begin{figure}[ht!]
    \centering
    \begin{subfigure}[b]{0.32\linewidth}
        \centering
    \begin{tikzpicture}
    \node[inner sep=0pt] (img) at (0,0)
    {\includegraphics[width=\linewidth]{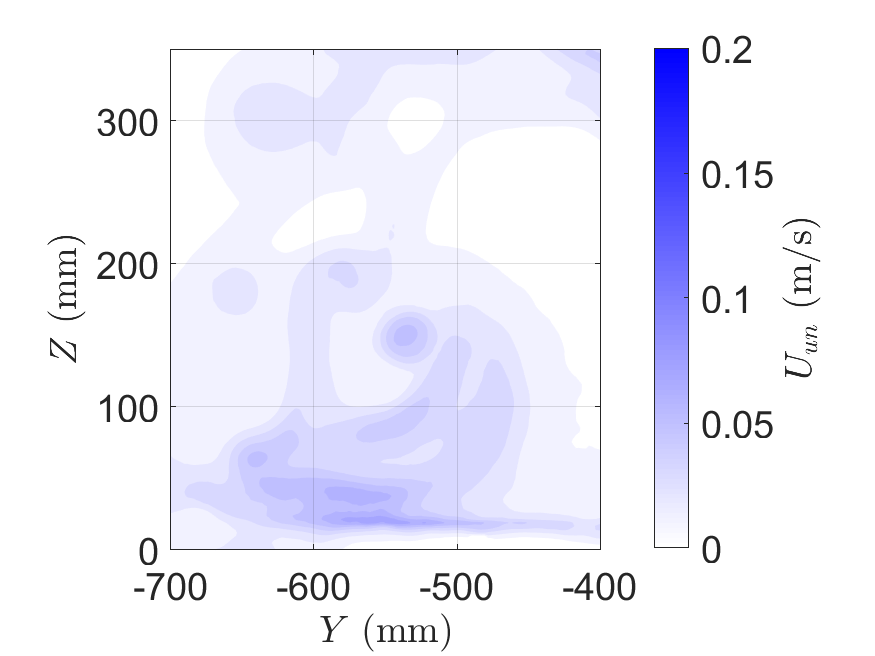}};
\end{tikzpicture}
        \caption{Plane 3: Uncertainty in U}
    \end{subfigure}
    \begin{subfigure}[b]{0.32\linewidth}
        \centering
    \begin{tikzpicture}
    \node[inner sep=0pt] (img) at (0,0)
    {\includegraphics[width=\linewidth]{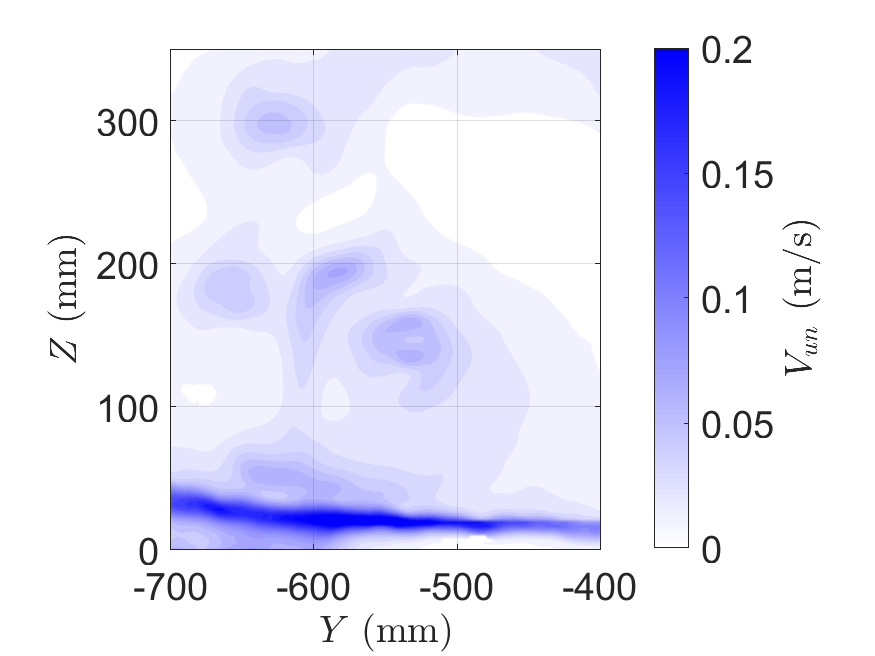}};
    \end{tikzpicture}
        \caption{Plane 3: Uncertainty in V}
    \end{subfigure}    
    \begin{subfigure}[b]{0.32\linewidth}
        \centering
    \begin{tikzpicture}
    \node[inner sep=0pt] (img) at (0,0)
    {\includegraphics[width=\linewidth]{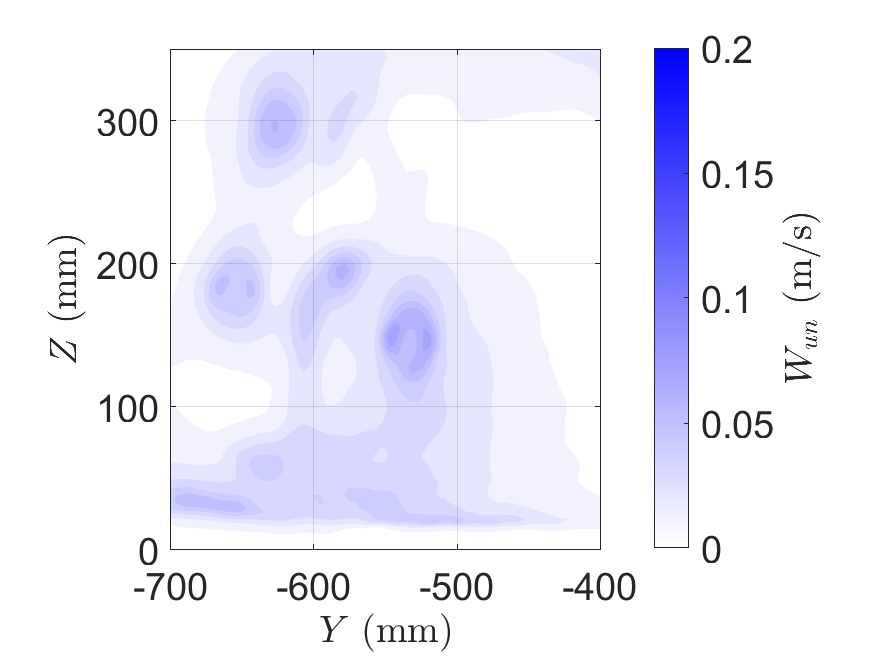}};
\end{tikzpicture}
        \caption{Plane 3: Uncertainty in W}
    \end{subfigure}
    \begin{subfigure}[b]{0.32\linewidth}
        \centering
    \begin{tikzpicture}
    \node[inner sep=0pt] (img) at (0,0)
    {\includegraphics[width=\linewidth]{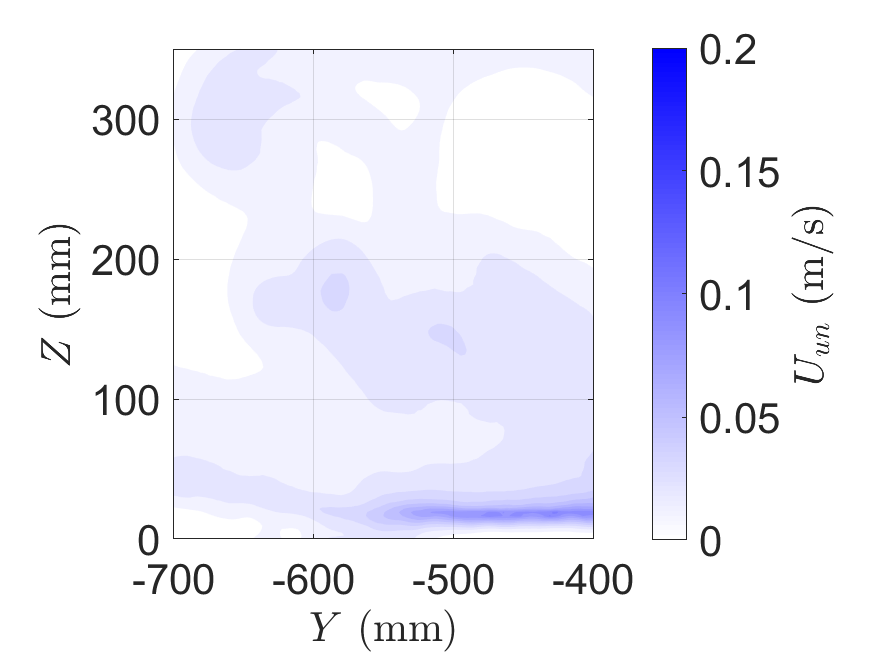}};
\end{tikzpicture}
        \caption{Plane 4 - U uncert.}
    \end{subfigure}
    \begin{subfigure}[b]{0.32\textwidth}
        \centering
    \begin{tikzpicture}
    \node[inner sep=0pt] (img) at (0,0)
    {\includegraphics[width=\linewidth]{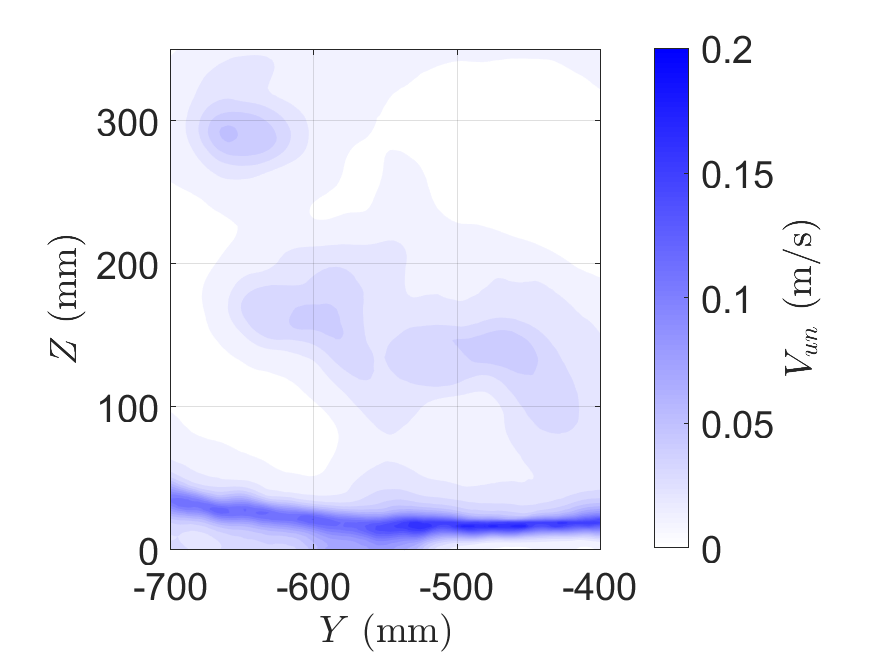}};
\end{tikzpicture}
        \caption{Plane 4 - V uncert.}
    \end{subfigure}    
    \begin{subfigure}[b]{0.32\textwidth}
        \centering
   \begin{tikzpicture}
    \node[inner sep=0pt] (img) at (0,0)
    {\includegraphics[width=\linewidth]{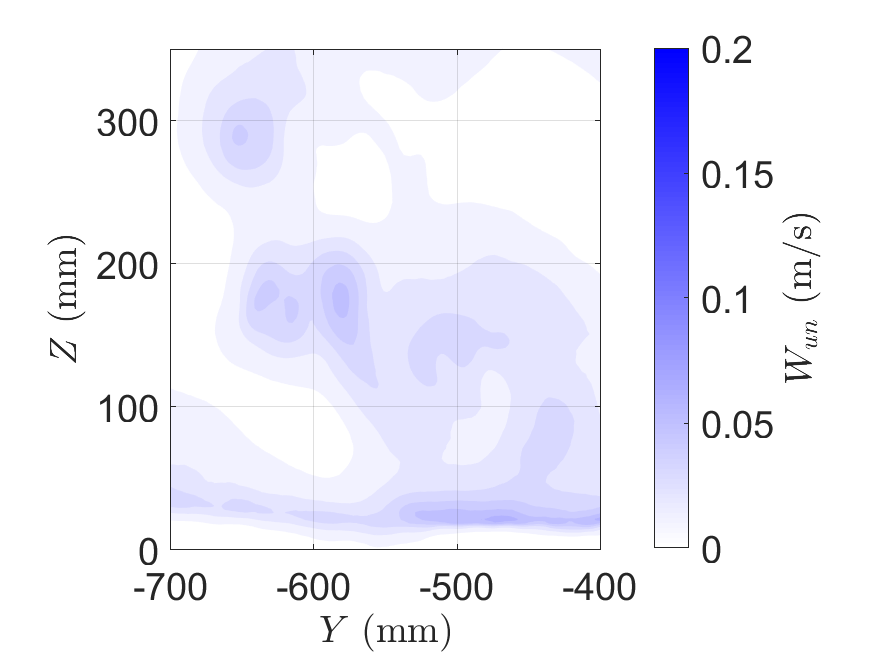}};
\end{tikzpicture}
        \caption{Plane 4 - W uncert. }
    \end{subfigure}
    \begin{subfigure}[b]{0.32\textwidth}
        \centering
    \begin{tikzpicture}
    \node[inner sep=0pt] (img) at (0,0)
    {\includegraphics[width=\linewidth]{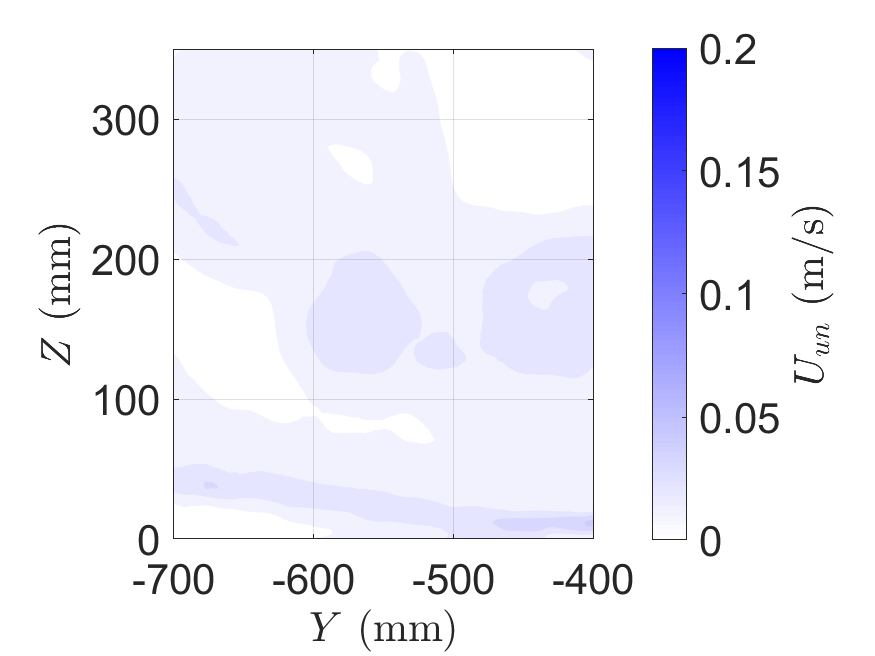}};
\end{tikzpicture}
        \caption{Plane 5 - U uncert. }
    \end{subfigure}
    \begin{subfigure}[b]{0.32\textwidth}
        \centering
    \begin{tikzpicture}
    \node[inner sep=0pt] (img) at (0,0)
    {\includegraphics[width=\linewidth]{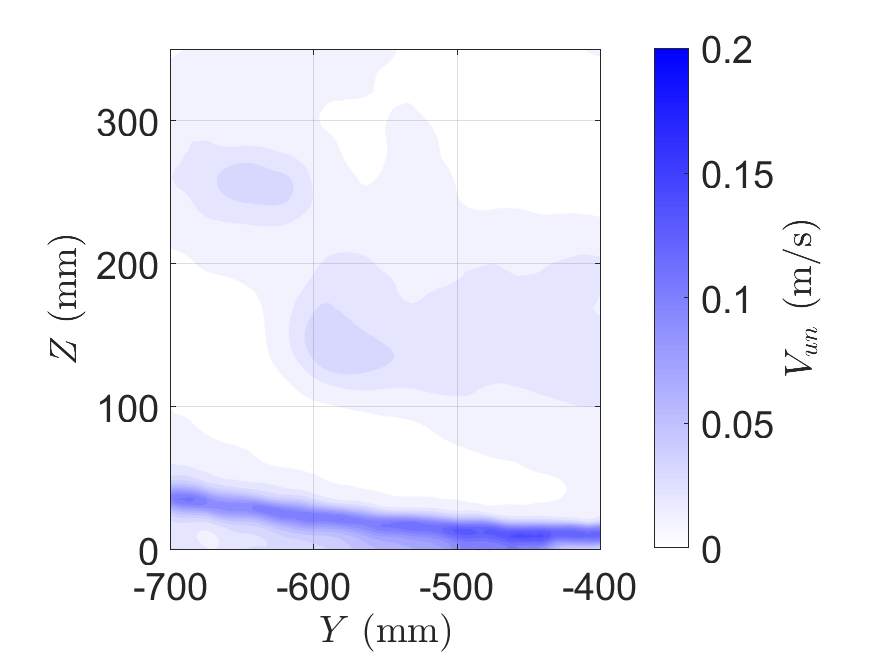}};
\end{tikzpicture}
        \caption{Plane 5 - V uncert. }
    \end{subfigure}    
    \begin{subfigure}[b]{0.32\textwidth}
        \centering
\includegraphics[width=\linewidth]{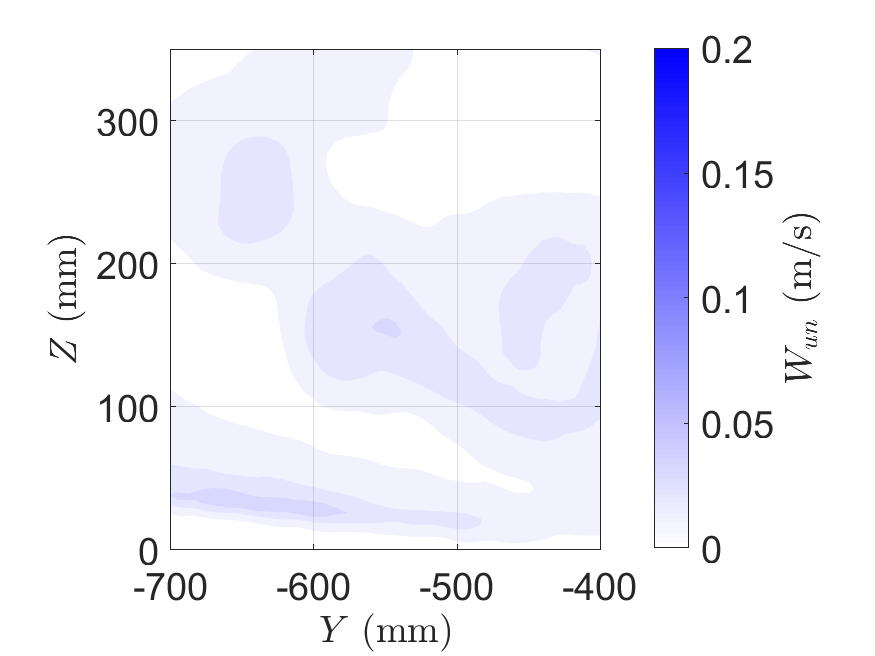}
        \caption{Plane 5 - W uncert. }
    \end{subfigure}
    \caption{Uncertainty in the velocity field from the experiment at Plane 3, 4 and 5.}
    \label{fig:Exp_uncert_Plane3to5}
\end{figure}

Figure \ref{fig:Exp_uncert_Plane3to5} shows the experimental uncertainty for each individual component calculated using Davis 10 based on the linear regression method described in \cite{janke2021uncertainty}.

\subsection{Numerical model complementary figures}
\begin{figure}[ht!]
\centering
\begin{subfigure}[t]{0.35\textwidth}
\begin{tikzpicture}
    \node[inner sep=0pt] (img) at (0,0) {\includegraphics[width=\textwidth]{{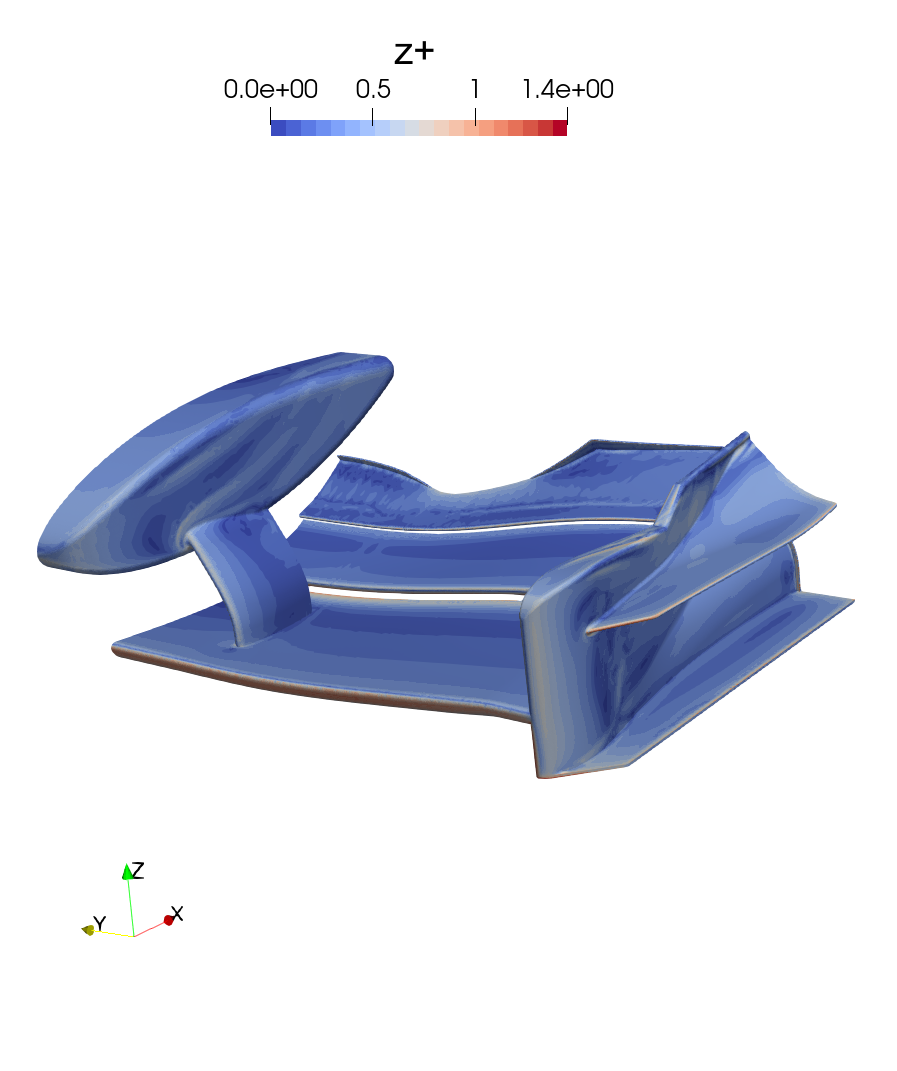}}};
\end{tikzpicture}
\caption{}
\end{subfigure}
\begin{subfigure}[t]{0.35\textwidth}
\begin{tikzpicture}
    \node[inner sep=0pt] (img) at (0,0) {\includegraphics[width=\textwidth]{{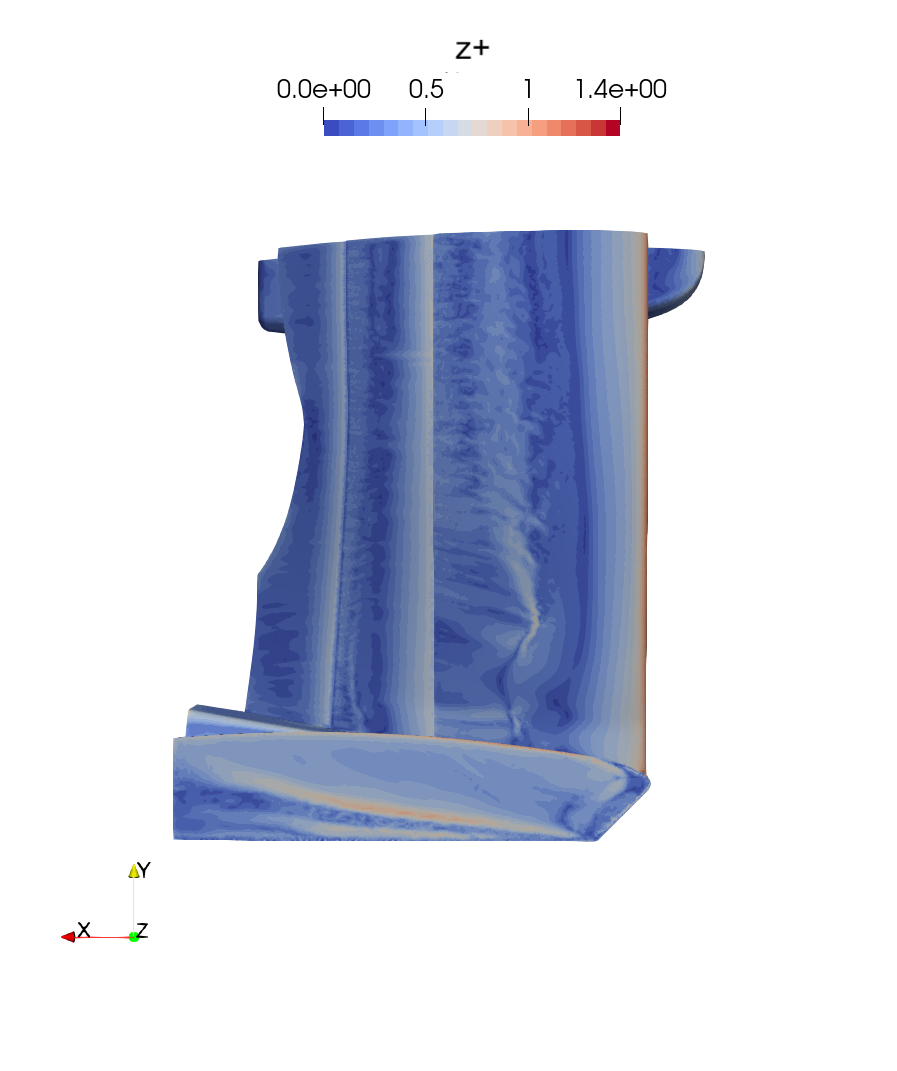}}};
\end{tikzpicture}
\caption{}
\end{subfigure}
\caption{\small $z^{+}$ values at the surface of the IFW: a) 3D view, b) Bottom view.}
\label{fig:yplus}
\end{figure}

Figure \ref{fig:yplus} shows the first value of $z^+$ at the surface on the IFW for the simulation.

\newpage
\ack{We would like to thank Lavision and the 10x5 team for the support during the experiment.}

\funding{
This project received funding from the European Union’s Horizon 2020 research and innovation programme under the Marie Skłodowska-Curie grant agreement No 955923.
The authors would like to thank EPSRC for the computational time made available on the UK supercomputing facility ARCHER/ARCHER2 (https://www.archer2.ac.uk) via the UK Turbulence Consortium (EP/X035484/1). Post-processing activities of this project leveraged the high-performance computing facility through the Imperial College Research Computing Service (doi: 10.14469/hpc/2232).
The experimental work was supported by the EPSRC Network grant (EPSRC, EP/X012069) and the Royal Society Industry Fellowship (Grant IF/R1/231049).}

\roles{Isabella Fumarola - experiment methodology and execution, data processing, simulation to experiment data comparison, writing and editing. Alexandra I. Liosi - simulation methodology, software implementation, data comparison, writing and reviewing. Parv Khurana - simulation methodology, software implementation, data comparison, editing, and reviewing. Adam Meziane - experimental data analysis. Isaac Balbolia - experiment methodology and execution, preliminary data analysis. Sherwin J. Spencer - simulation coordinator, concept development, software implementation, reviewing. Jonathan F. Morrison - experiment coordinator, reviewing.}

\data{The data will be available on the NWTF experimental database \cite{NWTF_database}.}


\bibliographystyle{plain}
\bibliography{Biblio.bib}

\end{document}